\pdfoutput=1
\documentclass[prl,reprint,notitlepage,noeprint]{revtex4-2}

\usepackage[english]{babel}
\usepackage{amsmath,amssymb,amsfonts,bbm,xcolor,graphicx,comment,txfonts}
\definecolor{darkblue}{rgb}{0.,0.,0.5}
\usepackage[colorlinks,linkcolor=darkblue,citecolor=darkblue,urlcolor=darkblue]{hyperref}

\newcommand{\hc}{\mathrm{H.c.}}
\newcommand{\id}{\mathbbm{1}}

\newcommand{\e}{e}
\newcommand{\diff}{d}
\newcommand{\imag}{i}
\newcommand{\ket}[1]{\lvert #1 \rangle}
\newcommand{\bra}[1]{\langle #1 \rvert}

\makeatletter
\newcommand*{\transpose}{%
  {\mathpalette\@transpose{}}%
}
\newcommand*{\@transpose}[2]{%
  \raisebox{\depth}{$\m@th#1\intercal$}%
}
\makeatother

\newcommand{\abs}[1]{\left\lvert #1 \right\rvert}

\newcommand{\tr}{\mathop{\mathrm{tr}}}

\newcommand{\atanh}{\mathop{\mathrm{arctanh}}}
\newcommand{\Pf}{\mathop{\mathrm{pf}}}
\renewcommand{\Re}{\mathop{\mathrm{Re}}}
\renewcommand{\Im}{\mathop{\mathrm{Im}}}

\makeatletter
\DeclareFontFamily{OMX}{MnSymbolE}{}
\DeclareSymbolFont{MnLargeSymbols}{OMX}{MnSymbolE}{m}{n}
\SetSymbolFont{MnLargeSymbols}{bold}{OMX}{MnSymbolE}{b}{n}
\DeclareFontShape{OMX}{MnSymbolE}{m}{n}{
    <-6>  MnSymbolE5
   <6-7>  MnSymbolE6
   <7-8>  MnSymbolE7
   <8-9>  MnSymbolE8
   <9-10> MnSymbolE9
  <10-12> MnSymbolE10
  <12->   MnSymbolE12
}{}
\DeclareFontShape{OMX}{MnSymbolE}{b}{n}{
    <-6>  MnSymbolE-Bold5
   <6-7>  MnSymbolE-Bold6
   <7-8>  MnSymbolE-Bold7
   <8-9>  MnSymbolE-Bold8
   <9-10> MnSymbolE-Bold9
  <10-12> MnSymbolE-Bold10
  <12->   MnSymbolE-Bold12
}{}

\let\llangle\@undefined
\let\rrangle\@undefined
\DeclareMathDelimiter{\llangle}{\mathopen}%
                     {MnLargeSymbols}{'164}{MnLargeSymbols}{'164}
\DeclareMathDelimiter{\rrangle}{\mathclose}%
                     {MnLargeSymbols}{'171}{MnLargeSymbols}{'171}
\makeatother

\usepackage{verbatim}

\begin{document}


\title{Relaxation to a Parity-Time Symmetric Generalized Gibbs Ensemble \\ after
  a Quantum Quench in a Driven-Dissipative Kitaev Chain}

\author{Elias Starchl}

\author{Lukas M. Sieberer}

\email{lukas.sieberer@uibk.ac.at}

\affiliation{Institute for Theoretical Physics, University of Innsbruck, 6020
  Innsbruck, Austria}

\begin{abstract}
  The construction of the generalized Gibbs ensemble, to which isolated
  integrable quantum many-body systems relax after a quantum quench, is based
  upon the principle of maximum entropy. In contrast, there are no universal and
  model-independent laws that govern the relaxation dynamics and stationary
  states of open quantum systems, which are subjected to Markovian drive and
  dissipation. Yet, as we show, relaxation of driven-dissipative systems after a
  quantum quench can, in fact, be determined by a maximum entropy ensemble, if
  the Liouvillian that generates the dynamics of the system has parity-time
  symmetry. Focusing on the specific example of a driven-dissipative Kitaev
  chain, we show that, similarly to isolated integrable systems, the approach to
  a parity-time symmetric generalized Gibbs ensemble becomes manifest in the
  relaxation of local observables and the dynamics of subsystem entropies. In
  contrast, the directional pumping of fermion parity, which is induced by
  nontrivial non-Hermitian topology of the Kitaev chain, represents a phenomenon
  that is unique to relaxation dynamics in driven-dissipative systems. Upon
  increasing the strength of dissipation, parity-time symmetry is broken at a
  finite critical value, which thus constitutes a sharp dynamical transition
  that delimits the applicability of the principle of maximum entropy. We show
  that these results, which we obtain for the specific example of the Kitaev
  chain, apply to broad classes of noninteracting fermionic models, and we
  discuss their generalization to a noninteracting bosonic model and an
  interacting spin chain.
\end{abstract}

\maketitle


\paragraph*{Introduction.}

After a quench, generic isolated quantum many-body systems relax locally to a
state that is determined, according to the fundamental postulates of statistical
mechanics~\cite{Schwabl2006}, by maximization of entropy, subject to the
constraints imposed by integrals of motion~\cite{Polkovnikov2011, Eisert2015,
  DAlessio2016}.  For integrable systems, which are characterized by an
extensive number of integrals of motion, the resultant equilibrium state is the
generalized Gibbs ensemble (GGE)~\cite{Rigol2007, Vidmar2016, Essler2016}.  The
principle of maximum entropy and, consequently, the structure of the GGE are
universal in the sense that model-dependent details affect only the specific
form of the integrals of motion and the numerical coefficients that enter the
GGE as Lagrange multipliers, and that are determined by the initial
state. Likewise, relaxation to the GGE is characterized by a set of universal
characteristic traits such as light-cone spreading of
correlations~\cite{Essler2016, Lieb1972, Calabrese2006} and linear growth and
volume-law saturation of the entropy of a finite subsystem~\cite{Calabrese2005,
  Alba2017, Alba2018, Calabrese2020}. In contrast to this scenario of
generalized thermalization of isolated integrable systems, open systems, which
are subjected to Markovian drive and dissipation~\cite{Breuer2002}, typically
evolve toward nonequilibrium steady states that are determined by the interplay
of internal Hamiltonian dynamics and the coupling to external reservoirs, and
that are, therefore, highly model-dependent~\cite{Diehl2008, Verstraete2009,
  Eisert2010, Sieberer2016a, Maghrebi2016a, Jin2016, Rota2017, Halati2022}. In
particular, the breaking of conservation laws due to the coupling to external
reservoirs entails the eventual loss of any memory of the initial state. That
is, the constraints that determine the GGE in isolated systems are lifted, and,
consequently, the notion of a maximum entropy ensemble appears to be rendered
meaningless. Therefore, the existence of model-independent principles that
govern the relaxation dynamics and stationary states of open systems is
seemingly ruled out.

How are these drastically different paradigms of relaxation in isolated and
driven-dissipative systems connected when $\gamma$, the strength of the coupling
to external reservoirs, is gradually diminished? For any $\gamma > 0$, an open
system eventually reaches a stationary state that is vitally determined by the
coupling to external reservoirs and takes the form of a GGE only in the limit
$\gamma \to 0$~\cite{Lange2017, Lange2018, Lenarcic2018, Reiter2021}. In this
Letter, however, we show that the universal principles that govern generalized
thermalization after a quantum quench in isolated systems can retain their
validity---in suitably generalized form---even for finite values of $\gamma$
that are comparable to characteristic energy scales of the system
Hamiltonian. This robustness is caused by parity-time (PT) symmetry of the
Liouvillian~\cite{Prosen2012, *Prosen2012a, Medvedyeva2016, VanCaspel2018,
  Minganti2019, Shibata2020, Huber2020, Huybrechts2020, Arkhipov2020,
  Curtis2021, Roccati2021, Roccati2022, Claeys2022, Nakanishi2022} that
generates the dynamics of the system. Focusing on the specific example of a
driven-dissipative generalization~\cite{VanCaspel2019, Lieu2020, Sayyad2021} of
the Kitaev chain~\cite{Kitaev2001}, we find that in the PT-symmetric phase, the
quadratic eigenmodes of the adjoint Liouvillian oscillate at different
frequencies, but crucially they all decay with the same rate. Consequently,
after factoring out exponential decay, dephasing~\cite{Essler2016, Barthel2008}
leads to local relaxation to a PT-symmetric GGE (PTGGE). In analogy to the GGE
for isolated noninteracting fermionic many-body systems and interacting
integrable systems that can be mapped to noninteracting
fermions~\cite{Rigol2007, Vidmar2016, Essler2016, Calabrese2011, Calabrese2012I,
  Calabrese2012II}, we specify the PTGGE in terms of eigenmodes of the generator
of the postquench dynamics. However, the PTGGE generalizes the GGE to account
for noncanonical statistics of these eigenmodes, and the nonconservation of the
associated mode occupation numbers renders the PTGGE intrinsically
time-dependent. We illustrate relaxation to the PTGGE in terms of the fermion
parity and the entropy of a finite subsystem. Thereby, we reveal the directional
pumping of fermion parity, which occurs for quenches from the topologically
trivial phase of the isolated Kitaev chain to the non-Hermitian topological
phase of the driven-dissipative Kitaev chain~\cite{Sayyad2021}, as a phenomenon
that is unique to driven-dissipative systems, and we establish the validity of a
dissipative quasiparticle picture~\cite{Alba2021, Carollo2022, Alba2022} for
values of $\gamma$ up to the sharply defined boundary of the PT-symmetric
phase. Going beyond the example of the Kitaev chain, we show that our results
apply to broad classes of noninteracting fermionic models and, in suitably
generalized form, also to models of noninteracting bosons and interacting spins.

\paragraph*{Model.}

We consider a Kitaev chain~\cite{Kitaev2001} of length $L$ with hopping matrix
element $J$, pairing amplitude $\Delta$, and chemical potential $\mu$, as
described by the Hamiltonian
\begin{equation}
  \label{eq:H-Kitaev}
  H = \sum_{l = 1}^L \left( -J c^{\dagger}_l c_{l + 1} + \Delta c_l c_{l + 1} + \hc
  \right) - \mu \sum_{l = 1}^L \left( c^{\dagger}_l c_l -\frac{1}{2} \right).
\end{equation}
The operators $c_l$ and $c_l^{\dagger}$ annihilate and create, respectively, a
fermion on lattice site $l$. Unless stated otherwise, we assume periodic
boundary conditions with $c_{L + 1} = c_1$. The system is prepared in the ground
state $\ket{\psi_0}$ for $J = \Delta$ and $\mu_0$. We focus on the limit
$\mu_0 \to - \infty$, such that the initial state is the topologically trivial
vacuum state, $\ket{\psi_0} = \ket{\Omega}$ with $c_l \ket{\Omega} = 0$, but our
results are not affected qualitatively by this choice. At $t = 0$, the chemical
potential is quenched to a finite value $\mu$, while $J = \Delta$ is kept
fixed. At the same time, the system is coupled to Markovian
reservoirs. Consequently, the postquench dynamics is described by a quantum
master equation for the system density matrix $\rho$~\cite{Gorini1976,
  Lindblad1976},
\begin{equation}
  \label{eq:master-equation}
  \imag \frac{\diff}{\diff t} \rho = \mathcal{L} \rho = [H, \rho] + \imag
  \sum_{l = 1}^L \left( 2 L_l \rho L_l^{\dagger} - \{ L_l^{\dagger} L_l, \rho \}
  \right),
\end{equation}
where we choose
$L_l = \sqrt{\gamma_l} c_l + \sqrt{\gamma_g}
c_l^{\dagger}$
as a coherent superposition of loss and gain at rates $\gamma_l$ and
$\gamma_g$, respectively~\cite{VanCaspel2019, Lieu2020,
  Sayyad2021}. The mean rate
$\gamma = (\gamma_l + \gamma_g)/2$ measures the overall
strength of dissipation, whereas the relative rate
$\delta = \gamma_l - \gamma_g$ is akin to an inverse
temperature: For $\delta = 0$, the system evolves for $t \to \infty$ toward a
steady state $\rho_{\mathrm{SS}}$ with infinite temperature,
$\rho_{\mathrm{SS}} = \rho_{\infty} = 1/2^L$~\cite{Sayyad2021}; In contrast, for
$\delta \to \infty$, the steady state is pure,
$\rho_{\mathrm{SS}} = \ket{\Omega} \bra{\Omega}$.

Since the initial state $\rho_0 = \ket{\psi_0} \bra{\psi_0}$ is Gaussian and the
Liouvillian $\mathcal{L}$ is quadratic and, therefore, preserves Gaussianity,
the time-evolved state $\rho(t) = \e^{\imag \mathcal{L} t} \rho_0$ is fully
determined by the covariance matrix
\begin{equation}
  \label{eq:g-l}
  g_{l - l'}(t) =
  \begin{pmatrix}
    \langle [c_l, c_{l'}^{\dagger}](t) \rangle & \langle [c_l, c_{l'}](t)
    \rangle \\ \langle [c_{l'}^{\dagger}, c_l^{\dagger}](t) \rangle & \langle
    [c_{l'}^{\dagger}, c_l](t) \rangle
  \end{pmatrix},
\end{equation}
where $\langle \cdots (t) \rangle = \tr( \cdots \rho(t))$. The Fourier transform
$g_k = -\imag \sum_{l = 1}^L \e^{-\imag k l} g_l$ obeys the equation of motion
$\diff g_k/\diff t = - \imag z_k g_k + \imag g_k z_k^{\dagger} - s_k$,
where $z_k$ and $s_k$ can be expressed in terms of Pauli matrices as
$z_k = - \imag 2 \gamma \id - 2 \sqrt{\gamma_l \gamma_g}
\sigma_x + 2 \Delta \sin(k) \sigma_y - \left( 2 J \cos(k) + \mu \right)
\sigma_z$
and $s_k = - 2 \delta \sigma_z$~\cite{supplemental_material}. For
$\gamma \to 0$, $z_k$ reduces to the Bogoliubov-de Gennes Hamiltonian of the
isolated Kitaev chain~\cite{Chiu2016}, and it has inversion symmetry,
$z_k = \sigma_z z_{-k} \sigma_z$, and time-reversal symmetry,
$z_k = z_{-k}^{*}$. These symmetries are broken when $\gamma > 0$. However, the
Liouvillian still has PT symmetry in the sense that the traceless part of $z_k$,
given by $z_k' = z_k + \imag 2 \gamma \id$, is symmetric under the combined
operation of inversion and time-reversal,
$z_k' = \sigma_z z_k^{\prime *} \sigma_z$. PT symmetry implies that there are
two types of eigenvectors and associated eigenvalues $\lambda_{\pm, k}$ of
$z_k$~\cite{supplemental_material}: PT-symmetric eigenvectors, which come in
pairs with eigenvalues $\Re(\lambda_{+, k}) = - \Re(\lambda_{-, k})$ and
$\Im(\lambda_{\pm, k}) = - 2 \gamma$; and PT-breaking eigenvectors, for which
$\Re(\lambda_{\pm, k}) = 0$ and
$\Im(\lambda_{+, k} + \imag 2 \gamma) = - \Im(\lambda_{-, k} + \imag 2 \gamma)$.
The PT-symmetric phase is defined by the eigenvectors of $z_k$ being
PT-symmetric for all momenta $k$, which is the case for
$2 \sqrt{\gamma_l \gamma_g} < \abs{2 J - \abs{\mu}}$.
Then, the eigenvalues of $z_k$ are given by
$\lambda_{\pm, k} = - \imag 2 \gamma \pm \omega_k$ with
$\omega_k^2 = \varepsilon_k^2 - 4 \gamma_l \gamma_g$ and
$\varepsilon_k^2 = \left( 2 J \cos(k) + \mu \right)^2 + 4 \Delta^2 \sin(k)^2$.
For strong dissipation with
$2 \sqrt{\gamma_l \gamma_g} > 2 J + \abs{\mu}$, all
eigenvectors are PT-breaking. Finally, in the PT-mixed phase at intermediate
dissipation, eigenvectors of both types exist.

\paragraph*{PT-symmetric GGE.}

We now focus on relaxation dynamics after a quench to the PT-symmetric phase,
which is best described in terms of the eigenmodes of the adjoint
Liouvillian~\cite{supplemental_material}. With the matrix $V_k$ that
diagonalizes $z_k$, these modes are given by
\begin{equation}
  \label{eq:d-k}  
  \begin{pmatrix}
    d_k \\ d_{-k}^{\dagger}
  \end{pmatrix}
  = V_k^{\dagger}
  \begin{pmatrix}
    c_k \\ c_{-k}^{\dagger}
  \end{pmatrix},
  \quad V_k =
  \begin{pmatrix}
    \cos \! \left( \frac{\theta_k + \phi_k}{2} \right) & \imag \sin \! \left(
      \frac{\theta_k - \phi_k}{2} \right) \\ \imag \sin \! \left( \frac{\theta_k
        + \phi_k}{2} \right) & \cos \! \left( \frac{\theta_k - \phi_k}{2}
    \right)
  \end{pmatrix},
\end{equation}
where $c_k = \frac{1}{\sqrt{L}} \sum_{l = 1}^L \e^{-\imag k l} c_l$,
$\tan(\theta_k) = - 2 \Delta \sin(k)/(2 J \cos(k) + \mu)$, and
$\tan(\phi_k) = 2 \sqrt{\gamma_l \gamma_g}/\omega_k$. For
$\gamma = 0$, $V_k$ reduces to the usual unitary Bogoliubov transformation. When
$\gamma > 0$, non-unitarity of $V_k$ is reflected in the statistics of the modes
$d_k$ as expressed through their anticommutation relations:
\begin{equation}
  \label{eq:d-k-statistics}  
  \begin{pmatrix}
    \{ d_k, d_{k'}^{\dagger} \} & \{ d_k, d_{-k'} \} \\ \{ d_{-k}^{\dagger},
    d_{k'}^{\dagger} \} & \{ d_{-k}^{\dagger}, d_{-k'} \}
  \end{pmatrix}
  = f_k \delta_{k, k'},
\end{equation}
where
$f_k = V_k^{\dagger} V_k = \id + 2 \sqrt{\gamma_l
  \gamma_g} \sigma_y/\varepsilon_k$.
To discuss the dynamics of the modes $d_k$, we consider their
commutators. Expectation values of normal commutators evolve
as~\cite{supplemental_material}
\begin{equation}
  \label{eq:d-k-normal-commutator}
  \langle [d_k, d_k^{\dagger}](t) \rangle = \e^{- 4 \gamma t} \langle [d_k,
  d_k^{\dagger}] \rangle_0 + \left( 1 - \e^{- 4 \gamma t} \right) \langle [d_k,
  d_k^{\dagger}] \rangle_{\mathrm{SS}},
\end{equation}
where $\langle \cdots \rangle_0 = \tr( \cdots \rho_0)$ and
$\langle \cdots \rangle_{\mathrm{SS}} = \tr( \cdots \rho_{\mathrm{SS}})$ denote
expectation values in the initial and steady state, respectively. For anomalous
commutators we find
\begin{multline}
  \label{eq:d-k-anomalous-commutator}
  \langle [d_k, d_{-k}](t) \rangle = \e^{- \imag 2 \left( \omega_k - \imag 2
      \gamma \right) t} \langle [d_k, d_{-k}] \rangle_0 \\ + \left( 1 - \e^{-
      \imag 2 \left( \omega_k - \imag 2 \gamma \right) t} \right) \langle [d_k,
  d_{-k}] \rangle_{\mathrm{SS}}.
\end{multline}
We first consider the case of balanced loss and gain, $\delta = 0$. Then,
heating to infinite temperature is reflected in the exponential decay and
vanishing in the steady state of the expectation values of both normal and
anomalous commutators. Crucially, in the PT-symmetric phase, the decay rate is
identical for all momentum modes. Thus, after factoring out exponential decay,
the system relaxes locally to a maximum entropy ensemble through dephasing of
modes with $\omega_k \neq \omega_{k'}$~\cite{Essler2016, Barthel2008}. Since the
decay of normal commutators is nonoscillatory, dephasing affects only anomalous
commutators. Therefore, we define the PTGGE as the maximum entropy
ensemble~\cite{Jaynes1957, *Jaynes1957a} that is compatible with the statistics
given in Eq.~\eqref{eq:d-k-statistics}, and the nondephasing expectation values
of normal commutators collected in the diagonal matrix
$\zeta_k(t) = \e^{- 4 \gamma t} \mathop{\mathrm{diag}}(\langle [d_k,
d_k^{\dagger}] \rangle_0, \langle [d_{-k}^{\dagger}, d_{-k}] \rangle_0)$.
We find, in terms of spinors
$D_k = \left( d_k, d_{-k}^{\dagger}
\right)^{\transpose}$~\cite{supplemental_material},
\begin{equation}
  \label{eq:rho-PTGGE}  
  \rho_{\mathrm{PTGGE}}(t) = \frac{1}{Z_{\mathrm{PTGGE}}(t)} \e^{- 2 \sum_{k \geq 0}
    D_k^{\dagger} f_k^{-1} \atanh \left( \zeta_k(t) f_k^{-1}
    \right) D_k},
\end{equation}
with normalization $Z_{\mathrm{PTGGE}}(t)$ such that
$\tr(\rho_{\mathrm{PTGGE}}(t)) = 1$. The PTGGE reduces to the conventional GGE
when $\gamma = 0$ such that $f_k = \id$ and $\zeta_k(t)$ becomes
time-independent. Relaxation to the PTGGE in the PT-symmetric phase stands in
stark contrast to the long-time dynamics in the PT-mixed and PT-broken phases,
which is determined by the single slowest-decaying mode. Therefore, the boundary
of the PT-symmetric phase corresponds to a sharp dynamical transition that
delimits the applicability of the principle of maximum entropy.

When $\delta \neq 0$, the PTGGE captures relaxation dynamics only up to a
crossover time scale $t_{\times}$ that is determined by the equivalence of
initial-state and steady-state contributions in
Eq.~\eqref{eq:d-k-normal-commutator},
$\e^{-4 \gamma t_{\times}} \lvert \langle [d_k, d_k^{\dagger}] \rangle_0 \rvert
= \left( 1 - \e^{-4 \gamma t_{\times}} \right) \lvert \langle [d_k,
d_k^{\dagger}] \rangle_{\mathrm{SS}} \rvert$.
Since $\langle [d_k, d_k^{\dagger}] \rangle_{\mathrm{SS}}$ is proportional to
$\delta$~\cite{supplemental_material}, this equation implies
$t_{\times} \sim \left( 1/\gamma \right) \abs{\ln(c_{\times} \abs{\delta})}$
with a constant coefficient $c_{\times} > 0$ for $\delta \to 0$. Consequently,
within the entire PT-symmetric phase, which includes values of $\gamma$ that are
comparable to Hamiltonian energy scales, $t_{\times}$ can be large enough such
that relaxation to the PTGGE can be observed if $\delta$ is sufficiently
small. The precise condition on the value of $\delta$ depends on the observable
under consideration. Below, we provide a quantitative discussion for the fermion
parity of a finite subsystem.

\paragraph*{Relaxation of subsystem parity.}

\begin{figure}
  \includegraphics[width=\linewidth]{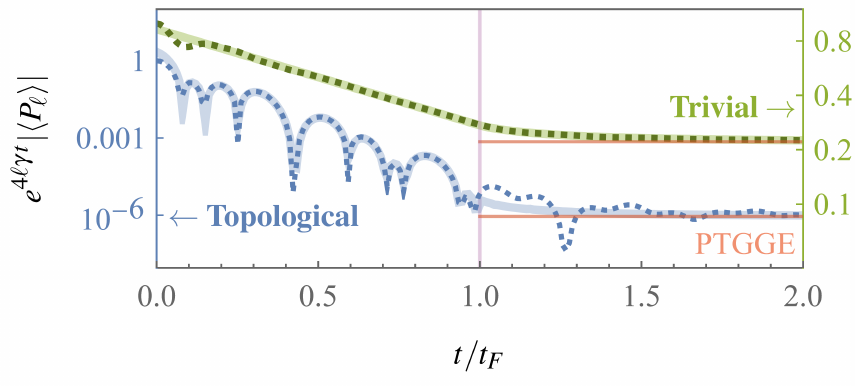}
  \caption{Subsystem parity after quenches to the trivial (green, $\mu = - 4 J$)
    and topological (blue, $\mu = - J$) PT-symmetric phases for
    $\gamma = 0.3 J$, $\delta = 0$, and $\ell = 20$. The solid lines are
    obtained from Eqs.~\eqref{eq:subsystem-parity-space-time-scaling-trivial}
    and~\eqref{eq:subsystem-parity-space-time-scaling-topological}, where we set
    $\alpha_+ = 0.08$ and $\alpha_- = 0.11$ to achieve best agreement with the
    numerical data shown as dashed lines. Straight vertical and horizontal lines
    indicate $t = t_F$ and the PTGGE predictions for the stationary
    values, respectively. In all figures, $L$ is chosen large enough to avoid
    finite-size effects.}
  \label{fig:subsystem-parity}
\end{figure}

To illustrate relaxation to the PTGGE, we consider the fermion parity of a
subsystem that consists of $\ell$ contiguous lattice sites,
$P_{\ell} = \e^{\imag \pi \sum_{l = 1}^{\ell} c_l^{\dagger} c_l}$. The
expectation value $\langle P_{\ell} \rangle = \Pf(\Gamma_{\ell})$ is given by
the Pfaffian of the reduced covariance matrix
$\Gamma_{\ell} = \left( \Gamma_{l, l'} \right)_{l, l' = 1}^{2
  \ell}$~\cite{Lieb1961,
  Barouch1970I, *Barouch1971II, *Barouch1971III}, where
$\Gamma = \imag R^{\dagger} G R$, $G$ is a block Toeplitz matrix built from the
$2 \times 2$ blocks $g_l$ in Eq.~\eqref{eq:g-l}, and
$R = \oplus_{l = 1}^{\ell} \frac{1}{\sqrt{2}} \left(\begin{smallmatrix} 1 & -
    \imag \\ 1 & \imag
\end{smallmatrix}\right)
$.
For the isolated Kitaev chain, a combined Jordan-Wigner~\cite{Jordan1928} and
Kramers-Wannier~\cite{Kramers1941, Fisher1995} transformation maps the subsystem
parity to order parameter correlations in the transverse field Ising
model~\cite{supplemental_material}. Based on the analytical results of Calabrese
et al.~\cite{Calabrese2011, Calabrese2012I, Calabrese2012II} for the relaxation
of order parameter correlations in the space-time scaling limit
$\ell, t \to \infty$ with $\ell/t$ fixed, in
Eqs.~\eqref{eq:subsystem-parity-space-time-scaling-trivial}
and~\eqref{eq:subsystem-parity-space-time-scaling-topological} below, we
formulate analytical conjectures for the time dependence of the subsystem parity
in the driven-dissipative Kitaev chain, which we find to be in excellent
agreement with numerical results.

First, we consider quenches to the topologically trivial~\cite{Sayyad2021}
PT-symmetric phase with $\abs{\mu} > 2 J$. Then, as shown in
Fig.~\ref{fig:subsystem-parity}, for $\delta = 0$, the behavior of the subsystem
parity in the space-time scaling limit is well described
by~\cite{supplemental_material}
\begin{equation}
  \label{eq:subsystem-parity-space-time-scaling-trivial}
  \langle P_{\ell}(t) \rangle \sim P_0 \e^{- 4 \ell \gamma t + \int_0^{\pi}
    \frac{\diff k}{2 \pi} \min(2 \abs{v_k} t, \ell) \tr \left( \ln \left(
        \abs{\zeta_k' f_k^{-1}} \right) \right)},
\end{equation}
where $\zeta_k' = \e^{4 \gamma t} \zeta_k(t)$ is time-independent. The value
$\gamma = 0.3 J$ chosen in Fig.~\ref{fig:subsystem-parity} leads to sizeable
modifications of statistics and dynamics of Liouvillian as compared to
Hamiltonian elementary excitations, which are accounted for in
Eq.~\eqref{eq:subsystem-parity-space-time-scaling-trivial} by the appearance of
$f_k$ and the definition of the velocity
$v_k = \diff \omega_k/\diff k$ in terms of $\omega_k$ rather than the
Hamiltonian dispersion relation $\varepsilon_k$. Relaxation to the PTGGE is best
revealed by considering the rescaled subsystem parity
$\e^{4 \ell \gamma t} \langle P_{\ell}(t) \rangle$, which decays up to the Fermi
time~\cite{Calabrese2012I} $t_F = \ell/(2 v_{\mathrm{max}})$ where
$v_{\mathrm{max}} = \max_k \abs{v_k}$, before it approaches a stationary
value. The prefactor $P_0$ in
Eq.~\eqref{eq:subsystem-parity-space-time-scaling-trivial} is obtained by
fitting the long-time limit of the rescaled subsystem parity to the PTGGE
prediction.

\begin{figure}
  \includegraphics[width=\linewidth]{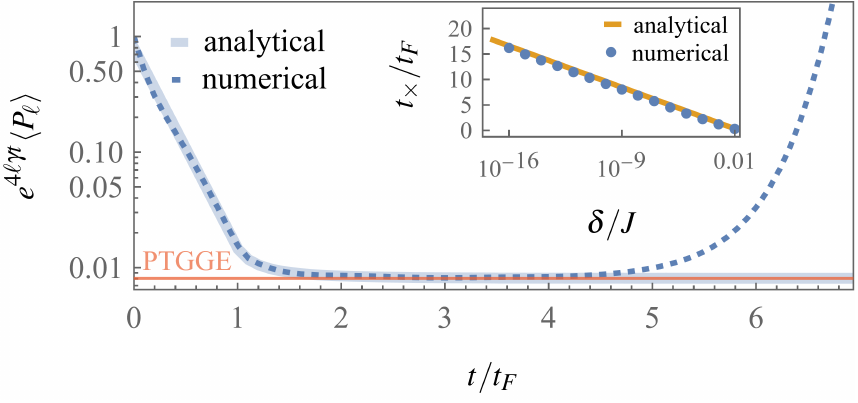}
  \caption{Deviation from the PTGGE due to $\delta \neq 0$ for $\mu = - 2.5 J$,
    $\gamma = 0.1 J$, $\delta = 10^{-7} J$, and $\ell = 20$. The rescaled
    subsystem parity (dashed line) follows
    Eq.~\eqref{eq:subsystem-parity-space-time-scaling-trivial} (solid line) up
    to the crossover time scale $t_{\times} \approx 5.7 t_F$ defined
    as
    $\lvert \langle P_{\ell}(t_{\times}) \rangle - \langle P_{\ell}(t_{\times})
    \rangle_{\mathrm{PTGGE}} \rvert = \langle P_{\ell}(t_{\times})
    \rangle_{\mathrm{PTGGE}}$.
    Inset: $t_{\times}$ diverges logarithmically for $\delta \to 0$. The
    numerical data is in good agreement with an analytical
    estimate~\cite{supplemental_material}.}
  \label{fig:t-cross}
\end{figure}

For small $\delta \neq 0$, we expect $\langle P_{\ell}(t) \rangle$ to deviate
from Eq.~\eqref{eq:subsystem-parity-space-time-scaling-trivial} after a
crossover time
$t_{\times} \sim \left( 1/\gamma \right) \ln(c_{\times} \abs{\delta})$. This
expectation is confirmed in Fig.~\ref{fig:t-cross}, where we also compare
numerical results for $t_{\times}$ with an analytical
estimate~\cite{supplemental_material}. The condition to observe relaxation of
the subsystem parity to the PTGGE, therefore, reads
$t_F < t_{\times}$.

\paragraph*{Directional parity pumping.}

\begin{figure}
  \includegraphics[width=\linewidth]{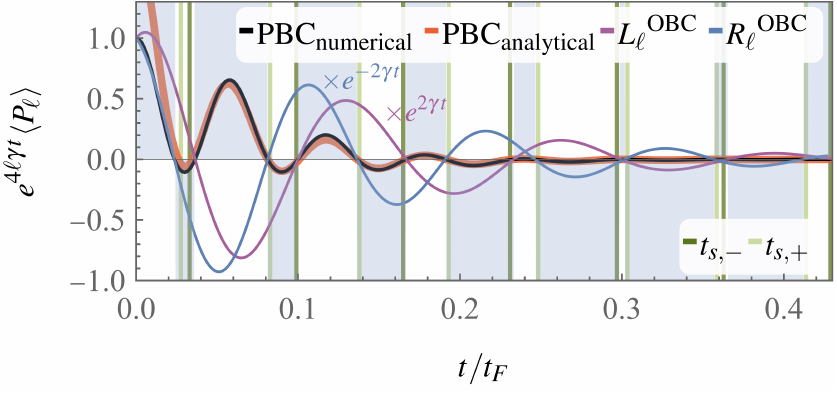}
  \caption{Directional pumping of subsystem parity for a quench to the
    topological PT-symmetric phase with $\mu=-0.5 J$, $\gamma=0.3J$,
    $\delta = 0$, and $\ell=30$. For periodic boundary conditions (PBC), the
    subsystem parity (black line: numerics; blue shading: sign of numerical
    data; red line:
    Eq.~\eqref{eq:subsystem-parity-space-time-scaling-topological} with
    $\alpha_+ = \alpha_- = 0.09$) crosses zero at multiples of both
    $t_{s, +}$ and $t_{s, -}$. In contrast, for open boundary
    conditions (OBC), zero crossings occur only at multiples of
    $t_{s, -}$ and $t_{s, +}$ for subsystems, respectively,
    $L_{\ell}$ (violet line) and $R_{\ell}$ (blue line). Factors
    $\e^{\pm 2 \gamma t}$ compensate for additional exponential decay (left end)
    and growth (right end) due to edge modes~\cite{supplemental_material}.}
  \label{fig:subsystem-parity-crossings}
\end{figure}

For quenches to the PT-symmetric phase with $\abs{\mu} < 2 J$, due to nontrivial
non-Hermitian topology of the Liouvillian~\cite{Sayyad2021}, the rescaled
subsystem parity repeatedly crosses zero before it relaxes to a stationary
value. Physically, these zero crossings can be interpreted as pumping of parity
between the subsystem and its complement. The period of the zero crossings is
determined by soft modes of the PTGGE, i.e., momenta $k_{s, \pm}$, for
which the exponent in Eq.~\eqref{eq:rho-PTGGE} vanishes. For the isolated Kitaev
chain~\cite{Calabrese2012I, Essler2016}, the soft modes
$k_{s, +} = - k_{s, -}$ are locked onto each other by
inversion symmetry~\cite{supplemental_material}, and the period of zero
crossings is given by
$t_{s} = \pi/(2 \varepsilon_{k_{s, +}}) = \pi/(2
\varepsilon_{k_{s, -}})$.
In contrast, for the PTGGE in Eq.~\eqref{eq:rho-PTGGE}, we find that due to the
breaking of inversion symmetry when $\gamma > 0$, there are two distinct soft
modes with
$k_{s, +} \neq - k_{s, -}$~\cite{supplemental_material}, and,
consequently two distinct time scales
$t_{s, \pm} = \pi/(2 \omega_{k_{s, \pm}})$. As shown in
Fig.~\ref{fig:subsystem-parity}, for $t < t_F$, the resulting
oscillatory decay of the subsystem parity is captured by the following modified
space-time scaling limit~\cite{supplemental_material}:
\begin{equation}
  \label{eq:subsystem-parity-space-time-scaling-topological}
  \langle P_{\ell}(t) \rangle \sim 2 \cos(\omega_{k_{s, +}} t +
  \alpha_+) \cos(\omega_{k_{s, -}} t + \alpha_-)
  \langle P_{\ell}(t)
  \rangle_{\mathrm{nonosc}},  
\end{equation}
where $\alpha_{\pm}$ are undetermined phase shifts and the nonoscillatory part
is given by Eq.~\eqref{eq:subsystem-parity-space-time-scaling-trivial}, which
also approximately describes the behavior of $\langle P_{\ell}(t) \rangle$ for
$t > t_F$.

The two timescales $t_{s, +}$ and $t_{s, -}$ have a clear
physical meaning in terms of the exchange of parity through, respectively, the
left and right boundaries of the subsystem. This is confirmed numerically in
Fig.~\ref{fig:subsystem-parity-crossings} by considering a chain with open
boundary conditions and subsystems $L_{\ell} = \{ 1, \dotsc, \ell \}$ and
$R_{\ell} = \{ L - \ell + 1, \dots, L \}$ located at the left and right ends of
the chain~\cite{supplemental_material}. Then, zero crossings of
$\langle P_{\ell}(t) \rangle$ occur only with period $t_{s, -}$ and
$t_{s, +}$, respectively. In contrast, for a chain with periodic
boundary conditions, $\langle P_{\ell}(t) \rangle$ exhibits zero crossings at
multiples of both $t_{s, +}$ and $t_{s, -}$. As we show in the
Supplemental Material~\cite{supplemental_material}, the occurrence of different
periods of parity pumping for subsystems at the left and right ends of the chain
requires both mixedness of the time-evolved state and breaking of inversion
symmetry and is, therefore, unique to driven-dissipative systems.

\paragraph*{Evolution of subsystem entropy.}

\begin{figure}
  \includegraphics[width=\linewidth]{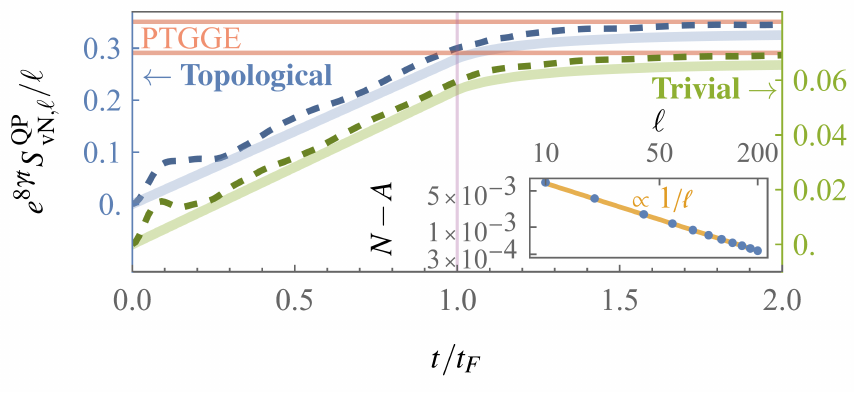}
  \caption{Quasiparticle-pair contribution to the subsystem entropy after
    quenches to the trivial (green, $\mu = - 4 J$) and topological (blue,
    $\mu = - J$) PT-symmetric phases for $\gamma = 0.3 J$, $\delta = 0$, and
    $\ell = 20$. The numerical data (dashed lines) is close to
    Eq.~\eqref{eq:S-vN-qp-space-time-scaling} (solid lines). Inset: For the
    trivial quench at $t = 2 t_F$, the difference between the
    numerical data and Eq.~\eqref{eq:S-vN-qp-space-time-scaling} (blue dots)
    vanishes as $1 / \ell$ (orange line).}
  \label{fig:entropy}
\end{figure}

In isolated systems, a key signature of thermalization is provided by the growth
and saturation of the von Neumann entropy of a finite subsystem,
$S_{\mathrm{vN}, \ell} = - \tr(\rho_{\ell} \ln(\rho_{\ell}))$. Here, we consider
a subsystem that consists of $\ell$ contiguous lattice sites, and whose density
matrix $\rho_{\ell}$ is obtained by taking the trace over the $L - \ell$
remaining sites, $\rho_{\ell} = {\tr}_{L - \ell}(\rho)$. Quantitative
predictions for the time dependence of $S_{\mathrm{vN}, \ell}$ in the space-time
scaling limit can be derived from a quasiparticle picture~\cite{Calabrese2005,
  Alba2017, Alba2018, Calabrese2020}, according to which the initial state acts
as source of pairs of entangled quasiparticles. The ballistic propagation of
quasiparticles leads to growth of the subsystem entropy in proportion to the
number of pairs of entangled quasiparticles that are shared between the
subsystem and its complement.

In open systems, the subsystem entropy
$S_{\mathrm{vN}, \ell} = S_{\mathrm{vN}, \ell}^{\mathrm{QP}} + \left( \ell/L
\right) S_{\mathrm{vN}}^{\mathrm{stat}}$
is the sum of two contributions~\cite{Maity2020, Alba2021, Carollo2022,
  Alba2022}: $S_{\mathrm{vN}, \ell}^{\mathrm{QP}}$ measures correlations due to
the propagation of quasiparticle pairs, and
$S_{\mathrm{vN}}^{\mathrm{stat}} = S_{\mathrm{vN}, L}$ is the statistical
entropy due to the mixedness of the time-evolved state. Based on results of
Refs.~\cite{Carollo2022, Alba2022} for weak dissipation $\gamma \sim 1/\ell$, we
conjecture that for quenches to the PT-symmetric phase and $\delta = 0$, the
quasiparticle-pair contribution $S_{\mathrm{vN}, \ell}^{\mathrm{QP}}$ obeys the
following space-time scaling limit~\cite{supplemental_material}:
\begin{equation}
  \label{eq:S-vN-qp-space-time-scaling}
  S_{\mathrm{vN}, \ell}^{\mathrm{QP}}(t) \sim \int_0^{\pi} \frac{\diff k}{2
    \pi} \min(2 \abs{v_k} t, \ell) \tr \! \left( S \! \left(
      \zeta_k(t) f_k^{-1} \right) - S(g_k(t))_d
  \right),
\end{equation}
where
$S(\xi) = - \frac{1 + \xi}{2} \ln \! \left( \frac{1 + \xi}{2} \right) - \frac{1
  - \xi}{2} \ln \! \left( \frac{1 - \xi}{2} \right)$.
The subscript ``\textit{d}'' in last term indicates that due to dephasing, only the
nonoscillatory components of the trace are required to capture the space-time
scaling limit. At long times $\gamma t \gg 1$, since
$\zeta_k(t), g_k(t) \sim \e^{-4 \gamma t}$, we can expand
$S(\xi) \sim \ln(2) - \xi^2/2$. Then, due to the cancellation of the leading
constant term in the difference in Eq.~\eqref{eq:S-vN-qp-space-time-scaling}, we
obtain $S_{\mathrm{vN}, \ell}^{\mathrm{QP}}(t) \sim \e^{- 8 \gamma t}$.
Therefore, in analogy to the subsystem parity, relaxation to the PTGGE becomes
visible by considering the rescaled quasiparticle-pair entropy
$\e^{8 \gamma t} S_{\mathrm{vN}, \ell}^{\mathrm{QP}}(t)$. As shown in
Fig.~\ref{fig:entropy} the rescaled quasiparticle-pair entropy grows up to the
Fermi time $t_F$ before it saturates to a stationary value predicted
by the PTGGE.

\paragraph*{Discussion.}

An important question concerns the validity of the PTGGE beyond the specific
example of the Kitaev chain. As we show in the Supplemental
Material~\cite{supplemental_material}, our results apply directly to
symmetry-preserving deformations of the Kitaev chain, and also to a class of
fermionic models with a particle-number conserving Hamiltonian, for which a
natural choice of dissipation is provided by incoherent loss and
gain. Furthermore, we find that for an interacting spin chain that can be mapped
to fermions but with quadratic jump operators, relaxation of a subset of
observables is described by the PTGGE. Finally, for a model of noninteracting
bosons, we demonstrate relaxation to an ensemble that generalizes the PTGGE for
fermions while maintaining the key property of conserving an extensive amount of
information about the initial state. It is intriguing to speculate whether PT
symmetry can affect also the dynamics of nonintegrable driven-dissipative
systems in a similar way so as to induce relaxation to a PT-symmetric Gibbs
ensemble.


\paragraph*{Acknowledgments.}

We thank Jinlong Yu for helpful discussions and acknowledge support from the
Austrian Science Fund (FWF) through the Project No.~P 33741-N.








%

\end{document}


\title{Supplemental Material: Relaxation to a Parity-Time Symmetric Generalized
  Gibbs \\ Ensemble after a Quantum Quench in a Driven-Dissipative Kitaev Chain}

\author{Elias Starchl}

\author{Lukas M. Sieberer} \email{lukas.sieberer@uibk.ac.at}

\affiliation{Institute for Theoretical Physics, University of Innsbruck, 6020
  Innsbruck, Austria}


\begin{abstract}
  In the Supplemental Material, we provide technical details of our analytical
  and numerical studies of quench dynamics in a driven-dissipative Kitaev chain,
  and we present additional results for PT-symmetric relaxation in quench
  dynamics of fermionic and bosonic SSH models with loss and gain and a
  dimerized XY spin chain with dephasing. We first discuss relevant properties
  of the isolated Kitaev chain, in particular, its symmetries and the mapping to
  the transverse field Ising model employed in the main text. Then, we summarize
  how these properties are affected by Markovian drive and dissipation, and how
  quench dynamics in the driven-dissipative Kitaev chain can be described
  analytically and numerically in terms of the time evolution of the covariance
  matrix. Focusing on the PT-symmetric phase, we present two complementary
  derivations of the PT-symmetric generalized Gibbs ensemble (PTGGE): The first
  one is based on the dephased form of the covariance matrix, and the second
  one, which is outlined in the main text, benefits from the clear picture of
  quench dynamics that is provided by a description in terms of eigenmodes of
  the Liouvillian. We then describe how the numerical results for the subsystem
  parity and entropy shown in the main text are obtained, and we motivate our
  conjectures for the time-dependence of these quantities in the
  space-time-scaling limit, which are summarized in Eqs.~(9), (10), and~(11) in
  the main text. Finally, we present our results for relaxation to the PTGGE in
  other models.
\end{abstract}

\maketitle

\tableofcontents

\section{Isolated Kitaev chain}

Here, we summarize aspects of the isolated Kitaev chain that are relevant for
our study of quench dynamics in a driven-dissipative Kitaev chain. This includes
the diagonalization and ground state of the Kitaev chain for open and periodic
boundary conditions, symmetries, and the mapping between the Kitaev chain and
the transverse field Ising model.

\subsection{Hamiltonian}
\label{sec:hamiltonian}

The Hamiltonian of the isolated Kitaev chain, as given in the main text, reads
\begin{equation}
  \label{eq:H-Kitaev}  
  H = \sum_{l = 1}^L \left( -J c^{\dagger}_l c_{l + 1} + \Delta c_l c_{l + 1} + \hc
  \right) - \mu \sum_{l = 1}^L \left( c^{\dagger}_l c_l -\frac{1}{2} \right).
\end{equation}
All of our results for quench dynamics presented in the main text are obtained
for $J = \Delta$, but we keep $J$ and $\Delta$ as distinct parameters in most
expressions below. However, we always assume that $J, \Delta \in \R_{>
  0}$.
Further, we consider both periodic boundary conditions with $c_{L + 1} = c_1$,
and open boundary conditions with $c_{L + 1} = 0$.

An alternative form of the Hamiltonian, which puts emphasis on translational
invariance and, therefore, proves especially convenient when we consider
periodic boundary conditions, is given by
\begin{equation}  
  H = \frac{1}{2} \sum_{l, l' = 1}^L C_l^{\dagger} h_{l - l'} C_{l'},
\end{equation}
where we collect fermionic annihilation and creation operators in spinors,
\begin{equation}
  \label{eq:C-l}
  C_l = \left( c_l, c_l^{\dagger} \right)^{\transpose},
\end{equation}
and where
\begin{equation}
  h_l = \mathbf{h}_l \cdot \boldsymbol{\sigma},
\end{equation}
with the vector of Pauli matrices,
\begin{equation}
  \boldsymbol{\sigma} = \left( \sigma_x, \sigma_y, \sigma_z \right)^{\transpose},
\end{equation}
and with coefficients given by
\begin{equation}
  \mathbf{h}_l = \left( 0, \imag \Delta \left( \delta_{l, 1} - \delta_{l, - 1}
    \right), - J \left( \delta_{l, 1} + \delta_{l, - 1} \right) - \mu \delta_{l,
    0} \right)^{\transpose}.
\end{equation}
Here and in the following, for periodic boundary conditions, each index of a
Kronecker delta is to be understood modulo the length of the chain $L$, e.g.,
$\delta_{l, l' + L} = \delta_{l, l'}$.

In addition to the above formulation of the Kitaev chain in terms of $L$ complex
or Dirac fermions $c_l$ with $l \in \{ 1, \dotsc, L \}$, it is often convenient
to work with $2 L$ Majorana fermions $w_l$, where $l \in \{ 1, \dotsc, 2 L \}$,
which are defined by
\begin{equation}
  \label{eq:Majorana-fermions}
  w_{2 l - 1} = c_l  + c^{\dagger}_l, \qquad w_{2 l} = \imag \left( c_l  -
    c^{\dagger}_l  \right),
\end{equation}
and obey the anticommutation relation $\{ w_l, w_{l'} \} = 2 \delta_{l,
  l'}$.
Further, it is often convenient to express the transformation to Majorana
fermions in terms of spinors of operators as follows:
\begin{equation}
  \label{eq:Majorana-transformation}
  C_l = \frac{1}{\sqrt{2}} r W_l,
\end{equation}
where $C_l$ is defined in Eq.~\eqref{eq:C-l},
\begin{equation}
  \label{eq:W-l}  
  W_l = \left( w_{2 l - 1}, w_{2 l} \right)^{\transpose},
\end{equation}
and the unitary matrix $r$ is given by
\begin{equation}
  \label{eq:r}
  r = \frac{1}{\sqrt{2}}
  \begin{pmatrix}
    1 & - \imag \\
    1 & \imag
  \end{pmatrix}.
\end{equation}
In terms of Majorana fermions, the Hamiltonian of the isolated Kitaev chain in
Eq.~\eqref{eq:H-Kitaev} can be written as
\begin{equation}
  H = \frac{\imag}{4} \sum_{l, l' = 1}^L W_l^{\transpose} a_{l - l'} W_{l'},
\end{equation}
where
\begin{equation}
  \label{eq:a-l}
  a_l = - \imag r^{\dagger} h_l r = \mathbf{a}_l \cdot \boldsymbol{\sigma},
\end{equation}
with
\begin{equation}
  \label{eq:a-vector-l}
  \mathbf{a}_l = \left( \Delta \left( \delta_{l, 1} - \delta_{l, - 1} \right),
    \imag \left[ J \left( \delta_{l, 1} + \delta_{l, -1} \right) + \mu
      \delta_{l, 0} \right], 0 \right)^{\transpose}.
\end{equation}
Here and in the following, the transformations of various quantities between the
bases of complex and Majorana fermions, such as the transformation from $h_l$ to
$a_l$, can be performed efficiently by noting that they correspond merely to
permutations of the components in expansions in terms of Pauli matrices, which
are here given by $\mathbf{h}_l$ and $\mathbf{a}_l$, respectively. In
particular, to calculate
\begin{equation}
  r^{\dagger} h_l r = \mathbf{h}_l \cdot \left( r^{\dagger} \boldsymbol{\sigma} r \right),
\end{equation}
we use
\begin{equation}
  \label{eq:r-dagger-sigma-r}
  r^{\dagger}
  \begin{pmatrix}
      \sigma_x \\ \sigma_y \\ \sigma_z
  \end{pmatrix}
  r =
  \begin{pmatrix}
      \sigma_z \\ \sigma_x \\ \sigma_y
  \end{pmatrix}
  = P^{\transpose}
  \begin{pmatrix}
      \sigma_x \\ \sigma_y \\ \sigma_z
  \end{pmatrix},
\end{equation}
where the permutation matrix $P$ is given by
\begin{equation}
  \label{eq:P}
  P =
  \begin{pmatrix}
    0 & 1 & 0 \\ 0 & 0 & 1 \\ 1 & 0 & 0
  \end{pmatrix}.
\end{equation}
We thus obtain Eqs.~\eqref{eq:a-l} and~\eqref{eq:a-vector-l}:
\begin{equation}
  r^{\dagger} h_l r = \mathbf{h}_l \cdot \left( P^{\transpose}
    \boldsymbol{\sigma} \right) = \left( P \mathbf{h}_l \right) \cdot
  \boldsymbol{\sigma} = \imag \mathbf{a}_l \cdot \boldsymbol{\sigma} = \imag
  a_l.
\end{equation}

For future reference, we define a real antisymmetric matrix $A$ as the block
Toeplitz matrix built from the $2 \times 2$ blocks $a_l$,
\begin{equation}
  \label{eq:A}
  A =
  \begin{pmatrix}
    a_0 & a_{-1} & \cdots & a_{1 - L} \\
    a_1 & a_0 & & \vdots \\
    \vdots & & \ddots & \vdots \\
    a_{L - 1} & \cdots & \cdots & a_0
  \end{pmatrix}.
\end{equation}
In terms of the matrix $A$, the Hamiltonian of the Kitaev chain can be written as
\begin{equation}
  H = \frac{\imag}{4} \sum_{l, l' = 1}^{2 L} w_l A_{l, l'} w_{l'}.
\end{equation}

\subsection{Covariance matrix}
\label{sec:covariance-matrix}

In our studies of quench dynamics, we always deal with Gaussian states $\rho$:
The initial state $\rho_0 = \ket{\psi_0} \bra{\psi_0}$, where $\ket{\psi_0}$ is
the ground state of the isolated Kitaev chain, is Gaussian, and the quadratic
Liouvillian we consider below preserves Gaussianity. Gaussian states are fully
determined by two-point functions, which are collected in the real and
antisymmetric covariance matrix,
\begin{equation}
  \label{eq:Gamma}
  \Gamma_{l, l'} = \frac{\imag}{2} \langle [w_l, w_{l'}] \rangle =
  \imag \left( \langle w_l w_{l'} \rangle - \delta_{l, l'} \right),
\end{equation}
where $\langle \dotsc \rangle = \tr( \dotsc \rho )$.
We often find it convenient to work with $2 \times 2$ blocks of $\Gamma$, which
can be specified in terms of the Majorana spinors defined in Eq.~\eqref{eq:W-l}:
\begin{equation}
  \label{eq:gamma-l}
  \gamma_{l, l'} =
  \begin{pmatrix}
    \Gamma_{2 l - 1, 2 l' - 1} & \Gamma_{2 l - 1, 2 l'} \\
    \Gamma_{2 l, 2 l' - 1} & \Gamma_{2 l, 2 l'}
  \end{pmatrix}
  = \imag \left( \langle W_l W_{l'}^{\transpose} \rangle - \delta_{l, l'} \id \right).
\end{equation}
Switching to the basis of complex fermions, we define a matrix $G$ through the
relation
\begin{equation}
  G = - \imag R \Gamma R^{\dagger},
\end{equation}
where the matrix $R$ is given in terms of the $2 \times 2$ matrix $r$ in
Eq.~\eqref{eq:r} by
\begin{equation}
  \label{eq:R}
  R = \bigoplus_{l = 1}^L r.
\end{equation}
We refer to $G$ as covariance matrix in the complex basis, or simply complex
covariance matrix. As done in Eq.~\eqref{eq:gamma-l} for the covariance matrix
in the Majorana basis, we define blocks of the complex covariance matrix,
\begin{equation}
  \label{eq:g-l}
  g_{l, l'} =
  \begin{pmatrix}
    G_{2 l - 1, 2 l' - 1} & G_{2 l - 1, 2 l'} \\
    G_{2 l, 2 l' - 1} & G_{2 l, 2 l'}
  \end{pmatrix}
  = - \imag r \gamma_{l, l'} r^{\dagger}.
\end{equation}
In terms of the complex spinors given in Eq.~\eqref{eq:C-l}, the blocks $g_{l,
  l'}$ can be written as
\begin{equation}
  g_{l, l'} = 2 \langle C_l C_{l'}^{\dagger} \rangle - \delta_{l, l'} \id =
  \begin{pmatrix}
    \langle [c_l, c_{l'}^{\dagger}] \rangle & \langle [c_l, c_{l'}] \rangle \\
    \langle [c_{l'}^{\dagger}, c_l^{\dagger}] \rangle & \langle
    [c_{l'}^{\dagger}, c_l] \rangle
  \end{pmatrix}.
\end{equation}

\subsection{Diagonalization and ground state}
\label{sec:diag-ground-state}

We proceed to explain how the Hamiltonian of the Kitaev chain in
Eq.~\eqref{eq:H-Kitaev} can be diagonalized and how its ground state can be
found. This can be done numerically for finite systems with open or periodic
boundary conditions. Analytical progress is possible for a system with periodic
boundary conditions. The ground state serves as the initial state in the quench
protocol discussed in the main text and further below.

\subsubsection{Numerical diagonalization}

The excitation spectrum and ground state of the Kitaev chain can be found
numerically as follows: By employing the algorithm described in
Ref.~\cite{Wimmer2012} (see also Ref.~\cite{Surace2021} for a discussion in the
context of fermionic Gaussian systems), one obtains a real special orthogonal
matrix $O$, which brings $A$ in Eq.~\eqref{eq:A} to the
following canonical block-diagonal form:
\begin{equation}
  \label{eq:A-Kitaev-canoncial-form}
  O A O^{\transpose} = \bigoplus\limits_{l = 1}^L
  \begin{pmatrix}
    0 & - \varepsilon_l \\ \varepsilon_l & 0
  \end{pmatrix},
\end{equation}
where $\varepsilon_l \geq 0$ are the energies of elementary
excitations. Further, the covariance matrix of the ground state is given
by~\cite{Kraus2009}
\begin{equation}
  \label{eq:Gamma-0-vacuum}
  \Gamma_0 = O^{\transpose} \left( \bigoplus\limits_{l = 1}^L \imag \sigma_y
  \right) O.
\end{equation}
Below and in the main text, we focus on quenches from the ground state of the
Kitaev chain for $\mu_0 \to - \infty$, i.e., the vacuum state $\ket{\Omega}$,
for which $c_l \ket{\Omega} = 0$ for all $l \in \{ 1, \dotsc L \}$. Then, $O$ is
simply the $2 L \times 2 L$ identity matrix.

\subsubsection{Analytical diagonalization for periodic boundary conditions}

For periodic boundary conditions, analytical progress can be made most
conveniently in the complex basis. In particular, we express the fermionic
operators $c_l$ in terms of momentum-space operators $c_k$ as
\begin{equation}
  c_l = \frac{1}{\sqrt{L}} \sum_{k \in \BZ} \e^{\imag k l} c_k, \quad c_k =
  \frac{1}{\sqrt{L}} \sum_{l = 1}^L \e^{-\imag k l} c_l,
\end{equation}
where the Brillouin zone $\BZ$ contains $L$ momenta with spacing
$\Delta k = 2 \pi/L$, that is,
$\BZ = \{ \pi + \Delta k, - \pi + \Delta k, \dotsc, \pi \}$. The spinors in
Eq.~\eqref{eq:C-l} transform to
\begin{equation}
  \label{eq:C-k}
  C_k =
  \begin{pmatrix}
    c_k \\ c_{-k}^{\dagger}
  \end{pmatrix}
  = \frac{1}{\sqrt{L}} \sum_{l = 1}^L \e^{-\imag k l} C_l,
\end{equation}
and the Hamiltonian of the Kitaev chain can be written as
\begin{equation}
  \label{eq:H-Kitaev-momentum-space}
  H = \sum_{k \geq 0} C_k^{\dagger} h_k C_k,
\end{equation}
with the Bogoliubov-De Gennes (BdG) Hamiltonian
\begin{equation}
  \label{eq:H-BdG}
  h_k = \sum_{l = 1}^L \e^{-\imag k l} h_l = \mathbf{h}_k \cdot \boldsymbol{\sigma},
\end{equation}
where
\begin{equation}
  \mathbf{h}_k = \left( 0, 2 \Delta \sin(k), - 2 J \cos(k) - \mu
  \right)^{\transpose}.
\end{equation}
Below, we find it convenient to parameterize the vector $\mathbf{h}_k$ in terms
of its magnitude,
\begin{equation}
  \label{eq:varepsilon-k}
  \varepsilon_k = \abs{\mathbf{h}_k} = \sqrt{\left( 2 J \cos(k) + \mu \right)^2
    + 4 \Delta^2 \sin(k)^2},
\end{equation}
and a unit vector
\begin{equation}
  \label{eq:n-hat-k}
  \unitvec{n}_k = \frac{\mathbf{h}_k}{\varepsilon_k} = \left( 0, \sin(\theta_k), \cos(\theta_k) \right)^{\transpose},
\end{equation}
where the angle $\theta_k$ is defined implicitly by the relation
\begin{equation}
  \label{eq:theta-k}
  \varepsilon_k \e^{\imag \theta_k} = - 2 J \cos(k) - \mu + \imag 2 \Delta \sin(k).
\end{equation}

The diagonalization of the BdG Hamiltonian corresponds geometrically to a
rotation of $\mathbf{h}_k$ to the $z$ axis, which can be accomplished by a
rotation around the $x$ axis by the angle $\theta_k$: In particular, with
\begin{equation}
  \label{eq:U-x-theta-k}
  U_{x, \theta_k} = \e^{\imag \left( \theta_k/2 \right) \sigma_x} =
  \begin{pmatrix}
    \cos(\theta_k/2) & \imag \sin(\theta_k/2) \\ \imag \sin(\theta_k/2) &
    \cos(\theta_k/2)
  \end{pmatrix},
\end{equation}
we obtain
\begin{equation}
  \label{eq:H-BdG-diagonal}
  U_{x, \theta_k}^{\dagger} h_k U_{x, \theta_k} = \varepsilon_k \sigma_z.
\end{equation}
The corresponding Bogoliubov transformation of fermionic operators reads
\begin{equation}
  \label{eq:Bogoliubov-quasiparticles}
  D_k =
  \begin{pmatrix}
    d_k \\ d_{-k}^{\dagger}
  \end{pmatrix}
  = U_{x, \theta_k}^{\dagger} C_k.
\end{equation}
In terms of the thusly defined fermionic modes $d_k$, the Hamiltonian becomes
\begin{equation}
  \label{eq:H-Kitaev-diagonal}
  H = \sum_{k \geq 0} \varepsilon_k \left( d_k^{\dagger} d_k - d_{-k}
    d_{-k}^{\dagger} \right),
\end{equation}
which shows that $\varepsilon_k$ determines the dispersion relation of
single-particle excitations, and that the ground state $\ket{\psi_0}$ is given
by the vacuum state of the operators $d_k$, i.e., $d_k \ket{\psi_0} = 0$ for all
$k \in \BZ$. To obtain an explicit expression for the ground state, we have to
treat momenta $k \in \{ 0, \pi \}$ and $k \in \BZ \setminus \{ 0, \pi \}$
separately. For momenta $k \in \{ 0, \pi \}$, Eq.~\eqref{eq:theta-k} implies
\begin{equation}
  \theta_0 =
  \begin{cases}
    0 & \text{for } \mu < - 2 J, \\
    \pi & \text{for } \mu > - 2 J,
  \end{cases}
\end{equation}
and
\begin{equation}
  \theta_{\pi} =
  \begin{cases}
    0 & \text{for } \mu < 2 J, \\
    \pi & \text{for } \mu > 2 J.
  \end{cases}
\end{equation}
By inserting these results in the Bogoliubov transformation
Eq.~\eqref{eq:Bogoliubov-quasiparticles}, we obtain
\begin{equation}  
  d_0 =
  \begin{cases}
    c_0 & \text{for } \mu < - 2 J, \\
    - \imag c_0^{\dagger} & \text{for } \mu > - 2 J,
  \end{cases}
\end{equation}
and, with $c_{\pi} = c_{-\pi}$,
\begin{equation}  
  d_{\pi} =
  \begin{cases}
    c_{\pi} & \text{for } \mu < 2 J, \\
    - \imag c_{\pi}^{\dagger} & \text{for } \mu > 2 J.
  \end{cases}
\end{equation}
Therefore, the conditions $d_0 \ket{\psi_0} = d_{\pi} \ket{\psi_0} = 0$ are
satisfied, if, in the state $\ket{\psi_0}$, the mode with momentum $k = 0$ is
occupied when $\mu > - 2 J$, and the mode with momentum $k = \pi$ is occupied
when $\mu > 2 J$.

For momenta $k \in \BZ \setminus \{ 0, \pi \}$, we employ the following
alternative form of the Bogoliubov transformation:
\begin{equation}
  d_k = \mathfrak{u}_k c_k \mathfrak{u}_k^{\dagger},
\end{equation}
where
\begin{equation}
  \begin{split}
    \mathfrak{u}_k & = \e^{\left( \theta_k/2 \right) \left( c_{-k} c_k -
        c_k^{\dagger} c_{-k}^{\dagger} \right)} \\ & = \e^{\imag
      \tan(\theta_k/2) c_k^{\dagger} c_{-k}^{\dagger}} \e^{\imag
      \sin(\theta_k/2) \cos(\theta_k/2) c_{-k} c_k} \e^{- \ln(\cos(\theta_k/2))
      \left( c_k^{\dagger} c_k + c_{-k}^{\dagger} c_{-k} - 1 \right)}.
  \end{split}
\end{equation}
From $c_k \ket{\Omega} = 0$ we obtain $d_k \mathfrak{u}_k \ket{\Omega} = 0$,
i.e., $\mathfrak{u}_k \ket{\Omega}$ is the vacuum of the Bogoliubov modes
$d_k$. Further, since $\mathfrak{u}_k$ is unitary, the transformed vacuum
$\mathfrak{u}_k \ket{\Omega}$ is normalized.

When we study quench dynamics below, we restrict ourselves to initial states
with $\mu < - 2 J$ such that, according to the above discussion, the modes with
$k \in \{ 0, \pi \}$ are not occupied. Then, the ground state is given by
\begin{equation}
  \begin{split}
    \ket{\psi_0} & = \prod_{k \in \BZ \setminus \{ 0, \pi \}} \mathfrak{u}_k
    \ket{\Omega} \\ & = \prod_{k \in \BZ \setminus \{ 0, \pi \}} \left(
      \cos(\theta_k/2) + \imag \sin(\theta_k/2) c_k^{\dagger} c_{-k}^{\dagger}
    \right) \ket{\Omega}.
  \end{split}
\end{equation}
Finally, we specify the covariance matrix in the ground state. To that end, we
first note that due to translational invariance in a system with periodic
boundary conditions, the $2 \times 2$ block in Eq.~\eqref{eq:g-l} depends only
on the difference of its indices, $g_{l, l'} = g_{l - l'}$. Then, its Fourier
transform is given by
\begin{equation}
  \label{eq:g-k}
  g_k = \sum_{l = 1}^L \e^{-\imag k l} g_l =  2 \langle C_k C_k^{\dagger} \rangle - \id =
  \begin{pmatrix}
    \langle [c_k, c_k^{\dagger}] \rangle & \langle [c_k, c_{-k}] \rangle \\
    \langle [c_{-k}^{\dagger}, c_k^{\dagger}] \rangle & \langle
    [c_{-k}^{\dagger}, c_{-k}] \rangle
  \end{pmatrix}
  .
\end{equation}
To obtain an explicit expression for the complex covariance matrix $g_k$ in the
ground state, we use the following ground-state expectation values, which follow
from the condition $d_k \ket{\psi_0} = 0$ that defines the ground state,
\begin{equation}
  \langle D_k D_k^{\dagger} \rangle_0 =
  \begin{pmatrix}
    \langle d_k d_k^{\dagger} \rangle_0 & \langle d_k d_{-k} \rangle_0 \\
    \langle d_{-k}^{\dagger} d_k^{\dagger} \rangle_0 & \langle d_{-k}^{\dagger}
    d_{-k} \rangle_0
    \end{pmatrix}
    = \frac{\id + \sigma_z}{2},
\end{equation}
where $\langle \dotsc \rangle_0 = \braket{\psi_0 | \dotsc | \psi_0}$. Further,
we use Eq.~\eqref{eq:H-BdG-diagonal}, according to which
$h_k/\varepsilon_k = \unitvec{n}_k \cdot \boldsymbol{\sigma} = U_{x, \theta_k}
\sigma_z U_{x, \theta_k}^{\dagger}$,
and Eq.~\eqref{eq:Bogoliubov-quasiparticles}. We thus find
\begin{equation}
  \label{eq:g-k-ground-state}
  g_{0, k} = 2 U_{x, \theta_k} \langle D_k D_k^{\dagger} \rangle_0 U_{x, \theta_k}^{\dagger} - \id
  = U_{x, \theta_k} \sigma_z U_{x, \theta_k}^{\dagger} = \unitvec{n}_k \cdot \boldsymbol{\sigma}.
\end{equation}
When we rotate this expression to the Majorana basis, we find it convenient to
include a prefactor $- \imag$ relative to Eq.~\eqref{eq:g-l}, i.e., we define the
momentum-space covariance matrix in the Majorana basis as
\begin{equation}
  \label{eq:gamma-k}
  \gamma_k = r^{\dagger} g_k r = - \imag \sum_{l = 1}^L \e^{-\imag k l}
  \gamma_l, \qquad \gamma_l = \frac{\imag}{L} \sum_{k \in \BZ} \e^{\imag k l}
  \gamma_k.
\end{equation}
With this convention, $\gamma_k = \gamma_k^{\dagger}$ is Hermitian. We note that
in terms of spinors of Majorana operators, which we define in analogy to
Eq.~\eqref{eq:C-k},
\begin{equation}
  \label{eq:W-k}
  W_k = \frac{1}{\sqrt{L}} \sum_{l = 1}^L \e^{-\imag k l} W_l = \sqrt{2}
  r^{\dagger} C_k,
\end{equation}
the covariance matrix can be written as
\begin{equation}
  \label{eq:gamma-k-W-k}
  \gamma_k = \langle W_k W_k^{\dagger} \rangle - \id.
\end{equation}
We obtain the covariance matrix in the ground state from
Eq.~\eqref{eq:g-k-ground-state} by using Eqs.~\eqref{eq:r-dagger-sigma-r} and~\eqref{eq:P},
\begin{equation}
  \label{eq:gamma-k-ground-state}
  \gamma_{0, k} = \unitvec{n}_k \cdot \left( r^{\dagger}
    \boldsymbol{\sigma} r \right) = \unitvec{n}_k \cdot \left( P^{\transpose} \boldsymbol{\sigma} \right)
  = \left( P \unitvec{n}_k \right) \cdot \boldsymbol{\sigma}
  = \unitvec{a}_k \cdot \boldsymbol{\sigma},
\end{equation}
where $\unitvec{a}_k = \mathbf{a}_k/\varepsilon_k$ and
\begin{equation}
  \label{eq:a-vector-k}
  \mathbf{a}_k = P \mathbf{h}_k = \left( 2 \Delta \sin(k), - 2 J \cos(k) - \mu,
    0 \right)^{\transpose}.
\end{equation}
For future reference, we define the Fourier transform of $a_l$ in
Eq.~\eqref{eq:a-l} including the imaginary unity as a prefactor:
\begin{equation}
  \label{eq:a-k}
  a_k = r^{\dagger} h_k r = \imag \sum_{l = 1}^L \e^{-\imag k l} a_l =
  \mathbf{a}_k \cdot \boldsymbol{\sigma}.
\end{equation}
Again, this convention leads to a Hermitian matrix $a_k = a_k^{\dagger}$.

\subsection{Symmetries of the isolated Kitaev chain}
\label{sec:symm-isol-kita}

We next discuss symmetries of the Kitaev chain, focusing on the BdG Hamiltonian
$h_k$ in Eq.~\eqref{eq:H-BdG}. To prepare for a comparison to the
driven-dissipative Kitaev chain introduced below, we state two equivalent forms
of each symmetry, where the second form follows from Hermiticity of the BdG
Hamiltonian, $h_k = h_k^{\dagger}$. Particle-hole symmetry (PHS), time-reversal
symmetry (TRS), and chiral symmetry (CS), which results from the combination of
PHS and TRS, are expressed by
\begin{align}
  \label{eq:H-BdG-PHS}
  & \text{PHS:} & h_k & = - \sigma_x h_{-k}^{\transpose} \sigma_x = - \sigma_x
                  h_{-k}^{*} \sigma_x, \\
  \label{eq:H-BdG-TRS}
  & \text{TRS:} & h_k & = h_{-k}^{\transpose} = h_{-k}^{*}, \\
  \label{eq:H-BdG-CS}
  & \text{CS:} & h_k & = - \sigma_x h_k \sigma_x = - \sigma_x h_k^{\dagger} \sigma_x.
\end{align}
We note that TRS is ensured by our choice $J \in \R_{> 0}$ (a nontrivial phase
of $\Delta$ can always be absorbed in a redefinition of the operators
$c_l$). Then, the time-reversal invariant Kitaev chain belongs to the
Altland-Zirnbauer class BDI~\cite{Altland1997}, which, in one spatial dimension,
is characterized by an integer-valued winding number~\cite{Hasan2010, Qi2011,
  Chiu2016}. The ground state of the Kitaev chain is topologically trivial with
a vanishing winding number for $\abs{\mu} > 2 J$, and topologically nontrivial,
as indicated by a nonvanishing value of the winding number, for
$\abs{\mu} < 2 J$.

In what follows, an important role is played by inversion symmetry (IS) and its
breaking due to Markovian drive and dissipation. For the isolated Kitaev chain,
inversion is represented by a unitary operator $I$ that is defined as
follows~\cite{Leumer2021}:
\begin{equation}
  \label{eq:inversion-second-quantized}
  \begin{aligned}
    & \text{open boundary conditions:} & I c_l I^{\dagger} & = \imag c_{L + 1 -
      l}, \\
    & \text{periodic boundary conditions:} & I c_l I^{\dagger} & = \imag c_{-
      l}.
  \end{aligned}
\end{equation}
The Hamiltonian in Eq.~\eqref{eq:H-Kitaev} is symmetric under inversion,
$H = I H I^{\dagger}$. In the transformation of the Hamiltonian, the factor
$\imag$ on the right-hand side of Eq.~\eqref{eq:inversion-second-quantized}
cancels in the hopping and chemical potential terms, but is required to leave
the pairing term invariant. For periodic boundary conditions, the spinors $C_k$
in Eq.~\eqref{eq:C-k} transform as
\begin{equation}
  \label{eq:C-k-IS}
  I C_k I^{\dagger} = \imag \sigma_z C_{-k}.
\end{equation}
Applying this transformation to the momentum-space representation of the
Hamiltonian in Eq.~\eqref{eq:H-Kitaev-momentum-space} yields, for the BdG
Hamiltonian Eq.~\eqref{eq:H-BdG}, the following conditions of IS and, combining
IS with TRS, parity-time symmetry (PTS):
\begin{align}
  \label{eq:H-BdG-IS}
  & \text{IS:} & h_k & = \sigma_z h_{-k} \sigma_z = \sigma_z h_{-k}^{\dagger}
                       \sigma_z, \\
  \label{eq:H-BdG-PTS}
  & \text{PTS:} & h_k & = \sigma_z h_k^{\transpose} \sigma_z = \sigma_z h_k^{*}
                        \sigma_z.
\end{align}

For future reference, we list symmetry properties that can be derived from the
symmetries of the BdG Hamiltonian. First, for $a_k$ defined in
Eq.~\eqref{eq:a-k}, we find
\begin{align}
  \label{eq:a-k-PHS}
  & \text{PHS:} & a_k & = - a_{-k}^{\transpose} = - a_{-k}^{*}, \\
  \label{eq:a-k-TRS}
  & \text{TRS:} & a_k & = \sigma_z a_{-k}^{\transpose} \sigma_z = \sigma_z
                        a_{-k}^{*} \sigma_z, \\
  \label{eq:a-k-CS}
  & \text{CS:} & a_k & = - \sigma_z a_k \sigma_z = - \sigma_z a_k^{\dagger}
                       \sigma_z, \\
  \label{eq:a-k-IS}
  & \text{IS:} & a_k & = \sigma_y a_{-k} \sigma_y = \sigma_y a_{-k}^{\dagger}
                       \sigma_y, \\
  \label{eq:a-k-PTS}
  & \text{PTS:} & a_k & = \sigma_x a_k^{\transpose} \sigma_x = \sigma_x a_k^{*} \sigma_x.
\end{align}
Further, PHS yields restrictions on the transformation $U_{x, \theta_k}$ in
Eq.~\eqref{eq:U-x-theta-k} that diagonalizes the BdG Hamiltonian. By combining
Eqs.~\eqref{eq:H-BdG-diagonal} and~\eqref{eq:H-BdG-PHS} we obtain
\begin{equation}
  \sigma_x U_{x, \theta_k}^{\transpose} \sigma_x h_{-k} \sigma_x U_{x, \theta_k}^{*} \sigma_x = \varepsilon_k
  \sigma_z.
\end{equation}
That is, we can identify $\varepsilon_k = \varepsilon_{-k}$ and
\begin{equation}
  \label{eq:U-x-theta-k-PHS}
  U_{x, \theta_k} = \sigma_x U_{x, \theta_{-k}}^{*} \sigma_x,
\end{equation}
which implies $\theta_k = - \theta_{-k}$. Alternatively, $\theta_k$ being an odd
function of $k$ can also be seen as a consequence of IS Eq.~\eqref{eq:H-BdG-IS},
which implies
\begin{equation}
  \label{eq:U-x-theta-k-IS}
  U_{x, \theta_k} = \sigma_z U_{x, \theta_{-k}} \sigma_z.
\end{equation}
Moreover, in analogy to the PHS of the BdG Hamiltonian $h_k$ and its counterpart
$a_k$ in the Majorana basis, one can check that also the covariance matrices
$g_k$ and $\gamma_k$, defined in Eqs.~\eqref{eq:g-k} and~\eqref{eq:gamma-k},
respectively, obey PHS,
\begin{equation}
  \label{eq:g-k-gamma-k-PHS}
  g_k = - \sigma_x g_{-k}^{\transpose} \sigma_x, \qquad \gamma_k = -
  \gamma_{-k}^{\transpose}.
\end{equation}
In particular, when we expand the covariance matrix in the basis of Pauli
matrices as
\begin{equation}
  \gamma_k = \gamma_{\id, k} \id + \boldsymbol{\gamma}_k \cdot \boldsymbol{\sigma},
\end{equation}
PHS implies
\begin{equation}
  \label{eq:gamma-id-k-PHS}
  \gamma_{\id, k} = - \gamma_{\id, -k},
\end{equation}
and
\begin{equation}
  \label{eq:gamma-vector-k-PHS}
  \gamma_{x, k} = - \gamma_{x, -k}, \quad \gamma_{y, k} =  \gamma_{y, -k},
  \quad \gamma_{z, k} = - \gamma_{z, -k}.
\end{equation}
Finally, Eq.~\eqref{eq:C-k-IS} yields, for an inversion-symmetric state
$\rho = I \rho I^{\dagger}$ in in Eqs.~\eqref{eq:g-k}
and~\eqref{eq:gamma-k-W-k}, the following conditions for the correlation and
covariance matrices:
\begin{equation}
  \label{eq:g-k-gamma-k-IS}
  g_k = \sigma_z g_{-k} \sigma_z, \qquad \gamma_k = \sigma_y \gamma_{-k} \sigma_y.
\end{equation}

\subsection{Relation to the transverse field Ising model}
\label{sec:relat-transv-field}

Through the Jordan-Wigner transformation, the isolated Kitaev chain can be
mapped to the transverse field Ising model, which serves as a paradigmatic model
system in studies of quench dynamics and generalized
thermalization~\cite{Calabrese2011, Calabrese2012I, Calabrese2012II}. To
establish the connection between our work and the existing literature on
quenches in the transverse field Ising model, we provide relevant details of
this mapping in the following. In Sec.~\ref{sec:subsystem-parity}, we exploit
this connection to formulate conjectures for the time dependence of the
subsystem parity.

Usually, the Jordan-Wigner transformation is applied directly to the Hamiltonian
of the transverse field Ising model, which is given by
\begin{equation}
  \label{eq:H-Ising}
  H_{\mathrm{Ising}} = -  J \sum_{l = 1}^L \left( \sigma_l^x \sigma_{l + 1}^x + h
    \sigma_l^z \right).
\end{equation}
Spin operators can be expressed in terms of fermionic operators through the
Jordan-Wigner transformation~\cite{Jordan1928}:
\begin{equation}
  \label{eq:Jordan-Wigner}
  \begin{split}
    \sigma_l^+ & = \e^{\imag \pi \sum_{l' = 1}^{l - 1} c_{l'}^{\dagger} c_{l'}} c_l, \\
    \sigma_l^- & = \e^{\imag \pi \sum_{l' = 1}^{l - 1} c_{l'}^{\dagger} c_{l'}}
    c_l^{\dagger}, \\ \sigma_l^z & = \e^{\imag \pi c_l^{\dagger} c_l}.
  \end{split}
\end{equation}
By means of the Jordan-Wigner transformation, the Hamiltonian of the transverse
field Ising model is mapped to two copies of the Kitaev chain
Eq.~\eqref{eq:H-Kitaev} in decoupled sectors with even and odd fermion parity,
and $J = \Delta$ and $\mu = - 2 J h$~\cite{Calabrese2012I}. In particular, when
we restrict ourselves to positive values of $h$ (and, therefore, negative values
of $\mu$), the ferromagnetically ordered phase of the Ising model with
$0 < h < 1$ maps to the topological phase of the Kitaev chain with
$\mu < - 2 J$, and the paramagnetic phase with $h > 1$ maps to the trivial phase
with $- 2 J < \mu < 0$. Long-range order in the ground state in the
ferromagnetic phase is revealed as usual by adding a longitudinal field
$J h_x \sum_{l = 1}^L \sigma_l^x$ to the Hamiltonian Eq.~\eqref{eq:H-Ising} and
considering the following combination of limits:
\begin{equation}
  \lim_{\abs{l - l'} \to \infty} \lim_{h_x \to 0} \lim_{L \to \infty}
  \bra{\psi_0} O_{l, l'} \ket{\psi_0} \neq 0,
\end{equation}
where, for $l' > l$,
\begin{equation}
  O_{l, l'} = \sigma_l^x \sigma_{l'}^x = \left( c_l + c_l^{\dagger} \right)
  \e^{\imag \pi \sum_{m = l}^{l' - 1} c_m^{\dagger} c_m} \left( c_{l'} +
    c_{l'}^{\dagger} \right).
\end{equation}
That is, longitudinal or order parameter correlations map to the string order
parameter $O_{l, l'}$~\cite{Budich2016} for the Kitaev chain.

We find it convenient to deviate from the route outlined above and employ the
Kramers-Wannier duality~\cite{Kramers1941, Fisher1995} of the transverse field
Ising model before applying the Jordan-Wigner transformation. The
Kramers-Wannier self-duality of the Ising model is established by defining new
spin operators through the relations
\begin{equation}
  \tau^z_l = \sigma^x_l \sigma^x_{l + 1}, \qquad \tau^x_l \tau^x_{l + 1} =
  \sigma^z_l.
\end{equation}
Then, the Hamiltonian maps onto itself with the global energy scale transforming
as $J \to J h$, and the transverse field as $h \to 1/h$, i.e., the ferromagnetic
phase of $\sigma$ spins maps to the paramagnetic phase of $\tau$
spins. Consequently, the ferromagnetic and paramagnetic phases of the Ising
model for $\tau$ spins map to the trivial and topological phases of the Kitaev
chain, respectively. In particular, the topologically trivial vacuum state,
which we choose as the initial state for our studies of quench dynamics,
corresponds to the ground state of the Ising model in the strongly ferromagnetic
limit. Longitudinal correlations of $\tau$ spins,
\begin{equation}
  P_{l, l'} = \tau_l^x \tau_{l' + 1}^x = \prod_{m = l}^{l'} \tau_m^x \tau_{m
    + 1}^x = \prod_{m = l}^{l'} \sigma_m^z,
\end{equation}
serve as a ``disorder parameter'' for $\sigma$ spins, and the expectation value
of $P_{l, l'}$ is maximized in the paramagnetic phase of $\sigma$ spins. Through
the Jordan-Wigner transformation Eq.~\eqref{eq:Jordan-Wigner}, $P_{l, l'}$ maps
to the fermion parity of a subsystem that consists of the lattice sites
$\{ l, \dotsc, l' \}$,
\begin{equation}  
  P_{l, l'} = \e^{\imag \pi \sum_{m = l}^{l'} c_m^{\dagger} c_m} = \prod_{m =
    l}^{l'} \left( \imag w_{2 m - 1} w_{2 m} \right).
\end{equation}
When the expectation value of $P_{l, l'}$ is taken in a translationally
invariant state, the result depends only on the difference $l' - l$. Then,
without loss of generality, we set $l = 1$ and $l' = \ell$, and define the
subsystem parity as
\begin{equation}
  \label{eq:subsystem-parity}
  P_{\ell} = \e^{\imag \pi \sum_{l = 1}^{\ell} c_l^{\dagger} c_l} = \prod_{l =
    1}^{\ell} \left( \imag w_{2 l - 1} w_{2 l} \right).
\end{equation}
As stated in the main text, the expectation value of the subsystem parity is
given by the Pfaffian of the reduced covariance matrix~\cite{Lieb1961,
  Barouch1970I, *Barouch1971II, *Barouch1971III},
\begin{equation}
  \label{eq:subsystem-parity-pfaffian}
  \langle P_{\ell} \rangle = \Pf(\Gamma_{\ell}),
\end{equation}
where
\begin{equation}
  \label{eq:Gamma-ell}
  \Gamma_{\ell} = \left( \Gamma_{l, l'} \right)_{l, l' = 1}^{2 \ell}.
\end{equation}
According to the above discussion, the ground-state expectation value
$\langle P_{\ell} \rangle_0 = \braket{\psi_0 | P_{\ell} | \psi_0}$ can be
regarded as a ``topological disorder parameter:'' The ground-state expectation
value $\langle P_{\ell} \rangle_0$ vanishes in the topological phase of the
Kitaev chain and is nonzero in the trivial phase. Accordingly, the results of
Refs.~\cite{Calabrese2011, Calabrese2012I, Calabrese2012II} for the time
evolution of order parameter correlations for quenches from the ordered to the
ordered and disordered phases of the transverse field Ising model, when
interpreted as applying to $\tau$ spins, describe the behavior of
$\langle P_{\ell} \rangle$ for quenches from the trivial to the trivial and
topological phases of the Kitaev chain, respectively.

\section{Driven-dissipative Kitaev chain}

We move on to consider quantum quenches in the driven-dissipative Kitaev
chain. As specified in the main text, we assume that the system is prepared in
the ground state $\ket{\psi_0}$ of the Hamiltonian of the isolated Kitaev chain
in Eq.~\eqref{eq:H-Kitaev} with parameters $J = \Delta$ and $\mu_0$. At $t = 0$,
the chemical potential is quenched to the value $\mu$, and, concurrently, the
system is coupled to Markovian reservoirs.

\subsection{Master equation and Liouvillian}

The time evolution of the driven-dissipative Kitaev chain after the quench is
described by a quantum master equation in Lindblad form~\cite{Gorini1976,
  Lindblad1976},
\begin{equation}
  \label{eq:master-equation}
  \imag \frac{d}{d t} \rho = \mathcal{L} \rho,
\end{equation}
where we define the Liouvillian $\mathcal{L}$ as the sum of the Hamiltonian
contribution $\mathcal{H}$ and the dissipator $\mathcal{D}$,
\begin{equation}
  \mathcal{L} = \mathcal{H} + \imag \mathcal{D},
\end{equation}
with
\begin{equation}
  \label{eq:H-adjoint-action}
  \mathcal{H} \rho = [H, \rho],
\end{equation}
and
\begin{equation}
  \mathcal{D} \rho = \sum_{l = 1}^L \left( 2 L_l \rho L_l^{\dagger} - \{ L_l^{\dagger}
    L_l, \rho \} \right).
\end{equation}
We consider local jump operators as given in the main text,
\begin{equation}
  \label{eq:L}
  L_l = \sqrt{\gamma_{l}} c_l + \sqrt{\gamma_{g}} c_l^{\dagger}. 
\end{equation}
In terms of Majorana operators $w_l$, the jump operators can be represented as
\begin{equation}
  \label{eq:L-Majorana-basis}
  L_l = \sum_{l' = 1}^{2 L} B_{l, l'} w_{l'},
\end{equation}
where the coefficients $B_{l, l'}$ form an $L \times 2L$ matrix. For future
reference, we define the bath matrix $M$ as the product
\begin{equation}
  \label{eq:M}
  M = B^{\transpose} B^{*}.
\end{equation}
According to its definition, the bath matrix is Hermitian; Hermiticity implies
the following symmetry conditions for the real and imaginary parts of $M$, which
we denote by $M_R$ and $M_I$, respectively:
\begin{equation}
  \label{eq:bath-matrix-hermitian}
  M_R = M_R^{\transpose}, \qquad M_I = - M_I^{\transpose}.
\end{equation}
In turn, these relations imply
\begin{equation}
  \label{eq:M-R-M-I}
  M + M^{\transpose} = 2 M_R, \qquad M - M^{\transpose} = \imag 2 M_I.
\end{equation}
Like the matrix $A$ in Eq.~\eqref{eq:A}, the bath matrix
$M$ is a block Toeplitz matrix, with $2 \times 2$ blocks $m_l$ given by
\begin{equation}
  \label{eq:m-l}
  m_l = m_{R, l} + \imag m_{I, l}.
\end{equation}
The real and imaginary parts, $m_{R, l}$ and $m_{I, l}$, respectively, read
\begin{equation}
  \label{eq:m-R-I-l}  
  m_{R, l} = \frac{\delta_{l, 0}}{2} \left[ \gamma \id +
    \sqrt{\gamma_{l} \gamma_{g}} \sigma_z \right], \qquad
  m_{I, l} = \frac{\delta_{l, 0}}{4} \delta \imag \sigma_y.
\end{equation}
Here, as in the main text, we introduce the mean rate $\gamma$, which represents
a measure of the overall strength of dissipation, and the difference of loss and
gain rates $\delta$,
\begin{equation}
  \gamma = \frac{\gamma_{l} + \gamma_{g}}{2}, \qquad \delta = \gamma_{l} - \gamma_{g}.
\end{equation}

\subsection{Time evolution of the covariance matrix}
\label{sec:time-evol-covar}

As stated in Sec.~\ref{sec:covariance-matrix}, for the quench protocol specified
in the main text, the system is in a Gaussian state at all times, and Gaussian
states are fully determined by the covariance matrix. Therefore, in the
following, we specify the equation of motion for the covariance matrix and its
numerical and analytical solution.

\subsubsection{Numerical time evolution}

The equation of motion for the covariance matrix reads
\begin{equation}
  \label{eq:Gamma-eom}
  \frac{d \Gamma}{d t} = - X \Gamma - \Gamma X^{\transpose} - Y,
\end{equation}
where
\begin{equation}
  \label{eq:X-Y}
  X = 4 \left( \imag H + M_R \right) = - A + 4 M_R, \quad Y = - 8 M_I,
\end{equation}
and its formal solution is given by~\cite{Prosen2011}
\begin{equation}
  \label{eq:Gamma-time-evolved}
  \Gamma(t) = \Gamma_1(t) + \Gamma_2(t),
\end{equation}
with
\begin{equation}
  \label{eq:Gamma-1-Gamma-2}
  \begin{split}
    \Gamma_1(t) & = \e^{- X t} \Gamma(0) \e^{- X^{\transpose} t}, \\ \Gamma_2(t)
    & = - \int_0^t d t' \, \e^{- X \left( t - t' \right)} Y \e^{-
      X^{\transpose} \left( t - t' \right)}.
  \end{split}
\end{equation}
The initial conditions and the steady state are encoded, respectively, in
$\Gamma_1(t)$ and $\Gamma_2(t)$. In particular, $\Gamma_1(0) = \Gamma(0)$ while
$\Gamma_2(0) = 0$, and $\Gamma_1(t) \to 0$ while
$\Gamma_2(t) \to \Gamma_{\mathrm{SS}}$ for $t \to \infty$. For all quenches
considered in the main text, the initial state is chosen as the ground state of
the Kitaev chain for $\mu_0 \to -\infty$, i.e., the vacuum state, and the
corresponding covariance matrix $\Gamma(0) = \Gamma_0$ is given in
Eq.~\eqref{eq:Gamma-0-vacuum}. The integral in the expression for $\Gamma_2(t)$
can be performed explicitly if we express the matrix $X$ as
$X = V \Lambda V^{-1}$, where $\Lambda$ is a diagonal matrix with the
eigenvalues $\lambda_l$ of $X$ on its diagonal. Then,
\begin{equation}
  \label{eq:Gamma-2-Hadamard}
  \Gamma_2(t) = - V \left[ \left( V^{-1} Y V^{-\transpose} \right) \circ K(t)
  \right] V^{\transpose},
\end{equation}
where we use the shorthand notation $V^{-\transpose}$ to denote the inverse of
the transpose, and where $A \circ B$ is the Hadamard product (i.e.,
element-wise) product of $A$ and $B$,
\begin{equation}
  \left( A \circ B \right)_{l, l'} = A_{l, l'} B_{l, l'}.
\end{equation}
Further, the matrix $K(t)$ is defined as
\begin{equation}
  K_{l, l'}(t) = \frac{1 - \e^{- \left( \lambda_l + \lambda_{l'} \right)
      t}}{\lambda_l + \lambda_{l'}}.
\end{equation}
Numerically, we find that $\lambda_l + \lambda_{l'} \neq 0$ for all $l, l'$ when
$\gamma > 0$. To obtain the numerical data shown in the plots in the main text,
we implement Eqs.~\eqref{eq:Gamma-1-Gamma-2} and~\eqref{eq:Gamma-2-Hadamard}.

\subsubsection{Analytical time evolution for periodic boundary conditions}

For periodic boundary conditions, due to translational invariance of the
Hamiltonian, the coupling to Markovian baths, and the initial state, all
matrices in Eq.~\eqref{eq:Gamma-eom} are block-circulant Toeplitz
matrices. Therefore, Eq.~\eqref{eq:Gamma-eom} can be rewritten as
\begin{equation}
  \frac{d \gamma_l}{d t} = - \sum_{l' = 1}^L
  \left( x_{l - l'} \gamma_{l'} + \gamma_{l - l'} x_{- l'}^{\transpose}
  \right) - y_l,
\end{equation}
where the $2 \times 2$ blocks $\gamma_l$ are defined in Eq.~\eqref{eq:gamma-l},
and $x_l$ and $y_l$ are defined analogously in terms of the matrices $X$ and $Y$
in Eq.~\eqref{eq:X-Y}, i.e.,
\begin{equation}
  \label{eq:x-y}
  x_l = - a_l + 4 m_{R, l}, \qquad y_l = - 8 m_{I, l},
\end{equation}
with $a_l$ and the real and imaginary parts of the blocks of the bath matrix
given in Eqs.~\eqref{eq:a-l} and~\eqref{eq:m-R-I-l}, respectively. In momentum
space, the equation of motion for the Fourier transform $\gamma_k$ in
Eq.~\eqref{eq:gamma-k} takes the form
\begin{equation}
  \label{eq:gamma-k-eom}
  \frac{d \gamma_k}{d t} = - \imag \left( x_k \gamma_k - \gamma_k
    x_k^{\dagger} \right) - y_k,
\end{equation}
where we define the Fourier transforms of $x_l$ and $y_l$ with a prefactor
$- \imag$ as
\begin{equation}
  \label{eq:x-k-y-k}
  \begin{split}
    x_k & = - \imag \sum_{l = 1}^L \e^{-\imag k l} x_l = x_{\id} \id +
    \mathbf{x}_k \cdot \boldsymbol{\sigma}, \\ y_k & = - \imag \sum_{l = 1}^L
    \e^{-\imag k l} y_l = \mathbf{y} \cdot \boldsymbol{\sigma}.
  \end{split}
\end{equation}
The coefficients in the expansions of $x_k$ and $y_k$ in terms of Pauli matrices
read
\begin{equation}
  \label{eq:x-k-components}
  x_{\id} = - \imag 2 \gamma, \quad \mathbf{x}_k =  \mathbf{a}_k + \imag
  \mathbf{b}, \quad \mathbf{b} = - 2 \sqrt{\gamma_{l}
    \gamma_{g}} \unitvec{e}_z,
\end{equation}
where $\mathbf{a}_k$ is given in Eq.~\eqref{eq:a-vector-k}, and
\begin{equation}
  \label{eq:y}
  \mathbf{y} = - 2 \delta \unitvec{e}_y.
\end{equation}
In the above expressions, we denote unit vectors in the $x$, $y$, and $z$
direction by $\unitvec{e}_{x, y, z}$. The solution to Eq.~\eqref{eq:gamma-k-eom}
is given by
\begin{equation}
  \label{eq:gamma-k-time-evolved}
  \gamma_k(t) = \gamma_{1, k}(t) + \gamma_{2, k}(t),
\end{equation}
where
\begin{equation}
  \label{eq:gamma-1-k-gamma-2-k}
  \begin{split}
    \gamma_{1, k}(t) & = \e^{-\imag x_k t} \gamma_k(0) \e^{\imag x_k^{\dagger}
      t}, \\
    \gamma_{2, k}(t) & = - \int_0^t d t' \, \e^{- \imag x_k \left( t - t'
      \right)} y_k \e^{\imag x_k^{\dagger} \left( t - t' \right)},
  \end{split}
\end{equation}
and the initial conditions are determined by
Eq.~\eqref{eq:gamma-k-ground-state},
$\gamma_k(0) = \unitvec{a}_{0, k} \cdot \boldsymbol{\sigma}$. In particular, for
$\mu_0 \to - \infty$, we find $\gamma_k(0) = \sigma_y$.

\subsubsection{Spectrum of the covariance matrix}
\label{sec:spectr-covar-matr}

For future reference, we briefly discuss the spectrum of the time-evolved
covariance matrix. To that end, it is convenient to define the coefficients
$\gamma_{\id, k}(t)$ and $\boldsymbol{\gamma}_k(t)$ in an expansion in terms of
Pauli matrices through the relation
\begin{equation}
  \label{eq:gamma-k-time-evolved-Pauli-expansion}
  \gamma_k(t) = \gamma_{\id, k}(t) \id + \boldsymbol{\gamma}_k(t) \cdot \boldsymbol{\sigma}.
\end{equation}
The eigenvalues of $\boldsymbol{\gamma}_k \cdot \boldsymbol{\sigma}$ are
$\pm \xi_k = \pm \lvert \boldsymbol{\gamma}_k \rvert$, where we omit the
explicit time dependence to lighten the notation. Therefore, the spectrum
$\sigma(\gamma_k)$ of the covariance matrix in
Eq.~\eqref{eq:gamma-k-time-evolved-Pauli-expansion} is given by
\begin{equation}
  \sigma(\gamma_k) = \{ \gamma_{\id, k} \pm \xi_k \}.
\end{equation}
We note that for an isolated system with unitary dynamics, time evolution
corresponds to precession of the vector $\boldsymbol{\gamma}_k(t)$ around
$\mathbf{a}_k$, with the explicit form of $\boldsymbol{\gamma}_k(t)$ given in
Eq.~\eqref{eq:gamma-vector-k-isolated} below. Then,
\begin{equation}  
  \gamma_{\id, k} = 0, \quad \xi_k = \abs{\unitvec{a}_{0, k}} = 1,
\end{equation}
and the spectrum is symmetric with respect to zero. This is, in general, not the
case for an open system. However, there are restrictions on the spectrum due to
PHS, which imply that the union of $\sigma(\gamma_k)$ and $\sigma(\gamma_{-k})$
is again symmetric with respect to zero. This can be seen as follows:
Equation~\eqref{eq:gamma-vector-k-PHS} implies that $\xi_k = \xi_{-k}$. Then,
with Eq.~\eqref{eq:gamma-id-k-PHS} we obtain
\begin{equation}
  \sigma(\gamma_{-k}) = \{ - \gamma_{\id, k} \pm \xi_k \}.
\end{equation}
Therefore, when we define
\begin{equation}
  \label{eq:xi-+-k-xi---k}
  \xi_{+, k} = \gamma_{\id, k} + \xi_k, \quad \xi_{-, k} =
  - \gamma_{\id, k} + \xi_k, 
\end{equation}
we can write
\begin{equation}
  \label{eq:gamma-k-gamma--k-spectrum}
  \sigma(\gamma_k) \cup \sigma(\gamma_{-k}) = \{ \pm \xi_{+, k}, \pm \xi_{-, k} \}.
\end{equation}

\subsection{Time evolution of the complex covariance matrix}

The equation of motion for the covariance matrix in the basis of complex
fermions, which is given in the main text, can be obtained by transforming
Eq.~\eqref{eq:gamma-k-eom} according to Eq.~\eqref{eq:gamma-k},
\begin{equation}
  \label{eq:g-k-eom}
  \frac{d g_k}{d t} = - \imag \left( z_k g_k - g_k
    z_k^{\dagger} \right) - s_k,
\end{equation}
where
\begin{equation}
  \label{eq:z-k-s-k}
  z_k = r x_k r^{\dagger} = z_{\id} \id + \mathbf{z}_k \cdot
  \boldsymbol{\sigma}, \qquad s_k = r y_k r^{\dagger} = \mathbf{s} \cdot \boldsymbol{\sigma}.
\end{equation}
By using Eqs.~\eqref{eq:r-dagger-sigma-r} and~\eqref{eq:P}, we obtain
\begin{equation}
  \label{eq:z-k-components-s-k}
  z_{\id} = - \imag 2 \gamma, \quad \mathbf{z}_k = \mathbf{h}_k - \imag 2
  \sqrt{\gamma_{l} \gamma_{g}} \unitvec{e}_x, \quad \mathbf{s} = - 2
  \delta \unitvec{e}_z.
\end{equation}
Rather than solving Eq.~\eqref{eq:g-k-eom} explicitly, we rotate
Eq.~\eqref{eq:gamma-k-time-evolved-Pauli-expansion} to the complex basis to
obtain the time-evolved covariance matrix in the form
\begin{equation}
  \label{eq:g-k-from-gamma-k}  
  g_k(t) = r \gamma_k(t) r^{\dagger} = g_{\id, k}(t) \id + \mathbf{g}_k(t)
  \cdot \boldsymbol{\sigma},
\end{equation}
where
\begin{equation}
  g_{\id, k}(t) = \gamma_{\id, k}(t), \quad \mathbf{g}_k(t) =
  P^{\transpose} \boldsymbol{\gamma}_k(t).
\end{equation}

\subsection{Symmetries of the Liouvillian}

Before we go into details of quench dynamics which are encoded in the time
dependence of the covariance matrix, we discuss relevant symmetries of the
Liouvillian $\mathcal{L}$ that generates the dynamics. According to
Eqs.~\eqref{eq:x-k-components} and~\eqref{eq:z-k-components-s-k}, the matrices
$x_k$ and $z_k$ can be regarded as generalizations of the matrices $a_k$ and the
BdG Hamiltonian $h_k$, respectively, to open systems. Therefore, by considering
$x_k$ and $z_k$, we can check directly how the symmetries of the isolated Kitaev
chain, which are summarized in Sec.~\ref{sec:symm-isol-kita}, are affected by
dissipation. We find that of the two versions of, e.g., PHS in
Eq.~\eqref{eq:H-BdG-PHS} for $\gamma = 0$, only one remains valid when
$\gamma > 0$. Further, IS and, therefore, PTS, apply only to the traceless parts
of $x_k$ and $z_k$, given by $x_k' = x_k + \imag 2 \gamma \id$ and
$z_k' = z_k + \imag 2 \gamma \id$, respectively. This is known as passive
PTS~\cite{Ornigotti2014, Joglekar2018, Roccati2022}. In particular, using the
terminology of Ref.~\cite{Kawabata2019}, the symmetries of $x_k$ are given by
\begin{align}
  \label{eq:x-k-PHS}
  & \text{PHS$^{\dagger}$:} & x_k & = - x_{-k}^{*}, \\
  \label{eq:x-k-TRS}
  & \text{TRS$^{\dagger}$:} & x_k & = \sigma_z x_{-k}^{\transpose} \sigma_z, \\
  \label{eq:x-k-CS}
  & \text{CS:} & x_k & = - \sigma_z x_k^{\dagger} \sigma_z, \\
  \label{eq:x-k-IS}
  & \text{IS$^{\dagger}$:} & x_k' & = \sigma_y x_{- k}^{\prime \dagger} \sigma_y, \\
  \label{eq:x-k-PTS}
  & \text{PTS:} & x_k' & = \sigma_x x_k^{\prime *} \sigma_x,
\end{align}
where PTS is the combination of TRS$^{\dagger}$ and IS$^{\dagger}$. For the sake
of completeness, we also list the symmetries of $z_k$, which result from
rotating the above relations to the complex basis:
\begin{align}
  \label{eq:z-k-PHS}
  & \text{PHS$^{\dagger}$:} & z_k & = - \sigma_x z_{-k}^{*} \sigma_x, \\
  \label{eq:z-k-TRS}
  & \text{TRS$^{\dagger}$:} & z_k & = z_{-k}^{\transpose}, \\
  \label{eq:z-k-CS}
  & \text{CS:} & z_k & = - \sigma_x z_k^{\dagger} \sigma_x, \\
  \label{eq:z-k-IS}
  & \text{IS$^{\dagger}$:} & z_k' & = \sigma_z z_{- k}^{\prime \dagger} \sigma_z, \\
  \label{eq:z-k-PTS}
  & \text{PTS:} & z_k' & = \sigma_z z_k^{\prime *} \sigma_z.
\end{align}
We next discuss implications of the above symmetry conditions that are relevant
to our studies of quench dynamics. First, we show that IS$^{\dagger}$ is not
sufficient to ensure that the time-evolved covariance matrix obeys IS as
specified in Eq.~\eqref{eq:g-k-gamma-k-IS}. To that end, we apply inversion to
the time-evolved covariance matrix Eq.~\eqref{eq:gamma-1-k-gamma-2-k}, whereby
we take into account that the initial state in
Eq.~\eqref{eq:gamma-k-ground-state}, as well as $y_k$ defined
Eq.~\eqref{eq:x-k-y-k}, obey inversion symmetry in the sense of
Eq.~\eqref{eq:g-k-gamma-k-IS},
\begin{equation}
  \begin{split}
    \sigma_y \gamma_{1, - k}(t) \sigma_y & = \e^{-\imag 2 x_{\id} t} \e^{-\imag
      \mathbf{x}_k^{\dagger} \cdot \boldsymbol{\sigma} t} \gamma_k(0) \e^{\imag
      \mathbf{x}_k \cdot \boldsymbol{\sigma} t}, \\
    \sigma_y \gamma_{2, - k}(t) \sigma_y & = - \int_0^t d t' \,
    \e^{-\imag 2 x_{\id} \left( t - t' \right)} \e^{- \imag
      \mathbf{x}_k^{\dagger} \cdot \boldsymbol{\sigma} \left( t - t' \right)}
    y_k \e^{\imag \mathbf{x}_k \cdot \boldsymbol{\sigma} \left( t - t' \right)}.
  \end{split}
\end{equation}
By comparing these expressions to Eq.~\eqref{eq:gamma-1-k-gamma-2-k}, we see
that inversion amounts to reversing the sign of the vector $\mathbf{b}$ in
Eq.~\eqref{eq:x-k-components}, which, in general, does not leave the state
invariant. For quenches to the PT-symmetric phase, this can be checked
explicitly in Eqs.~\eqref{eq:gamma-1-id-k-PT-symmetric},
\eqref{eq:gamma-1-vector-k-PT-symmetric}, \eqref{eq:gamma-2-id-k-PT-symmetric},
and~\eqref{eq:gamma-2-vector-k-PT-symmetric} below.

Further, as stated in the main text, and as we explain in detail in the
following, PTS has important consequences for the spectrum of
$z_k'$~\cite{Bender1998, *Bender2008}. Since
$z_k' = z_k + \imag 2 \gamma \id = \mathbf{z}_k \cdot \boldsymbol{\sigma}$, the
eigenvalues of $z_k'$ are given by
$\lambda_{\pm, k}' = \pm \sqrt{\mathbf{z}_k \cdot \mathbf{z}_k}$. We denote the
corresponding eigenvectors, which are also eigenvectors of $z_k$ with
eigenvalues $\lambda_{\pm, k} = \lambda_{\pm, k}' - \imag 2 \gamma$, by
$\ket{\psi_{\pm, k}}$. Then, the condition of PTS leads to
\begin{equation}
  \label{eq:PTS}
  z_k' \sigma_z \ket{\psi_{+, k}}^{*} = \lambda_{+, k}^{\prime *}
  \sigma_z \ket{\psi_{+, k}}^{*}.
\end{equation}
That is, $\sigma_z \ket{\psi_{+, k}}^{*}$ is an eigenvector of $z_k'$, and the
corresponding eigenvalue is given by $\lambda_{+, k}^{\prime *}$. Now, there are
two possibilities: (i)~$\lambda_{+, k}^{\prime *} = \lambda_{+, k}' \in \R$.
Then, $\sigma_z \ket{\psi_{+, k}}^{*} = \ket{\psi_{+, k}}$ (potentially, up to a
phase that can be absorbed in a redefiniton of $\ket{\psi_{+, k}}$). Further,
also $\lambda_{-, k}' = - \lambda_{+, k}' \in \R$, and
$\sigma_z \ket{\psi_{-, k}}^{*} = \ket{\psi_{-, k}}$. For the eigenvalues of
$z_k$, this implies $\Re(\lambda_{+, k}) = - \Re(\lambda_{-, k})$ and
$\Im(\lambda_{\pm, k}) = - 2 \gamma$. (ii)
$\lambda_{+, k}^{\prime *} = - \lambda_{+, k}' = \lambda_{-, k}' \in \imag \R$,
and $\sigma_z \ket{\psi_{\pm, k}}^{*} = \ket{\psi_{\mp, k}}$. For the
eigenvalues $z_k$, this implies $\Re(\lambda_{\pm, k}) = 0$ and
$\Im(\lambda_{+, k}) = - \Im(\lambda_{-, k}) - 4 \gamma$. We are, therefore,
lead to distinguish three phases: In the PT-symmetric phase, (i) applies to all
values of $k \in \BZ$, while (ii)~applies to all $k$ in the PT-broken
phase. Finally, in the PT-mixed phase, both PT-symmetric and PT-breaking
momentum modes exist.

Before we study how these phases are realized in the driven-dissipative Kitaev
chain, we briefly describe the form of PTS in position space, which is relevant,
in particular, for the analysis of systems with open boundary conditions. By
taking the inverse Fourier transform of Eqs.~\eqref{eq:x-k-TRS},
\eqref{eq:x-k-IS}, and~\eqref{eq:x-k-PTS} according to Eq.~\eqref{eq:x-k-y-k},
we obtain
\begin{align}
  & \text{TRS$^{\dagger}$:} & x_l & = \sigma_z x_{-l}^{\transpose} \sigma_z, \\
  & \text{IS$^{\dagger}$:} & x_l' & = - \sigma_y x_l^{\prime \transpose} \sigma_y, \\
  & \text{PTS:} & x_l' & = - \sigma_x x_{-l}' \sigma_x,
\end{align}
where we define $x_l' = x_l - 2 \gamma \id$. For the matrix $X$, which is
constructed from the $2 \times 2$ blocks $x_l$ like the matrix $A$ in
Eq.~\eqref{eq:A} is constructed from $a_l$, and the shifted matrix $X' = X - 2
\gamma \id$, we find
\begin{align}
  & \text{TRS$^{\dagger}$:} & X & = \Sigma_{z, L} X^{\transpose} \Sigma_{z, L}, \\
  & \text{IS$^{\dagger}$:} & X' & = - \Sigma_{y, L} \Xi_L X^{\prime \transpose} \Xi_L
                                  \Sigma_{y, L}, \\
  \label{eq:X-prime-PTS}
  & \text{PTS:} & X' & = - \Sigma_{x, L} \Xi_L X^{\prime} \Xi_L \Sigma_{x, L}.
\end{align}
where, for $\mu \in \{ x, y, z \}$,
\begin{equation}
  \label{eq:Sigma-mu-ell}
  \Sigma_{\mu, \ell} = \bigoplus_{l = 1}^{\ell} \sigma_{\mu},
\end{equation}
and $\Xi_{\ell}$ is a $2 \ell \times 2 \ell$ block anti-diagonal matrix, with
$2 \times 2$ identity matrices on the anti-diagonal,
\begin{equation}
  \label{eq:Xi-ell}
  \Xi_{\ell} =
  \begin{pmatrix}
    & & \id \\
    & \iddots & \\
    \id & &
  \end{pmatrix}.  
\end{equation}
According to our convention for the Fourier transform in Eq.~\eqref{eq:x-k-y-k},
for a system with periodic boundary conditions, the union of the spectra of the
matrices $x_k$ for all $k \in \BZ$ yields the spectrum of $- \imag X$.
Therefore, we focus on the consequences of PTS for the spectrum of the
latter. PTS in the form given in Eq.~\eqref{eq:X-prime-PTS} implies that for
each nonzero eigenvalue $\lambda'$ of
$- \imag X' = - \imag \left( X - 2 \gamma \id \right)$, which obeys the
eigenvalue equation $- \imag X' \ket{\psi} = \lambda' \ket{\psi}$, also
$- \lambda'$ is an eigenvalue, and the corresponding eigenvector is given by
$\Sigma_{x, L} \Xi_L \ket{\psi}$.  Further, since $- \imag X'$ is purely
imaginary, its eigenvalues $\lambda'$ are either purely imaginary or come in
pairs $\lambda'$ and $- \lambda^{\prime *}$.  Therefore, for each eigenvalue
$\lambda'$, there are three possibilities: (i)~There is a pair of real
eigenvalues given by $\pm \lambda' \in \R$.  (ii)~There is a pair of purely
imaginary eigenvalues given by $\pm \lambda' \in \imag \R$.  (iii)~If
$\Re(\lambda') \neq 0$ and $\Im(\lambda') \neq 0$, there are four eigenvalues
given by $\pm \lambda'$ and $\pm \lambda^{\prime *}$. Numerically, we find that
only cases (i)~and (ii)~are realized for the driven-dissipative Kitaev
chain. That is, the eigenmodes of
$- \imag X = - \imag \left( X' + 2 \gamma \id \right)$ are either
(i)~PT-symmetric, with eigenvalues
$\lambda = \lambda' - \imag 2 \gamma \in \R - \imag 2 \gamma$, or
(ii)~PT-breaking, with $\lambda \in \imag \R$.

\subsection{Spectrum of the Liouvillian}

As can be shown by employing the formalism of third
quantization~\cite{Prosen2008, Prosen2010}, for a system with periodic boundary
conditions, the spectrum of the Liouvillian is determined by the spectrum of the
matrix $x_k$ in Eq.~\eqref{eq:x-k-y-k}, or, equivalently, the matrix $z_k$
defined in Eq.~\eqref{eq:z-k-s-k}, similarly to the spectrum of the Hamiltonian
Eq.~\eqref{eq:H-Kitaev-momentum-space} that can be obtained by populating the
single-particle energy levels determined by the BdG Hamiltonian
Eq.~\eqref{eq:H-BdG}. The eigenvalues of $z_k$ are given by
\begin{equation}
  \label{eq:lambda-pm-k}
  \lambda_{\pm, k} = - \imag 2 \gamma \pm \omega_k,
\end{equation}
where
\begin{equation}
  \label{eq:omega-k}
  \omega_k = \sqrt{\varepsilon_k^2 - 4 \gamma_{l} \gamma_{g}}.
\end{equation}
As stated in the main text, the PT-symmetric phase, in which
$\omega_k \in \R_{> 0}$, is realized for
\begin{equation}
  2 \sqrt{\gamma_{l} \gamma_{g}} < \abs{2 J - \abs{\mu}}.
\end{equation}
Then, the bands $\lambda_{\pm, k}$ are separated by a real line
gap~\cite{Kawabata2019, Sayyad2021}. We note that for a system with open
boundary conditions, we find numerically that the spectrum of $- \imag X$ obeys
$\sigma(- \imag X) \in - \imag 2 \gamma + \R$ as suggested by
Eq.~\eqref{eq:lambda-pm-k} only for $\abs{\mu} > 2 J$. In contrast, for
$\abs{\mu} < 2 J$ there are PT-breaking edge modes with purely imaginary
eigenvalues in addition to fully PT-symmetric bulk of the
system~\cite{Sayyad2021}. For $\delta = 0$, the eigenvalues corresponding to the
edge modes are $0$ and $- \imag 4 \gamma$. Increasing the strength of
dissipation leads to the occurrence of PT-breaking modes also in the bulk. In
particular, in the PT-broken phase for
\begin{equation}
  2 \sqrt{\gamma_{l} \gamma_{g}} > 2 J + \abs{\mu},
\end{equation}
we write $\omega_k = \imag \kappa_k \in \imag \R_{> 0}$ with
\begin{equation}
  \label{eq:kappa-k}
  \kappa_k = \sqrt{4 \gamma_{l} \gamma_{g} - \varepsilon_k^2},
\end{equation}
such that Eq.~\eqref{eq:lambda-pm-k} becomes
\begin{equation}
  \lambda_{\pm, k} = - \imag \left( 2 \gamma \mp \kappa_k \right).
\end{equation}
In this phase, the bands $\lambda_{\pm, k}$ are separated by an imaginary line
gap. Finally, for intermediate strengths of dissipation, the spectrum is
gapless. In the gapless or PT-mixed phase, the bands $\lambda_{\pm, k}$ cross at
exceptional points at $k = \pm k_{*}$, at which $\omega_k$ defined in
Eq.~\eqref{eq:omega-k} vanishes, and which separate PT-symmetric from
PT-breaking modes.

For future reference, we note that $z_k$ can be diagonalized in the
PT-symmetric phase by combining an ordinary rotation with a hyperbolic
rotation, i.e., a rotation by an imaginary angle. In particular, with
\begin{equation}
  \label{eq:U-k}
  U_k =  U_{x, \theta_k} U_{y, - \imag \beta_k} = \e^{\imag \left( \theta_k/2
    \right) \sigma_x} \e^{\left( \beta_k/2 \right) \sigma_y},
\end{equation}
where
\begin{equation}
  \label{eq:beta-k}
  \beta_k = - \atanh(2 \sqrt{\gamma_{l}
    \gamma_{g}}/\varepsilon_k),
\end{equation}
we obtain
\begin{equation}
  z_k = - \imag 2 \gamma \id + \omega_k U_k \sigma_z U_k^{-1}.
\end{equation}
By rotating this relation to the real basis, we find that $x_k$ can be
diagonalized as (note the appearance of $\sigma_y$ instead of $\sigma_z$)
\begin{equation}
  \label{eq:x-k-diagonal-PT-symmetric}
  x_k = - \imag 2 \gamma \id + \omega_k S_k \sigma_y S_k^{-1},
\end{equation}
where
\begin{equation}
  \label{eq:S-k}
  S_k = r^{\dagger} U_k r = U_{z, \theta_k} U_{x, - \imag \beta_k} = \e^{\imag
    \left( \theta_k/2 \right) \sigma_z} \e^{\left( \beta_k/2 \right) \sigma_x}.
\end{equation}

The symmetries of $z_k$ impose restrictions on the transformation
$U_k$. Analogously to the derivation of Eq.~\eqref{eq:U-x-theta-k-IS}, on can
show that inversion symmetry leads to the relation
\begin{equation}
  \label{eq:U-k-IS}
  U_k = \sigma_z U_{-k}^{-\dagger} \sigma_z,
\end{equation}
which implies $\theta_k = - \theta_{-k}$ and $\beta_k = \beta_{-k}$.

\section{Quench dynamics in the PT-symmetric phase and relaxation to the PTGGE}

We now focus on quench dynamics in the PT-symmetric phase and, in particular, on
relaxation to the PT-symmetric generalized Gibbs ensemble (PTGGE). In the main
text, we derive the PTGGE, which describes the late-time dynamics in the
PT-symmetric phase, as the maximum-entropy ensemble that is compatible with
given dephased expectation values of commutators of eigenmodes of the
Liouvillian, and with the modified statistics of these modes as encoded in their
anticommutators. Below, we present details of this calculation. But first, we
present a derivation of the PTGGE that is based on the explicit form of the
time-evolved covariance matrix Eq.~\eqref{eq:gamma-k-time-evolved}.

\subsection{Dephasing and relaxation to the PTGGE}
\label{sec:deph-relax-bgge}

The most direct way to derive the PTGGE is to determine simplifications of the
time-evolved covariance matrix Eq.~\eqref{eq:gamma-k-time-evolved} in the
late-time limit due to dephasing. Technical details of the time evolution
generated by the matrix $x_k$, which can be interpreted as a non-Hermitian
Hamiltonian, are presented in the \hyperref[sec:time-evol-nH]{Appendix}.

\subsubsection{The covariance matrix in the PT-symmetric phase}

First, we provide explicit expressions for the components $\gamma_{1, k}(t)$ and
$\gamma_{2, k}(t)$ of the time-evolved covariance matrix defined in
Eq.~\eqref{eq:gamma-1-k-gamma-2-k}. With the initial condition
$\gamma_k(0) = \unitvec{a}_{0, k} \cdot \boldsymbol{\sigma}$ specified in
Eq.~\eqref{eq:gamma-k-ground-state}, we write $\gamma_{1, k}(t)$ in the form
\begin{equation}
  \label{eq:gamma-1-k}
  \gamma_{1, k}(t) = \e^{-\imag x_k t} \unitvec{a}_{0, k} \cdot
  \boldsymbol{\sigma} \e^{\imag x_k^{\dagger} t} = \gamma_{1, \id, k}(t) \id +
  \boldsymbol{\gamma}_{1, k}(t) \cdot \boldsymbol{\sigma}.
\end{equation}
Then, Eq.~\eqref{eq:gamma-id-PT-symmetric} leads to
\begin{equation}
  \label{eq:gamma-1-id-k-PT-symmetric}
  \gamma_{1, \id, k}(t) = - \frac{\e^{- 4 \gamma t}}{\omega_k^2} \left( 1 - \cos(2 \omega_k t)
  \right) \unitvec{a}_{0, k} \cdot \left( \mathbf{a}_k \times \mathbf{b}
  \right),
\end{equation}
and from Eq.~\eqref{eq:gamma-vector-PT-symmetric} we obtain
\begin{multline}
  \label{eq:gamma-1-vector-k-PT-symmetric}
  \boldsymbol{\gamma}_{1, k}(t) = \e^{- 4 \gamma t} \left[ \unitvec{a}_{0, k} -
    \frac{\varepsilon_k^2}{\omega_k^2} \left( 1 - \cos(2 \omega_k t) \right)
    \mathbf{a}_{\perp, k} \right. \\ \left. + \frac{\varepsilon_k}{\omega_k}
    \sin(2 \omega_k t) \mathbf{a}_{o, k} \right],
\end{multline}
where, with $\unitvec{a}_k = \mathbf{a}_k/\varepsilon_k$, the parallel,
perpendicular, and out-of-plane components are defined as
\begin{equation}
  \begin{split}
    \mathbf{a}_{\parallel, k} & = \left( \unitvec{a}_{0, k} \cdot \unitvec{a}_k
    \right) \unitvec{a}_k, \\ \mathbf{a}_{\perp, k} & = \unitvec{a}_{0, k} -
    \mathbf{a}_{\parallel, k}, \\ \mathbf{a}_{o, k} & = -
    \unitvec{a}_{0, k} \times \unitvec{a}_k.
  \end{split}
\end{equation}
Analogously, for
\begin{equation}
  \gamma_{2, k}(t) = \gamma_{2, \id, k}(t) \id + \boldsymbol{\gamma}_{2, k}(t)
  \cdot \boldsymbol{\sigma},
\end{equation}
we find
\begin{multline}
  \label{eq:gamma-2-id-k-PT-symmetric}
  \gamma_{2, \id, k}(t) = \int_0^t d t' \, \frac{\e^{- 4 \gamma \left( t - t'
      \right)}}{\omega_k^2} \\ \times \left( 1 - \cos(2 \omega_k \left( t - t'
    \right)) \right) \mathbf{y} \cdot \left( \mathbf{a}_k \times \mathbf{b}
  \right),
\end{multline}
and
\begin{multline}
  \label{eq:gamma-2-vector-k-PT-symmetric}
  \boldsymbol{\gamma}_{2, k}(t) = - \int_0^t d t' \, \e^{- 4 \gamma
    \left( t - t' \right)} \\ \times \left[ \mathbf{y} -
    \frac{\varepsilon_k^2}{\omega_k^2} \left( 1 - \cos(2 \omega_k \left( t - t'
      \right)) \right) \mathbf{y}_{\perp, k} \right. \\ \left. +
    \frac{\varepsilon_k}{\omega_k} \sin(2 \omega_k \left( t - t' \right))
    \mathbf{y}_{o, k} \right],
\end{multline}
where
\begin{equation}
  \begin{split}
    \mathbf{y}_{\parallel, k} & = \left( \mathbf{y} \cdot \unitvec{a}_k \right)
    \unitvec{a}_k, \\ \mathbf{y}_{\perp, k} & = \mathbf{y} -
    \mathbf{y}_{\parallel, k}, \\ \mathbf{y}_{o, k} & = - \mathbf{y}
    \times \unitvec{a}_k.
  \end{split}
\end{equation}
The integration in Eqs.~\eqref{eq:gamma-2-id-k-PT-symmetric}
and~\eqref{eq:gamma-2-vector-k-PT-symmetric} is elementary and we omit the
lengthy results.

\subsubsection{Late-time limit and dephasing}
\label{sec:late-time-limit-dephasing}

In isolated systems, the defining signature of generalized thermalization after
a quantum quench is local observables assuming stationary expectation values
that are determined by the GGE. For noninteracting and translationally invariant
systems, generalized thermalization occurs through dephasing of momentum modes
that oscillate at different frequencies $\varepsilon_k \neq \varepsilon_{k'}$
for $k \neq k'$~\cite{Essler2016, Barthel2008}. In particular, for the isolated
Kitaev chain, the component of the time-evolved covariance matrix with momentum
$k$ can be obtained by setting $\gamma = 0$ in
Eqs.~\eqref{eq:gamma-1-id-k-PT-symmetric},
\eqref{eq:gamma-1-vector-k-PT-symmetric}, \eqref{eq:gamma-2-id-k-PT-symmetric},
and~\eqref{eq:gamma-2-vector-k-PT-symmetric}, which yields
$\gamma_{1, \id, k}(t) = \gamma_{2, k}(t) = 0$, such that
$\gamma_k(t) = \boldsymbol{\gamma}_k(t) \cdot \boldsymbol{\sigma}$ with
\begin{equation}
  \label{eq:gamma-vector-k-isolated}
  \boldsymbol{\gamma}_k(t) = \mathbf{a}_{\parallel, k} + \cos(2
  \varepsilon_k t) \mathbf{a}_{\perp, k} + \sin(2
  \varepsilon_k t) \mathbf{a}_{o, k}.
\end{equation}
The long-time behavior that is established through dephasing is captured by
setting the oscillatory components to zero. Thus, with subscript ``\textit{d}''
for dephased,
\begin{equation}
  \label{eq:gamma-vector-k-isolated-asymptotic}
  \boldsymbol{\gamma}_{d, k} = \mathbf{a}_{\parallel, k} = \left(
    \unitvec{a}_{0, k} \cdot \unitvec{a}_k \right) \unitvec{a}_k,
\end{equation}
and the generalized Gibbs ensemble (GGE) is the Gaussian state that is uniquely
determined by the dephased covariance matrix
$\gamma_{d, k} = \boldsymbol{\gamma}_{d, k} \cdot
\boldsymbol{\sigma}$.
Since $\boldsymbol{\gamma}_{d, k} \parallel \unitvec{a}_k$, the
dephased covariance matrix, and, therefore, the GGE, is diagonal in the
eigenbasis of the postquench Hamiltonian. The diagonal elements are determined
by the conserved mode occupation numbers of the eigenmodes $d_k$ in
Eq.~\eqref{eq:Bogoliubov-quasiparticles} of the postquench Hamiltonian in the
initial state $\ket{\psi_0}$,
\begin{equation}
  \label{eq:conserved-occupations}
  \unitvec{a}_{0, k} \cdot \unitvec{a}_k =  \braket{\psi_0 | [d_k,
    d_k^{\dagger}] | \psi_0} = 1 - 2 \braket{\psi_0 | d_k^{\dagger} d_k | \psi_0},
\end{equation}
as follows from setting $\gamma = 0$ in~Eq.~\eqref{eq:eta-k-0} below. For
quenches in the driven-dissipative Kitaev chain, this picture is modified in two
fundamental ways:

(i)~The time dependence of the covariance matrix is described by renormalized
oscillation frequencies $\omega_k$ given in Eq.~\eqref{eq:omega-k} and,
crucially, there is additional exponential decay. However, due to PTS, there is
only a single relaxation rate, which, for the driven-dissipative Kitaev chain,
is determined by the imaginary part $- \Im(\lambda_{\pm, k}) = 2 \gamma$ of the
eigenvalues of $z_k$ in Eq.~\eqref{eq:lambda-pm-k}. After factoring out the
resulting trivial exponential time dependence, the dispersiveness of the real
part $\Re(\lambda_{\pm, k}) = \pm \omega_k$ induces dephasing and, consequently,
relaxation dynamics in analogy to isolated systems. In contrast, the dynamics of
open systems without PTS is typically governed by a range of exponential
relaxation rates, and the slowest of these rates determines the late-time
behavior, while dephasing does not play an important role. For the PT-symmetric
driven-dissipative Kitaev chain, as in
Eq.~\eqref{eq:gamma-vector-k-isolated-asymptotic}, dephasing can be accounted
for by setting oscillatory contributions to zero. Therefore, the late-time
behavior is described by
\begin{equation}
  \label{eq:gamma-k-PT-symmetric-asymptotic}
  \gamma_k(t) \sim \gamma_{d, k}(t) = \gamma_{d, 1, k}(t) +
  \gamma_{d, 2, k}(t) + \gamma_{\mathrm{SS}, k},
\end{equation}
with
\begin{equation}
  \label{eq:gamma-d-1-k}
  \gamma_{d, 1, k}(t) = \e^{- 4 \gamma t} \gamma_{d, 1, k}', \quad
  \gamma_{d, 1, k}' = \gamma_{d, 1, \id, k}' \id +
  \boldsymbol{\gamma}_{d, 1, k}' \cdot \boldsymbol{\sigma},
\end{equation}
where
\begin{equation}
  \label{eq:gamma-1-id-k-PT-symmetric-asymptotic}
  \gamma_{d, 1, \id, k}' = - \frac{1}{\omega_k^2} \unitvec{a}_{0, k}
  \cdot \left( \mathbf{a}_k \times \mathbf{b} \right) = \frac{2 \varepsilon_k
    \sqrt{\gamma_{l} \gamma_{g}}}{\omega_k^2} \sin(\Delta
  \theta_k),
\end{equation}
and
\begin{equation}
  \label{eq:gamma-1-vector-k-PT-symmetric-asymptotic}
\begin{split}
  \boldsymbol{\gamma}_{d, 1, k}' & = \unitvec{a}_{0, k} -
  \frac{\varepsilon_k^2}{\omega_k^2} \mathbf{a}_{\perp, k} \\ & = \cos(\Delta
  \theta_k) \unitvec{a}_k + \frac{4 \gamma_{l}
    \gamma_{g}}{\omega_k^2} \sin(\Delta \theta_k) \unitvec{e}_z \times
  \unitvec{a}_k.
\end{split}
\end{equation}
In these expressions, $\Delta \theta_k$ is the angle between $\unitvec{a}_{0,
  k}$ and $\unitvec{a}_k$, 
\begin{equation}
  \label{eq:Delta-theta-k}
  \cos(\Delta \theta_k) = \unitvec{a}_{0, k} \cdot \unitvec{a}_k.
\end{equation}
Analogously,
\begin{equation}
  \gamma_{d, 2, k}(t) = \e^{- 4 \gamma t} \gamma_{d, 2, k}',
\end{equation}
where
\begin{equation}
  \label{eq:gamma-d-2-k-prime}
  \gamma_{d, 2, k}' = - \frac{1}{4 \gamma} \left[ \frac{1}{\omega_k^2} \mathbf{y} \cdot
    \left( \mathbf{a}_k \times \mathbf{b} \right) \id - \left(
      \mathbf{y} - \frac{\varepsilon_k^2}{\omega_k^2} \mathbf{y}_{\perp, k}
    \right) \cdot \boldsymbol{\sigma} \right].
\end{equation}
Finally,
\begin{multline}
  \label{eq:gamma-ss-k}
  \gamma_{\mathrm{SS}, k} = \frac{1}{4 \gamma} \left\{ \frac{1}{\varepsilon_k^2
      + \delta^2} \mathbf{y} \cdot \left( \mathbf{a}_k \times \mathbf{b} \right)
    \id \right. \\ \left. - \left[ \mathbf{y} -
      \frac{\varepsilon_k}{\varepsilon_k^2 + \delta^2} \left( \varepsilon_k
        \mathbf{y}_{\perp, k} - 2 \gamma \mathbf{y}_{o, k} \right)
    \right] \cdot \boldsymbol{\sigma} \right).
\end{multline}

(ii)~The contribution $\gamma_{2, k}(t)$ in Eq.~\eqref{eq:gamma-k-time-evolved},
which describes the approach to the steady state $\rho_{\mathrm{SS}}$ defined by
$\mathcal{L}(\rho_{\mathrm{SS}}) = 0$ and given explicitly in terms of the
covariance matrix in Eq.~\eqref{eq:gamma-ss-k}, does not have a counterpart in
isolated systems. To understand the physical significance of $\gamma_{2, k}(t)$,
it is instructive to consider once more the isolated Kitaev chain for
comparison. There, due to the conservation of mode occupation numbers
Eq.~\eqref{eq:conserved-occupations}, a memory of the initial state is kept in
Eq.~\eqref{eq:gamma-vector-k-isolated-asymptotic}, and, therefore, in the
GGE. In the case of the driven-dissipative Kitaev chain, this memory of the
initial state is contained in $\gamma_{1, k}(t)$, and fades away due to the
exponential damping factor $\e^{-4 \gamma t}$ in
Eqs.~\eqref{eq:gamma-1-id-k-PT-symmetric}
and~\eqref{eq:gamma-1-vector-k-PT-symmetric}. At the same time, as time
progresses, the contribution $\gamma_{1, k}(t)$ is being overwritten by
$\gamma_{2, k}(t)$, which, for $t \to \infty$, approaches the stationary form
given in Eq.~\eqref{eq:gamma-ss-k}. This picture shows that the PTGGE, as a
generalization of the GGE, is captured by the dephased form of
$\gamma_{1, k}(t)$ given in Eq.~\eqref{eq:gamma-d-1-k},
\begin{equation}
  \label{eq:gamma-PTGGE-k}
  \gamma_{\mathrm{PTGGE}, k}(t) = \e^{-4 \gamma t} \gamma_{\mathrm{PTGGE}, k}',
\end{equation}
where $\gamma_{\mathrm{PTGGE}, k}' = \gamma_{d, 1 , k}'$, while the
growth and eventual dominance of $\gamma_{2, k}(t)$ restricts the time frame
during which relaxation to the PTGGE can be observed. As pointed out in the main
text, $\gamma_{2, k}(t)$ vanishes for
$\delta = \gamma_{l} - \gamma_{g} = 0$. Then, there is no
temporal constraint on the validity of the PTGGE. In contrast, when
$\delta \neq 0$, the PTGGE captures the late-time relaxation of local observables
up to a crossover time scale $t_{\times}$. As stated in the main text and
detailed below, the precise value of $t_{\times}$ depends on the observable
under consideration.

\subsubsection{From the covariance matrix to the density matrix}

According to the above discussion, the PTGGE is uniquely and fully determined by
the dephased form of the contribution to the covariance matrix
$\gamma_{1, k}(t)$ given in Eq.~\eqref{eq:gamma-d-1-k}. For the sake of
completeness, we describe in the following how the explicit form of the density
matrix corresponding to a given covariance matrix $\gamma_k$ can be found.

The dynamics generated by the Liouvillian $\mathcal{L}$ in
Eq.~\eqref{eq:master-equation} couples only modes with momenta $\pm k$, and we
can consider the subspaces corresponding to these modes separately. In one such
subspace, for $k \notin \{ 0, \pi \}$ such that $k \neq -k$, a complete set of
fermionic operators is given by the doubled spinors of complex $\mathsf{C}_k$
and Majorana $\mathsf{W}_k$ fermions:
\begin{equation}
  \mathsf{C}_k =
  \begin{pmatrix}
    c_k \\ c_k^{\dagger} \\ c_{-k} \\ c_{-k}^{\dagger}
  \end{pmatrix}, \qquad \mathsf{W}_k =
  \begin{pmatrix}
    w_{1, k} \\ w_{2, k} \\ w_{3, k} \\ w_{4, k}
  \end{pmatrix},
\end{equation}
which are related by
\begin{equation}
  \mathsf{C}_k = \frac{1}{\sqrt{2}} \mathsf{R} \mathsf{W}_k, \qquad \mathsf{R} =
  r \oplus r.
\end{equation}
We note that the operators $\mathsf{W}_k$ describe true Majorana fermions, i.e.,
they are their own antiparticles, in contrast to the Fourier transform of the
Majorana modes defined in Eq.~\eqref{eq:W-k}. In analogy to
Eq.~\eqref{eq:Gamma}, we define a covariance matrix of $\mathsf{W}_k$ fermions
as
\begin{equation}
  \label{eq:y-k}
  \mathsf{y}_k = \imag \left( \langle \mathsf{W}_k \mathsf{W}_k^{\transpose}
    \rangle - \id \right) = \imag \mathsf{R}^{\dagger} \left( 2 \langle
    \mathsf{C}_k \mathsf{C}_k^{\dagger} \rangle - \id \right) \mathsf{R}.
\end{equation}
To establish the relation between $\mathsf{y}_k$ and the covariance matrix
$\gamma_k$ in Eq.~\eqref{eq:gamma-k} as well as the complex covariance matrix
$g_k$ in Eq.~\eqref{eq:g-k}, we note that the spinors in Eq.~\eqref{eq:C-k} can
be expressed in terms of the doubled spinors $\mathsf{C}_k$ as
\begin{equation}
  \label{eq:C-double-spinor}
  \begin{pmatrix}
    C_k \\ C_{-k}
  \end{pmatrix}
  =
  \begin{pmatrix}
    c_k \\ c_{-k}^{\dagger} \\ c_{-k} \\ c_k^{\dagger}
  \end{pmatrix}
  = \mathsf{P} \mathsf{C}_k, \quad \mathsf{P} =
  \begin{pmatrix}
    1 & 0 & 0 & 0 \\ 0 & 0 & 0 & 1 \\ 0 & 0 & 1 & 0 \\ 0 & 1 & 0 & 0
  \end{pmatrix}.
\end{equation}
Therefore,
\begin{equation}
  \label{eq:y-k-gamma-k}
  \begin{split}
    \mathsf{y}_k & = \imag \mathsf{R}^{\dagger} \mathsf{P} \left( 2
      \begin{pmatrix}
        \langle C_k C_k^{\dagger} \rangle & \langle C_k C_{-k}^{\dagger} \rangle
        \\ \langle C_{-k} C_k^{\dagger} \rangle & \langle C_{-k}
        C_{-k}^{\dagger} \rangle
      \end{pmatrix}
      - \id \right) \mathsf{P} \mathsf{R} \\ & = \imag \mathsf{R}^{\dagger}
    \mathsf{P}
    \begin{pmatrix}
        g_k & 0 \\ 0 & g_{-k}
      \end{pmatrix}
      \mathsf{P} \mathsf{R} = \imag \mathsf{Q}
      \begin{pmatrix}
        \gamma_k & 0 \\ 0 & \gamma_{-k}
      \end{pmatrix}
      \mathsf{Q},
  \end{split}
\end{equation}
where
\begin{equation}
  \mathsf{Q} = \mathsf{R}^{\dagger} \mathsf{P} \mathsf{R}.
\end{equation}
The corresponding density matrix is given by~\cite{Peschel2003, Peschel2009,
  Vidal2003}
\begin{equation}
  \label{eq:rho-k}
  \rho_k = \frac{1}{Z_k} \e^{\frac{\imag}{2} \mathsf{W}_k^{\transpose}
    \atanh(\mathsf{y}_k) \mathsf{W}_k} = \frac{1}{Z_k} \e^{- 2 C_k^{\dagger}
    \atanh(g_k) C_k + \tr(g_k)},
\end{equation}
where the normalization $Z_k$ is to be chosen such that $\tr(\rho_k) = 1$, which
leads to
\begin{equation}
  \label{eq:density-matrix-from-covariance-matrix}
  \rho_k = \frac{1}{\det \! \left( \id + \e^{-2 \atanh(g_k)} \right)} \e^{-2
    C_k^{\dagger} \atanh(g_k) C_k}.
\end{equation}
This form applies also to $k \in \{ 0, \pi \}$ when we define, for these
momenta, $C_k = c_k$ and $g_k = \langle [c_k, c_k^{\dagger}] \rangle$. Then, the
full density matrix is given by
\begin{equation}
  \rho = \prod_{k > 0} \rho_k.
\end{equation}
According to the discussion in Sec.~\ref{sec:late-time-limit-dephasing}, we
obtain the PTGGE by inserting
$g_k(t) = r \gamma_{\mathrm{PTGGE}, k}(t) r^{\dagger}$, with
$\gamma_{\mathrm{PTGGE}, k}(t)$ given in Eq.~\eqref{eq:gamma-PTGGE-k}. In the
next section, we establish the equivalence between the resulting expression for
$\rho_{\mathrm{PTGGE}}(t)$ and the form given in Eq.~(8) in the main text in
terms of eigenmodes of the adjoint Liouvillian.

\subsection{Derivation of the PTGGE from the principle of maximum entropy}
\label{sec:maximum-entropy-principle}

As described in the main text, the PTGGE can be obtained as the maximum entropy
ensemble that is compatible with the modified statistics and dephased
expectation values of commutators of eigenmodes of the Liouvillian. In the
following, we present details of this derivation.

\subsubsection{Eigenmodes of the adjoint Liouvillian}

Our goal is to find eigenmodes of the Liouvillian, which generalize the
Bogoliubov quasiparticle operators defined in
Eq.~\eqref{eq:Bogoliubov-quasiparticles} to the driven-dissipative
setting. Specifically, for the isolated Kitaev chain, an operator $O$ evolves in
the Heisenberg picture according to $d O/d t = \imag [H, O]$,
i.e., the dynamics is generated by the superoperator $\mathcal{H} O = [H, O]$.
The Bogoliubov quasiparticle operators $d_k$ are eigenmodes of $\mathcal{H}$
in the sense that they obey the eigenvalue equation
$\mathcal{H} d_k = - \varepsilon_k d_k$. As we explain below, this definition of
eigenmodes of the generator of dynamics cannot directly be generalized to
fermionic open systems. Instead of fermionic mode operators $d_k$ themselves, we
consider bilinear forms of operators,
\begin{equation}
  \label{eq:eta-k-chi-k}
  \eta_k = [d_k, d_k^{\dagger}], \qquad \chi_k = [d_k, d_{-k}],
\end{equation}
which in the isolated Kitaev chain obey $\mathcal{H} \eta_k = 0$ and
$\mathcal{H} \chi_k = - 2 \varepsilon_k \chi_k$ such that
$\langle \eta_k(t) \rangle = \langle \eta_k \rangle_0$ is conserved, while
$\langle \chi_k(t) \rangle$ is purely oscillatory and vanishes via dephasing.
We generalize this structure to open systems in three steps: (i)~We specify the
generator of dynamics for operators, and thereby clarify which eigenvalue
equation we actually want to solve. (ii)~To find a solution to the eigenvalue
equation, we make a bilinear ansatz in the form of $\eta_k$, which leads to an
explicit expression for the mode operators $d_k$. We find that the expectation
value $\langle \eta_k(t) \rangle$ is not conserved but nonoscillatory, i.e., not
subject to dephasing. (iii)~The form of $\eta_k$ and, in particular, the mode
operators $d_k$, completely determines $\chi_k$. As we show, the expectation
value $\langle \chi_k(t) \rangle$ is the sum of constant and purely oscillatory
contributions. The constant part vanishes when $\delta = 0$, such that, as in
the isolated Kitaev chain, $\langle \chi_k(t) \rangle$ does not contribute to
the PTGGE.

\paragraph*{(i) Adjoint Liouvillian.} For any operator $O$, the expectation
value $\langle O \rangle = \tr(O \rho)$ obeys the equation of motion
\begin{equation}
  \label{eq:O-eom}
  \frac{d}{d t} \langle O \rangle = \tr \! \left(
    O \frac{d \rho}{d t} \right) = \imag \tr \!
  \left( O \mathcal{L} \rho \right) = \imag \tr \! \left( \rho
    \mathcal{L}^{\dagger} O \right) = \imag \langle \mathcal{L}^{\dagger}
  O \rangle,
\end{equation}
with the adjoint Liouvillian,
\begin{equation}
  \mathcal{L}^{\dagger} = \mathcal{H} - \imag \mathcal{D}^{\dagger}, 
\end{equation}
where $\mathcal{H}$ is given in Eq.~\eqref{eq:H-adjoint-action}, and
\begin{equation}
  \label{eq:dissipator-dagger}
  \begin{split}
    \mathcal{D}^{\dagger} O & = \sum_{l = 1}^L \left( 2 L_l^{\dagger} O L_l - \{
      L_l^{\dagger} L_l, O \} \right) \\ & = \sum_{l = 1}^L \left( L_l^{\dagger}
      [O, L_l] + [L_l^{\dagger}, O] L_l \right).
  \end{split}
\end{equation}
As a general remark, we note that the adjoint Liouvillian generates the dynamics
of operators only in the sense of expectation values, while the Heisenberg
picture for open systems is defined in terms of quantum Langevin
equations~\cite{Gardiner2000, Wiseman2010}.

Now, if $O$ is an eigenmode of $\mathcal{L}^{\dagger}$ in the
following sense:
\begin{equation}
  \label{eq:O-eigenvalue-equation}
  \mathcal{L}^{\dagger} O = \lambda^{*} \left( O - O_{\mathrm{SS}} \right),
\end{equation}
where $O_{\mathrm{SS}} \in \R$, then,
\begin{equation}
  \frac{d}{d t} \langle O \rangle = \imag \lambda^{*} \left( \langle
    O \rangle - O_{\mathrm{SS}} \right),
\end{equation}
which is solved by
\begin{equation}
  \label{eq:O-time-evolved}
  \langle O(t) \rangle = O_{\mathrm{SS}} + \e^{\imag \lambda^{*} t} \left( \langle
    O(0) \rangle - O_{\mathrm{SS}} \right).
\end{equation}
For $\Im(\lambda^{*}) > 0$, we obtain
$\langle O(t) \rangle \to O_{\mathrm{SS}} = \langle O_{\mathrm{SS}} \rangle =
\tr(O \rho_{\mathrm{SS}})$
for $t \to \infty$. As we show in the following, nonoscillatory eigenmodes, for
which $\lambda$ is purely imaginary, $\lambda^{*} = \imag \kappa$, can be found
for operators $O$ in the form of $\eta_k$ in Eq.~\eqref{eq:eta-k-chi-k}, while
an ansatz in the form of $\chi_k$ leads to oscillatory eigenmodes. Here and in
the following, we use the term eigenmodes both for the bilinears $\eta_k$ and
$\chi_k$ and the fermionic operators $d_k$.

\paragraph*{(ii) Nonoscillatory normal commutators.} To find nonoscillatory
eigenmodes of the adjoint Liouvillian, we make an ansatz in the form of $\eta_k$
in Eq.~\eqref{eq:eta-k-chi-k}. In fact, for the time being, we do not assume
translational invariance, i.e., we do not assume that the eigenmodes can be
labelled by the momentum $k$. Then, $\eta = [d, d^{\dagger}]$ can be written in
the form
\begin{equation}
  \label{eq:eta-ansatz}
  \eta = \frac{\imag}{4} \sum_{l, l' = 1}^{2 L} w_l Q_{l, l'} w_{l'},
\end{equation}
where, without loss of generality, $Q$ can be assumed to be antisymmetric,
$Q = - Q^{\transpose}$. Further, $\eta$ is Hermitian if $Q$ is real, and $\eta$
is traceless by construction. We proceed to insert our ansatz for $\eta$ in
Eq.~\eqref{eq:O-eigenvalue-equation}. The contribution from the Hamiltonian is
given by
\begin{equation}  
  \mathcal{H} \eta = - \frac{1}{4} \sum_{l, l' = 1}^{2 L}
  w_l [A, Q]_{l, l'} w_{l'},
\end{equation}
and applying the adjoint dissipator to $\eta$ yields
\begin{equation}
  \begin{split}
    \mathcal{D}^{\dagger} \eta & = \sum_{m, m' = 1}^{2 L} M_{m', m}
    \left( w_m [\eta, w_{m'}] + [w_m, \eta] w_{m'} \right)
    \\ & = - \imag \sum_{l, l' = 1}^{2 L} w_l \{ M_R, Q \}_{l, l'} w_{l'} + 2 \tr(M_I Q).
  \end{split}
\end{equation}
where the bath matrix $M = M_R + \imag M_I$ is defined in Eq.~\eqref{eq:M}. We
identify the last, constant term in the above expression with the steady-state
expectation value $\eta_{\mathrm{SS}}$,
\begin{equation}
  \label{eq:eta-ss}
  \eta_{\mathrm{SS}} = - \frac{2}{\kappa} \tr(M_I Q),
\end{equation}
where we also anticipate that $\lambda^{*} = \imag \kappa$ is purely
imaginary. Then, Eq.~\eqref{eq:O-eigenvalue-equation} takes the form
\begin{equation}
  \sum_{l, l' = 1}^{2 L} w_l \left( [A, Q] + 4 \{ M_R, Q \} \right)_{l, l'}
  w_{l'} = \kappa \sum_{l, l' = 1}^{2 L} w_l Q_{l, l'} w_{l'},
\end{equation}
Next, we multiply this equation by $w_l w_{l'}$ and use
\begin{equation}
  \frac{1}{2^{L + 1}} \tr(w_{l'} w_l w_m w_{m'}) = \frac{1}{2} \left(
    \delta_{l, l'} \delta_{m, m'} - \delta_{l, m'} \delta_{l', m} + \delta_{l,
      m} \delta_{l', m'} \right),
\end{equation}
to obtain
\begin{equation}
  \label{eq:Q-eigenvalue-equation}
  X^{\transpose} Q + Q X = \kappa Q,
\end{equation}
where matrix $X$ is defined in Eq.~\eqref{eq:X-Y}. Now that we have illustrated
the derivation of the eigenvalue equation for the bilinear $\eta$, let us
briefly comment on the possibility of finding eigenmodes of
$\mathcal{L}^{\dagger}$ that are linear in fermionic operators. In general, such
linear eigenmodes cannot be found because there is no closed eigenvalue equation
in the space of linear operators. As a simple example, consider a lattice model
without Hamiltonian dynamics, $H = 0$, and incoherent loss,
$L_l = \sqrt{\gamma_{l}} c_l$.  Then, by applying the adjoint
dissipator Eq.~\eqref{eq:dissipator-dagger} to the fermionic operator $c_l$
yields
\begin{equation}
  \mathcal{D}^{\dagger} c_l = - \gamma_{l} \left( 1 + 4 \sum_{l' = 1}^L
    c_{l'}^{\dagger} c_{l'} \right) c_l,
\end{equation}
i.e., a sum of linear and cubic terms. In contrast, the second equality in
Eq.~\eqref{eq:dissipator-dagger} shows that, if $O$ is bilinear and $L_l$ is
linear in fermionic operators, also $\mathcal{D}^{\dagger} O$ is bilinear, and,
therefore, a closed eigenvalue equation can always be found for a bilinear
ansatz. The situation is very different for bosons: Then, the second equality in
Eq.~\eqref{eq:dissipator-dagger} shows that, if both $O$ and $L_l$ are linear in
bosonic operators, also $\mathcal{D}^{\dagger} O$ is linear.

To proceed, we exploit translational invariance of the driven-dissipative Kitaev
chain, which implies that there is an eigenmode $\eta_k$, expressed in terms of
a matrix $Q_k$, for each momentum mode $k$. We refine our ansatz in
Eq.~\eqref{eq:eta-ansatz} as follows:
\begin{equation}
  \label{eq:eta-k-W-l}
  \eta_k = \frac{\imag}{4} \sum_{l, l' = 1}^L W_l^{\transpose} q_{k, l - l'} W_{l'},
\end{equation}
where the spinors $W_l$ of Majorana operators are defined in Eq.~\eqref{eq:W-l},
and where we assume that $Q_k$ is a block Toeplitz matrix with $2 \times 2$
blocks
\begin{equation}
  \label{eq:q-k-l}
  q_{k, l - l'} =
  \begin{pmatrix}
    Q_{k, 2 l - 1, 2 l' - 1} & Q_{k, 2 l - 1, 2 l'} \\ Q_{k, 2 l, 2 l' - 1} &
    Q_{k, 2 l, 2 l'}
  \end{pmatrix}.
\end{equation}
We define the Fourier transform of $q_{k, l}$ as
\begin{equation}
  \label{eq:q-k-k-prime}
  q_{k, k'} = \imag \sum_{l = 1}^L \e^{-\imag k' l} q_{k, l}.
\end{equation}
For future reference, we note that, since $Q$ is real and antisymmetric, the
Fourier transform $q_{k, k'}$ of $q_{k, l}$ obeys the relations
\begin{equation}
  \label{eq:q-k-symmetries}
    q_{k, k'} = - q_{k, -k'}^{*}, \qquad q_{k, k'} = - q_{k, - k'}^{\transpose},
\end{equation}
which imply $q_{k, k'} = q_{k, k'}^{\dagger}$. By inserting
Eqs.~\eqref{eq:q-k-l} and~\eqref{eq:q-k-k-prime} in
Eq.~\eqref{eq:Q-eigenvalue-equation} we obtain
\begin{equation}
  \label{eq:q-k-eigenvalue-equation}
  x_{k'}^{\dagger} q_{k, k'} - q_{k, k'} x_{k'} = \imag \kappa q_{k, k'}.
\end{equation}
In the PT-symmetric phase, according to Eq.~\eqref{eq:lambda-pm-k}, the spectrum
of $x_k$ is given by $\sigma(x_k) = \{ - \imag 2 \gamma \pm \omega_k \}$, and
$x_k$ can be diagonalized as specified in
Eq.~\eqref{eq:x-k-diagonal-PT-symmetric}. This leads to
\begin{equation}
  \omega_{k'} \left( S_{k'}^{- \dagger} \sigma_y S_{k'}^{\dagger} q_{k, k'} - q_{k, k'} S_{k'}
    \sigma_y S_{k'}^{-1} \right) = \imag \left( \kappa - 4 \gamma \right) q_{k, k'},
\end{equation}
where we use the shorthand notation
$S_k^{- \dagger} = \left( S_k^{\dagger} \right)^{-1}$. On the right-hand side,
we identify $\kappa = 4 \gamma$, and are thus led to the following commutation
relation:
\begin{equation}
  \left[ \sigma_y, S_{k'}^{\dagger} q_{k, k'} S_{k'} \right] = 0.
\end{equation}
The matrices that commute with $\sigma_y$ are $\id$ and $\sigma_y$
itself. Therefore, the general solution for $q_{k, k'}$, which obeys the above
commutation relation, reads
\begin{equation}
  \label{eq:q-k-k-prime-general-solution}
  q_{k, k'} = S_{k'}^{- \dagger} \left( \alpha_{+, k, k'} \sigma_y + \alpha_{-, k, k'} \id \right) S_{k'}^{-1}, 
\end{equation}
with undetermined coefficients $\alpha_{+, k, k'}$ and $\alpha_{-, k, k'}$. This
solution can be seen to obey the conditions Eq.~\eqref{eq:q-k-symmetries} if
$\alpha_{+, k, k'}$ and $\alpha_{-, k, k'}$ are, respectively, symmetric and
antisymmetric functions of $k'$, and by using that PHS$^{\dagger}$ of $x_k$ in
Eq.~\eqref{eq:x-k-PHS} implies
\begin{equation}
  \label{eq:S-k-PHS}
  S_k = S_{-k}^{*},
\end{equation}
in analogy to Eq.~\eqref{eq:U-x-theta-k-PHS}. We proceed to fix the coefficients
$\alpha_{+, k, k'}$ and $\alpha_{-, k, k'}$ such that $\eta_k$ can be written in
terms of mode operators $d_k$ as $\eta_k = [d_k, d_k^{\dagger}]$. To this end,
we first express $\eta_k$ in terms of complex fermions. By inserting
Eq.~\eqref{eq:W-k} in Eq.~\eqref{eq:eta-k-W-l} we obtain
\begin{equation}
  \label{eq:eta-k-W-k-C-k}
  \begin{split}
    \eta_k & = \frac{1}{4} \sum_{k' \in \BZ} W_{k'}^{\dagger} q_{k, k'} W_{k'} =
    \frac{1}{2} \sum_{k' \in \BZ} C_{k'}^{\dagger} r q_{k, k'} r^{\dagger}
    C_{k'} \\ & = \frac{1}{2} \sum_{k' \in \BZ} C_{k'}^{\dagger} U_{k'}^{-
      \dagger} \left( \alpha_{+, k, k'} \sigma_z + \alpha_{-, k, k'} \id \right)
    U_{k'}^{-1} C_{k'},
  \end{split}
\end{equation}
where, in the last line, we use Eq.~\eqref{eq:S-k}. For solutions $\eta_k$ with
fixed momentum $k$, the sum over $k'$ should collapse to the two terms
$k' = \pm k$. Symmetric and antisymmetric combinations of these components are
given by
\begin{equation}
  \label{eq:alpha-s-a}
  \alpha_{\pm, k, k'} = - \frac{\alpha_{0, k}}{2} \left( \delta_{k', k} \pm
    \delta_{k', - k} \right),
\end{equation}
with the coefficient $\alpha_{0, k}$ to be determined below. Then, with
\begin{equation}
  P_{z, \pm} = \frac{1}{2} \left( \id \pm \sigma_z \right),
\end{equation}
we find
\begin{equation}
  \label{eq:eta-k-C-k}
  \begin{split}
    \eta_k & = \frac{\alpha_{0, k}}{2} \left( C_{-k}^{\dagger} U_{-k}^{-\dagger}
      P_{z, -} U_{-k}^{-1} C_{-k} - C_k^{\dagger} U_k^{-\dagger} P_{z, +}
      U_k^{-1} C_k \right) \\ & = \frac{\alpha_{0, k}}{2} \left(
      C_k^{\transpose} U_k^{-\transpose} P_{z, +} U_k^{- *} C_k^{\dagger
        \transpose} - C_k^{\dagger} U_k^{-\dagger} P_{z, +} U_k^{-1} C_k
    \right),
  \end{split}
\end{equation}
where in the second equality we use that $U_k$, which is defined in
Eq.~\eqref{eq:U-k}, obeys the same PHS as $U_{x, \theta_k}$ in
Eq.~\eqref{eq:U-x-theta-k-PHS}. $\eta_k$ takes its final form,
$\eta_k =[d_k, d_k^{\dagger}]$, if we define the eigenmodes of the Liouvillian
as
\begin{equation}
  \label{eq:D-k-V-k}
  D_k =
  \begin{pmatrix}
    d_k \\ d_{-k}^{\dagger}
  \end{pmatrix}
  = V_k^{\dagger} C_k,
\end{equation}
where
\begin{equation}
  \label{eq:V-k}
  V_k = \sqrt{\frac{2}{\tr \! \left( U_k^{-1} U_k^{-\dagger} \right)}}
    U_k^{-\dagger} =
  \begin{pmatrix}
    \cos \! \left( \frac{\theta_k + \phi_k}{2} \right) & \imag \sin \! \left(
      \frac{\theta_k - \phi_k}{2} \right) \\ \imag \sin \! \left( \frac{\theta_k
        + \phi_k}{2} \right) & \cos \! \left( \frac{\theta_k - \phi_k}{2}
    \right)
  \end{pmatrix}.
\end{equation}
The angle $\theta_k$ is defined in Eq.~\eqref{eq:theta-k}, and $\phi_k$ is
defined through the equality
\begin{equation}
  \label{eq:phi-k}
  \varepsilon_k \e^{\imag \phi_k} = \omega_k + \imag 2 \sqrt{\gamma_{l}
    \gamma_{g}},
\end{equation}
and is related to $\beta_k$ defined in Eq.~\eqref{eq:beta-k} via
\begin{equation}
  \label{eq:phi-k-beta-k}
  \tan(\phi_k) = - \frac{\varepsilon_k}{\omega_k} \tanh(\beta_k).
\end{equation}
When we insert Eq.~\eqref{eq:D-k-V-k} in Eq.~\eqref{eq:eta-k-C-k}, the
normalization in Eq.~\eqref{eq:V-k} cancels if we set
\begin{equation}
  \label{eq:alpha-0-k}
  \alpha_{0, k} = \tr \! \left( U_k^{-1} U_k^{-\dagger} \right) = 2 \cosh(\beta_k).
\end{equation}
The normalization in Eq.~\eqref{eq:V-k} is chosen such that
$\{ d_k, d_k^{\dagger} \} = 1$. As in the main text, we collect the
anticommutators of the modes $d_k$ in a matrix,
\begin{equation}
  \begin{pmatrix}
    \{ d_k, d_{k'}^{\dagger} \} & \{ d_k, d_{-k'} \} \\ \{ d_{-k}^{\dagger},
    d_{k'}^{\dagger} \} & \{ d_{-k}^{\dagger}, d_{-k'} \}
  \end{pmatrix}
  = f_k \delta_{k, k'},  
\end{equation}
where
\begin{equation}
  \label{eq:f-k}
  f_k =
  \begin{pmatrix}
    \{ d_k, d_{k}^{\dagger} \} & \{ d_k, d_{-k} \} \\ \{ d_{-k}^{\dagger},
    d_{k}^{\dagger} \} & \{ d_{-k}^{\dagger}, d_{-k} \}
  \end{pmatrix}
  = V_k^{\dagger} V_k = \id + \sin(\phi_k) \sigma_y.
\end{equation}

Next, we evaluate Eq.~\eqref{eq:eta-ss}. A short calculation yields
\begin{equation}
  \label{eq:eta-ss-k}
  \eta_{\mathrm{SS}, k} = \langle [d_k, d_k^{\dagger}] \rangle_{\mathrm{SS}} = -
  \frac{\delta \omega_k}{4 \varepsilon_k \gamma} \left( \cos(\theta_k) - \frac{2
      \sqrt{\gamma_{l} \gamma_{g}}}{\omega_k} 
    \sin(\theta_k) \right).
\end{equation}

Finally, to completely specify the time evolution of the expectation value of
$\eta_k$, we need to know the expectation value $\langle \eta_k \rangle_0$ in
the initial state. Starting from Eq.~\eqref{eq:eta-k-C-k}, we use
Eqs.~\eqref{eq:g-k} to find
\begin{equation}  
  \langle [d_k, d_k^{\dagger}] \rangle = \frac{\omega_k}{\varepsilon_k} \tr \!
  \left( U_k^{-\dagger} P_{z, +} U_k^{-1} g_k \right).  
\end{equation}
For the ground state, with covariance matrix given in
Eq.~\eqref{eq:g-k-ground-state}, we obtain
\begin{equation}
  \label{eq:eta-k-0}
  \langle [d_k, d_k^{\dagger}] \rangle_0 = \cos(\Delta \theta_k - \phi_k),
\end{equation}
where $\Delta \theta_k$ is defined in Eq.~\eqref{eq:Delta-theta-k}. Putting
everything together, we obtain the result stated in the main text:
\begin{equation}
  \langle [d_k, d_k^{\dagger}](t) \rangle = \e^{- 4 \gamma t} \langle [d_k,
  d_k^{\dagger}] \rangle_0 + \left( 1 - \e^{- 4 \gamma t} \right) \langle [d_k,
  d_k^{\dagger}] \rangle_{\mathrm{SS}},
\end{equation}
where $\langle [d_k, d_k^{\dagger}] \rangle_0$ and  $\langle [d_k,
d_k^{\dagger}] \rangle_{\mathrm{SS}}$ are given in Eqs.~\eqref{eq:eta-k-0}
and~\eqref{eq:eta-ss-k}, respectively.

\paragraph*{(iii) Oscillatory anomalous commutators.} We next show that also
$\chi_k = [d_k, d_{-k}]$ is an eigenmode of the Liouvillian in the sense of
Eq.~\eqref{eq:O-eigenvalue-equation}, however, where $\Re(\lambda^{*}) \neq 0$,
such that the nonconstant term in Eq.~\eqref{eq:O-time-evolved} is
oscillatory. To that end, in analogy to Eq.~\eqref{eq:eta-k-W-k-C-k}, we write
$\chi_k$ as
\begin{equation}
  \chi_k = \frac{1}{2} \sum_{k' \in \BZ} C_{k'}^{\dagger} r p_{k, k'} r^{\dagger} C_{k'},  
\end{equation}
where we define
\begin{equation}
  p_{k, k'} = - 2 \left( \delta_{k, k'} - \delta_{k, - k'} \right) r^{\dagger}
  V_{k'} \sigma_- V_{k'}^{\dagger} r.
\end{equation}
A straightforward calculation shows that $p_{k, k'}$ satisfies an eigenvalue
equation similar to Eq.~\eqref{eq:q-k-eigenvalue-equation},
\begin{equation}
  \label{eq:p-k-eigenvalue-equation}
  x_{k'}^{\dagger} p_{k, k'} - p_{k, k'} x_{k'} = - 2 \left( \omega_k - \imag 2
    \gamma \right) p_{k, k'},
\end{equation}
which leads us to identify $\lambda^{*} = -2 \left( \omega_k - \imag 2 \gamma
\right)$. Consequently, as stated in the main text, we find
\begin{multline}
  \langle [d_k, d_{-k}](t) \rangle = \e^{- \imag 2 \left( \omega_k - \imag 2
      \gamma \right) t} \langle [d_k, d_{-k}] \rangle_0 \\ + \left( 1 - \e^{-
      \imag 2 \left( \omega_k - \imag 2 \gamma \right) t} \right) \langle [d_k,
  d_{-k}] \rangle_{\mathrm{SS}},
\end{multline}
where the steady-state expectation value is given by
\begin{equation}
  \label{eq:chi-ss-k}
  \begin{split}
    \chi_{\mathrm{SS}, k} & = \langle [d_k, d_{-k}] \rangle_{\mathrm{SS}} \\ & =
    \frac{\imag \delta}{4 \left( \omega_k - \imag 2 \gamma \right)} \sum_{k' \in
      \BZ} \tr(\sigma_y p_{k, k'}) = \frac{\delta \sin(\theta_k)}{\omega_k -
      \imag 2 \gamma}.
  \end{split}
\end{equation}

\subsubsection{Maximum entropy ensemble}
\label{sec:maxim-entr-ensemble}

The PTGGE describes the late-time behavior for $\delta = 0$, such that, as can be
seen explicitly in Eqs.~\eqref{eq:eta-ss-k} and~\eqref{eq:chi-ss-k},
\begin{equation}
  \langle [d_k, d_k^{\dagger}] \rangle_{\mathrm{SS}} = \langle [d_k, d_{-k}]
  \rangle_{\mathrm{SS}} = 0,
\end{equation}
and when the contribution from the anomalous commutators becomes negligible due
to dephasing,
\begin{equation}
  \langle [d_k, d_{-k}](t) \rangle \to 0.
\end{equation}
To summarize, the PTGGE is characterized by the following expectation values:
\begin{equation}
  \label{eq:PTGGE-commutators}
  \begin{pmatrix}
    \langle [d_k, d_k^{\dagger}](t) \rangle_{\mathrm{PTGGE}} & \langle [d_k,
    d_{-k}](t) \rangle_{\mathrm{PTGGE}} \\ \langle [d_{-k}^{\dagger},
    d_k^{\dagger}](t) \rangle_{\mathrm{PTGGE}} & \langle [d_{-k}^{\dagger}, d_{-k}](t) \rangle_{\mathrm{PTGGE}}
  \end{pmatrix}
  = \zeta_k(t),
\end{equation}
where
\begin{equation}
  \label{eq:zeta-k}
  \zeta_k(t) = \e^{-4 \gamma t} \zeta_k',
\end{equation}
with
\begin{equation}
  \label{eq:zeta-k-prime}
  \begin{split}
    \zeta_k' & =
  \begin{pmatrix}
    \langle [d_k, d_k^{\dagger}] \rangle_0 & 0 \\ 0 & \langle [d_{-k}^{\dagger}, d_{-k}] \rangle_0
  \end{pmatrix}
  \\ & =
  \begin{pmatrix}
    \cos(\Delta \theta_k - \phi_k) & 0 \\ 0 & - \cos(\Delta \theta_k + \phi_k)
  \end{pmatrix}.
  \end{split}
\end{equation}
Specifically, we define the PTGGE as the maximum entropy ensemble that is
compatible with the expectation values of commutators summarized in
$\zeta_k(t)$, and with the statistics of the modes $d_k$ as determined by their
anticommutators in Eq.~\eqref{eq:f-k}. The same information is stored in the
covariance matrix in the complex basis, which, according to Eqs.~\eqref{eq:g-k}
and~\eqref{eq:D-k-V-k}, is given by
\begin{equation}
  \label{eq:g-PTGGE-k}
  g_{\mathrm{PTGGE}, k}(t) = V_k^{- \dagger} \zeta_k(t) V_k^{-1} = \e^{- 4 \gamma
    t} g_{\mathrm{PTGGE}, k}',
\end{equation}
where
\begin{equation}
  g_{\mathrm{PTGGE}, k}' = V_k^{-\dagger} \zeta_k' V_k^{-1}.
\end{equation}
For given two-point functions, the entropy is maximized for a Gaussian state,
which is determined uniquely by
Eq.~\eqref{eq:density-matrix-from-covariance-matrix}. Therefore, with
\begin{equation}
  C_k^{\dagger} \atanh \! \left( g_{\mathrm{PTGGE}, k}(t) \right) C_k = D_k
  f_k^{-1} \atanh \!\left( \zeta_k(t) f_k^{-1} \right) D_k,
\end{equation}
we obtain the PTGGE as stated in the main text:
\begin{equation}
  \rho_{\mathrm{PTGGE}}(t) = \frac{1}{Z_{\mathrm{PTGGE}}(t)} \e^{- 2 \sum_{k \geq 0}
    D_k^{\dagger} f_k^{-1} \atanh \left( \zeta_k(t) f_k^{-1}
    \right) D_k}.
\end{equation}
Finally, a straightforward calculation shows that
$\gamma_{\mathrm{PTGGE}, k}(t) = r^{\dagger} g_{\mathrm{PTGGE}, k}(t) r$ for
$\gamma_{\mathrm{PTGGE}, k}(t)$ given in Eq.~\eqref{eq:gamma-PTGGE-k} and that,
therefore, the derivations of the PTGGE in Secs.~\ref{sec:deph-relax-bgge}
and~\ref{sec:maximum-entropy-principle} yield the same result. However, the key
benefit of the derivation presented in the current section is that it clarifies
how the notions of eigenmodes of the generator of the dynamics and of a maximum
entropy ensemble generalize to PT-symmetric open systems. In particular, this
derivation shows that conservation of occupation numbers of eigenmodes is not
necessary for the PTGGE to capture the late-time limit of quench
dynamics. Instead, nonoscillatory decay is sufficient.

\subsection{Subsystem parity}
\label{sec:subsystem-parity}

In the following, we first describe how to obtain the data for the subsystem
fermion parity in Eq.~\eqref{eq:subsystem-parity-pfaffian} that is shown in
Figs.~1 and~2 in the main text, and the PTGGE prediction that describes the
late-time behavior of the subsystem parity and is shown as horizontal lines in
the figures. Then, we fill in details of the derivation of the conjectured
analytical expressions for the space-time scaling limit of the subsystem parity
given in Eqs.~(9) and~(10) in the main text. Next, we derive the estimate for
the crossover time $t_{\times}$ that is shown in the inset of Fig.~2, and
finally, we derive explicit expressions for the soft modes
$k_{s, \pm}$, and we argue that the existence of two distinct time
scales $t_{s, +}$ and $t_{s, -}$ is a unique feature of
driven-dissipative systems with broken inversion symmetry.

\subsubsection{Numerical time evolution}

In the PT-symmetric phase, it is convenient to define a rescaled covariance
matrix as
\begin{equation}
  \label{eq:Gamma-prime}
  \Gamma'(t) = \e^{4 \gamma t} \Gamma(t),
\end{equation}
where $\Gamma(t)$ is given in Eqs.~\eqref{eq:Gamma-1-Gamma-2}
and~\eqref{eq:Gamma-2-Hadamard}. Through this rescaling, one can avoid having to
perform numerical calculations with extremely small quantities when
$\gamma t \gg 1$.  In particular, we obtain the rescaled subsystem parity, which
is shown in Figs.~1 and~2 in the main text, directly as
\begin{equation}
  \e^{4 \ell \gamma t} \langle P_{\ell}(t) \rangle = \Pf(\Gamma'(t)),
\end{equation}
where we use the following property of the Pfaffian: for a
$2 \ell \times 2 \ell$ matrix $A$ and $\alpha \in \C$,
$\Pf(\alpha A) = \alpha^{\ell} \Pf(A)$.

\subsubsection{Long-time behavior}

As we show in Sec.~\ref{sec:maximum-entropy-principle}, the PTGGE is equivalent
to the dephased covariance matrix $\gamma_{\mathrm{PTGGE}, k}(t)$ in
Eq.~\eqref{eq:gamma-PTGGE-k}. The corresponding covariance matrix in position
space can be found through an inverse Fourier transform according to
Eq.~\eqref{eq:gamma-k} and by using Eq.~\eqref{eq:gamma-l}. In analogy to
Eq.~\eqref{eq:gamma-PTGGE-k}, we obtain the late-time behavior in the form
\begin{equation}
  \label{eq:Gamma-PTGGE}
  \Gamma(t) \sim \Gamma_{\mathrm{PTGGE}}(t) = \e^{-4 \gamma t}
  \Gamma_{\mathrm{PTGGE}}',
\end{equation}
where $\Gamma_{\mathrm{PTGGE}}'$ is time-independent. The late-time behavior of
the subsystem parity, which according to
Eq.~\eqref{eq:subsystem-parity-pfaffian} is just the Pfaffian of the reduced
covariance matrix Eq.~\eqref{eq:Gamma-ell}, is thus given by
\begin{equation}
  \label{eq:subsystem-parity-PTGGE}
  \langle P_{\ell}(t) \rangle \sim \langle P_{\ell}(t) \rangle_{\mathrm{PTGGE}} =
  \e^{- 4 \ell \gamma t} \Pf \! \left( \Gamma_{\mathrm{PTGGE}, \ell}' \right).
\end{equation}
We obtain the horizontal lines in Figs.~1 and~2 of the main text by evaluating
the right-hand side of the above equation numerically. In particular, we
reconstruct $\Gamma_{\mathrm{PTGGE}, \ell}'$ from $\gamma_{\mathrm{PTGGE}, k}'$
defined in Eq.~\eqref{eq:gamma-PTGGE-k} with the aid of Eqs.~\eqref{eq:gamma-l}
and~\eqref{eq:gamma-k}. In the latter, we impose the thermodynamic limit
$L \to \infty$ through
\begin{equation}
  \gamma_l = \frac{\imag}{L} \sum_{k \in \BZ} \e^{\imag k l} \gamma_k \sim \imag
  \int_{-\pi}^{\pi} \frac{d k}{2 \pi} \, \e^{\imag k l} \gamma_k.
\end{equation}

\subsubsection{Space-time scaling limit}
\label{sec:subsystem-parity-space-time-scaling}

As pointed out in Sec.~\ref{sec:relat-transv-field}, for the isolated Kitaev
chain, the subsystem fermion parity can be mapped to order parameter
correlations in the transverse field Ising model through the combination of a
Jordan-Wigner transformation and the Kramers-Wannier duality. The relaxation of
order parameter correlations after a quantum quench is studied in
Refs.~\cite{Calabrese2011, Calabrese2012I, Calabrese2012II}. In particular, for
a quench to the topologically trivial phase of the isolated Kitaev chain with
$\abs{\mu} > 2 J$, which corresponds to a quench to the ferromagnetic phase of
the Ising model, Eq.~(19) of Ref.~\cite{Calabrese2012I} yields the following
prediction for the behavior of the subsystem parity in the space-time scaling
limit $\ell, t \to \infty$ with $\ell/t$ fixed:
\begin{equation}
  \label{eq:subsystem-parity-space-time-scaling-trivial-Calabrese}
  \langle P_{\ell}(t) \rangle \sim P_0 \e^{\int_0^{\pi} \frac{d k}{\pi}
    \min(2 \abs{v_k} t, \ell) \ln(\abs{\cos(\Delta \theta_k)})},
\end{equation}
where $v_k = d \varepsilon_k/d k$. The first step to obtain
our conjecture in Eq.~(9) in the main text for the subsystem parity in the
driven-dissipative Kitaev chain is to rewrite the above equation in terms of the
covariance matrix in the complex basis:
\begin{equation}
  \begin{split}
    \ln(\abs{\cos(\Delta \theta_k)}) & = \frac{1}{2} \left( \ln(\abs{\cos(\Delta
        \theta_k)}) + \ln(\abs{\cos(\Delta \theta_{-k})}) \right) \\ & =
    \frac{1}{2} \left( \ln \! \left( \abs{\langle [d_k^{\dagger}, d_k]
          \rangle_0} \right) + \ln \! \left( \abs{\langle [ d_{-k},
          d_{-k}^{\dagger} ] \rangle_0} \right) \right) \\ & = \frac{1}{2} \tr
    \! \left( \ln \! \left(\abs{g_{\mathrm{GGE}, k}} \right) \right),
  \end{split}
\end{equation}
where by taking the limit $\gamma \to 0$ in Eqs.~\eqref{eq:PTGGE-commutators},
\eqref{eq:zeta-k}, \eqref{eq:zeta-k-prime}, and~\eqref{eq:g-PTGGE-k}, we find
that $g_{\mathrm{GGE}, k}$ is given by
\begin{equation}
  \label{eq:g-GGE-k}
  g_{\mathrm{GGE}, k} =
  U_{x, \theta_k}
  \begin{pmatrix}
    \langle [d_k, d_k^{\dagger}] \rangle_0 & 0 \\ 0 & \langle [d_{-k}^{\dagger}, d_{-k}] \rangle_0
  \end{pmatrix}
  U_{x, \theta_k}^{\dagger}.
\end{equation}
Then, Eq.~\eqref{eq:subsystem-parity-space-time-scaling-trivial-Calabrese} can
be written as
\begin{equation}
  \langle P_{\ell}(t) \rangle \sim P_0 \e^{\int_0^{\pi} \frac{d k}{2
      \pi} \min(2 \abs{v_k} t, \ell) \tr \left( \ln \left(\abs{g_{\mathrm{GGE},
            k}} \right) \right)}.
\end{equation}
To generalize this relation to the driven-dissipative Kitaev chain, we note that
as in Eq.~\eqref{eq:subsystem-parity-PTGGE}, the exponential decay of
$g_{\mathrm{PTGGE}, k}(t)$ in Eq.~\eqref{eq:g-PTGGE-k} leads to the appearance of
a prefactor $\e^{- 4 \ell \gamma t}$. Further, we set
$v_k = d \omega_k/d k$, and we replace $g_{\mathrm{GGE}, k}$
by $g_{\mathrm{PTGGE}, k}'$. Finally, by using the cyclic invariance of the trace
and Eq.~\eqref{eq:f-k}, we obtain
\begin{equation}
  \tr \! \left( \ln \! \left( \abs{g_{\mathrm{PTGGE}, k}'} \right) \right) = \tr
  \! \left( \ln \! \left( \abs{\zeta_k' f_k^{-1}} \right) \right),
\end{equation}
and are thus lead to Eq.~(9) of the main text. As stated there, the constant
prefactor $P_0$ can be determined by fitting the late-time limit,
\begin{equation}
  \label{eq:subsystem-parity-infinite-time}
  \lim_{t \to \infty} \e^{4 \ell \gamma t} \langle P_{\ell} \rangle \sim P_0
  \e^{\ell \int_0^{\pi} \frac{d k}{2 \pi} \tr \left( \ln \left(
        \abs{\zeta_k' f_k^{-1}} \right) \right)},
\end{equation}
to the numerical result in Eq.~\eqref{eq:subsystem-parity-PTGGE}.

Next, we consider quenches to the topological phase of the isolated Kitaev chain
for $\abs{\mu} < 2 J$, corresponding to quenches to the paramagnetic phase of
the transverse field Ising model. According to Eq.~(32) of
Ref.~\cite{Calabrese2012I}, for $t < t_F$, the oscillatory decay of
the subsystem parity in the space-time scaling limit is given by
\begin{equation}
  \langle P_{\ell}(t) \rangle \sim P_0 \left( 1 + \cos(2
    \varepsilon_{k_{s}} t + \alpha) \right) \e^{2 t \int_0^{\pi}
  \frac{d k}{\pi} \abs{v_k} \ln(\abs{\cos(\Delta \theta_k)})},
\end{equation}
where in the isolated system
$k_{s} = k_{s, +} = k_{s, -}$. To generalize this
prediction to the driven-dissipative Kitaev chain, we proceed as above, and,
additionally, we note that the oscillatory prefactor is a complete square due to
the identity $2 \cos(x)^2 = 1 + \cos(2 x)$. Based on this simple observation, we
conjecture that the proper generalization of the prefactor reads as follows:
\begin{equation}
  2 \cos(\varepsilon_{k_{s}} t + \alpha/2)^2 \to 2
  \cos(\omega_{k_{s, +}} t + \alpha_+) \cos(\omega_{k_{s, -}}
  t + \alpha_-),
\end{equation}
where, for $\gamma \to 0$, we set $\alpha_+ = \alpha_- = \alpha/2$. Thereby we
obtain Eq.~(10) of the main text.

\subsubsection{Condition to observe relaxation to the PTGGE}

The condition for the existence of a finite time window during which relaxation
to the PTGGE can be observed depends on the observable under
consideration. Here, we focus on the subsystem parity given in
Eq.~\eqref{eq:subsystem-parity-pfaffian}, and on quenches that originate from
the trivial phase of the isolated Kitaev chain. For $\delta = 0$, when
$\Gamma_2(t) = 0$ in Eq.~\eqref{eq:Gamma-time-evolved}, the behavior of the
subsystem parity is given by
$\langle P_{\ell}(t) \rangle = \Pf(\Gamma_{1, \ell}(t))$, and, as shown in
Figs.~1 and~2 of the main text, $\langle P_{\ell}(t) \rangle$ relaxes on a time
scale $t_F$ to the PTGGE prediction given in
Eq.~\eqref{eq:subsystem-parity-PTGGE}~\footnote{More precisely, at $t = t_F$,
  the initial fast relaxation of \unexpanded{$\langle P_{\ell}(t) \rangle$}
  turns into much slow decay towards the PTGGE prediction, which is reached
  after a time $t \sim \ell^{4/3} > t_F \sim \ell$. We note that for quenches
  from the trivial phase, we expect the subsystem parity to relax to the PTGGE
  prediction on an even longer time scale that is exponentially large in
  $\ell$~\cite{Calabrese2012II}}. If $\delta \neq 0$, for relaxation of
$\langle P_{\ell}(t) \rangle$ to the PTGGE prediction to be observable at all,
the contribution to $\langle P_{\ell}(t) \rangle$ due to $\Gamma_2(t) \neq 0$
has to be small at $t = t_F$. Then, for $t > t_F$, $\langle P_{\ell}(t) \rangle$
obeys the PTGGE prediction up to the crossover time scale $t_{\times}$. To
estimate $t_{\times}$, we expand the Pfaffian to lowest nontrivial order,
\begin{equation}
  \langle P_{\ell} \rangle = \Pf \! \left(
    \Gamma_{1, \ell} + \Gamma_{2, \ell} \right) \sim \Pf \!
  \left( \Gamma_{1, \ell} \right) \left( 1 + \frac{1}{2} \tr \! \left(
      \frac{\Gamma_{2, \ell}}{\Gamma_{1, \ell}} \right)
  \right),
\end{equation}
where we omit the explicit time dependence. Therefore, the condition for
relaxation to the PTGGE to be observable can be stated as
\begin{equation}
  \label{eq:subsystem-parity-PTGGE-relaxation-condition}
  \frac{1}{2} \abs{\tr \! \left( \frac{\Gamma_{2, \ell}}{\Gamma_{1, \ell}}
    \right)} \ll 1 \quad \text{for } t = t_F.
\end{equation}
At $t = t_F$, we may estimate $\Gamma_{1, \ell}$ and
$\Gamma_{2, \ell}$ by their ``dephased'' expressions, i.e., by keeping only
nonoscillatory contributions. According to
Eq.~\eqref{eq:gamma-k-PT-symmetric-asymptotic}, and with Eqs.~\eqref{eq:gamma-k}
and~\eqref{eq:gamma-l}, we can write
\begin{equation}
  \Gamma_1(t) \sim \Gamma_{d, 1}(t), \qquad \Gamma_2(t) \sim
  \Gamma_{d, 2}(t) + \Gamma_{\mathrm{SS}},
\end{equation}
where
\begin{equation}
  \Gamma_{d, 1}(t) \sim \e^{-4 \gamma t} \Gamma_{d, 1}', \qquad
  \Gamma_{d, 2}(t) \sim \e^{-4 \gamma t} \Gamma_{d, 2}'.
\end{equation}
Then, Eq.~\eqref{eq:subsystem-parity-PTGGE-relaxation-condition} becomes
\begin{equation}
  \frac{1}{2} \abs{\tr \! \left( \frac{\Gamma_{d, 2,
          \ell}'}{\Gamma_{d, 1, \ell}'} \right) + \e^{4 \gamma
      t_F} \tr \! \left( \frac{\Gamma_{\mathrm{SS},
          \ell}}{\Gamma_{d, 1, \ell}'} \right)} \ll 1.
\end{equation}
A sufficient condition for the validity of this inequality can be obtained by
using the triangle inequality. We find
\begin{equation}
  \frac{1}{2} \left( \abs{\tr \! \left( \frac{\Gamma_{d, 2,
            \ell}'}{\Gamma_{d, 1, \ell}'} \right)} + \e^{4 \gamma
      t_F} \abs{\tr \! \left( \frac{\Gamma_{\mathrm{SS},
            \ell}}{\Gamma_{d, 1, \ell}'} \right)} \right) \ll 1,
\end{equation}
which is satisfied if $t_F$ is small in comparison to the crossover
time scale $t_{\times}$, for which the left-hand side of the above inequality is
equal to one,
\begin{equation}
  t_F \ll t_{\times} = \frac{1}{4 \gamma} \ln \! \left( \frac{2 -
      \abs{\tr(\Gamma_{d, 2, \ell}'/\Gamma_{d, 1,
          \ell}')}}{\abs{\tr(\Gamma_{\mathrm{SS}, \ell}/\Gamma_{d, 1,
          \ell}')}} \right).
\end{equation}
For $\delta \to 0$, by taking into account that $\Gamma_{\mathrm{SS}, \ell}$ and
$\Gamma_{d, 2, \ell}'$ vanish linearly in $\delta$ according to
Eqs.~\eqref{eq:gamma-d-2-k-prime}, \eqref{eq:gamma-ss-k}, and~\eqref{eq:y},
while $\Gamma_{d, 1, \ell}'$ is finite, we find
\begin{equation}
  \label{eq:t-cross}
  t_{\times} \sim \frac{1}{4 \gamma} \abs{\ln(c_{\times} \abs{\delta})}, \quad
  c_{\times} = \lim_{\delta \to 0} \frac{1}{2 \delta} \abs{\tr \! \left(
      \frac{\Gamma_{\mathrm{SS}, \ell}}{\Gamma_{d, 1, \ell}'} \right)}.
\end{equation}
The limit $\delta \to 0$ can be taken explicitly in Eq.~\eqref{eq:gamma-ss-k},
\begin{multline}
  \lim_{\delta \to 0} \frac{\gamma_{\mathrm{SS}, k}}{\delta} = - \frac{1}{2
    \gamma} \left( \frac{1}{\varepsilon_k^2} \unitvec{e}_y \cdot \left(
      \mathbf{a}_k \times \mathbf{b} \right) \id \right. \\ \left. - \left\{
      \unitvec{e}_y - \left[ \varepsilon_k \mathbf{e}_{y, \perp, k} - \frac{2
          \gamma}{\varepsilon_k} \mathbf{e}_{y, o, k} \right] \right\}
    \cdot \boldsymbol{\sigma} \right),
\end{multline}
where we set $\mathbf{b} \to - 2 \gamma \unitvec{e}_z$ for $\delta \to 0$, and
where
\begin{equation}
  \begin{split}    
    \mathbf{e}_{y, \parallel, k} & = \left( \unitvec{e}_y \cdot \unitvec{a}_k
    \right) \unitvec{a}_k, \\ \mathbf{e}_{y, \perp, k} & = \unitvec{e}_y -
    \mathbf{e}_{y, \parallel, k}, \\ \mathbf{e}_{o, k} & = -
    \unitvec{e}_y \times \unitvec{a}_k.
  \end{split}
\end{equation}
This allows us to reconstruct
$\lim_{\delta \to 0} \Gamma_{\mathrm{SS}, \ell}/\delta$ numerically by employing
Eqs.~\eqref{eq:gamma-k} and~\eqref{eq:gamma-l}. Similarly, we can obtain
$\Gamma_{d, 1, \ell}'$ from $\gamma_{d, 1, k}'$, and, thereby,
calculate $c_{\times}$. This leads to the semianalytical estimate in the inset
of Fig.~2.

\subsubsection{Oscillatory dynamics from zero modes}

According to Eq.~\eqref{eq:zeta-k-prime}, soft modes of the PTGGE are determined
by the condition
\begin{equation}
  \label{eq:soft-mode-equation}
  \cos(\Delta \theta_k - \phi_k) = 0.
\end{equation}
For $\phi_k = 0$, due to $\Delta \theta_k = - \Delta \theta_{-k}$, which
according to Eq.~\eqref{eq:U-x-theta-k-IS} is a consequence of IS, the existence
of a soft mode $k_{s, +} > 0$ implies the existence of another
inversion conjugate soft mode $k_{s, -} = - k_{s, +}$. Due to
$\varepsilon_k = \varepsilon_{-k}$, the associated oscillation periods
$t_{s, \pm} = \pi/(2 \varepsilon_{k_{s}, \pm})$ are
identical. Interestingly, IS$^{\dagger}$ as stated in Eq.~\eqref{eq:z-k-IS}
implies, via Eqs.~\eqref{eq:U-k-IS} and~\eqref{eq:phi-k-beta-k}, that
$\phi_k = \phi_{-k}$. However, the sum $\Delta \theta_k - \phi_k$ that appears
in Eq.~\eqref{eq:soft-mode-equation} does not have definite transformation
behavior under inversion, just as the sums and differences of $\theta_k$ and
$\phi_k$ that appear in the Liouvillian eigenmode transformation matrix $V_k$ in
Eq.~\eqref{eq:V-k}. Consequently, the solutions $k_{s, \pm}$ of
Eq.~\eqref{eq:soft-mode-equation} are not related by inversion. In particular,
for quenches with pre- and postquench parameters given by, respectively,
$J_0 = \Delta_0$ and $\mu_0$, and $J = \Delta$ and $\mu$, we find
\begin{multline}
  k_{s, \pm} = \mp \sgn(\mu) \arccos \! \left( - \frac{1}{2 \left( J_0
        \mu + J \mu_0 \right)^2} \right. \\ \times \left\{ 4 J^2 J_0 \mu_0 + J_0
    \left( \mu^2 - 4 \gamma_{l} \gamma_{g} \right) \mu_0 + J
    \mu \left( 4 J_0^2 + \mu_0^2 \right) \mp 2 \sgn(\mu)
    \vphantom{\sqrt{\gamma_{l} \gamma_{g} \left[ 4 J_0^4 \mu^2
          - J_0^2 \left( 4 J^2 + \mu^2 - 4 \gamma_{l}
            \gamma_{g} \right) \mu_0^2 + J^2
          \mu_0^4 \right]}} \right. \\
  \left. \left. \times \sqrt{\gamma_{l} \gamma_{g} \left[ 4
          J_0^4 \mu^2 - J_0^2 \left( 4 J^2 + \mu^2 - 4 \gamma_{l}
            \gamma_{g} \right) \mu_0^2 + J^2 \mu_0^4 \right]} \right\}
  \right).
\end{multline}
In the limit $\mu_0 \to - \infty$ as in the examples considered in the main
text, the above general form reduces to
\begin{equation}
  k_{s,\pm}= \mp \sgn(\mu) \arccos \! \left( - \frac{1}{2 J} \left( \mu
      \mp \sgn(\mu) 2 \sqrt{\gamma_{l} \gamma_{g}} \right) \right).
\end{equation}

\subsubsection{Necessary conditions for $t_{s, +} \neq t_{s, -}$}

As claimed in the main text, different periods of parity pumping for subsystems
at the left and right ends of the chain require both mixedness of the
time-evolved state and breaking of inversion symmetry and are, therefore, unique
to driven-dissipative systems. To see that this is the case, consider, for
simplicity, an equal bipartition into subsystems
$L_{L/2} = \{ 1, \dotsc, L/2 \}$ and $R_{L/2} = \{ L/2 + 1, \dotsc, L \}$,
corresponding to the left and right halves of the chain. Then, for a pure state,
the eigenvalues of $\Gamma_L$ and $\Gamma_R$, which determine the entanglement
spectra for subsystems $L_{L/2}$ and $R_{L/2}$, are identical~\footnote{For a
  bipartition into subsystems of different sizes, e.g., $L$ smaller than $R$,
  the spectrum of $\Gamma_R$ is given by the spectrum of $\Gamma_L$ and
  additional pairs of eigenvalues $\pm \imag$.}. Consequently, due to
$\det(\Gamma_{\ell}) = \Pf(\Gamma_{\ell})^2 = \langle P_{\ell} \rangle^2$, zero
crossings of $\langle P_L \rangle$ and $\langle P_R \rangle$ occur at the same
time. In contrast, for a mixed state, the spectra of $\Gamma_L$ and $\Gamma_R$
are, in general, different. However, if the state is symmetric under inversion,
with inversion center in the middle of the chain, the matrices $\Gamma_L$ and
$\Gamma_R$ are, as we show below, unitarily equivalent. Consequently, even for
mixed states, the entanglement spectra and the subsystem parities, given by the
Pfaffians, are the same. Therefore, as stated above, both mixedness of the state
and inversion symmetry breaking are necessary for
$t_{s, +} \neq t_{s, -}$.

To show that $\Gamma_L$ and $\Gamma_R$ are unitarily equivalent for an
inversion-symmetric state, we first note that
Eq.~\eqref{eq:inversion-second-quantized} for open boundary conditions implies
the following transformation behavior of the spinors of Majorana operators
defined in Eq.~\eqref{eq:W-l}:
\begin{equation}  
  I W_l I^{\dagger} = \imag \sigma_y W_{L + 1 - l}.
\end{equation}
Therefore, with Eq.~\eqref{eq:gamma-l}, we find that the $2 \times 2$ blocks of
the covariance matrix of an inversion-symmetric state
$\rho = I \rho I^{\dagger}$ obey the condition
\begin{equation}
  \gamma_{l, l'} = \sigma_y \gamma_{L + 1 - l, L + 1 - l'} \sigma_y,
\end{equation}
which implies that the full covariance matrix behaves as
\begin{equation}
  \label{eq:Gamma-IS}
  \Gamma = \Sigma_{y, L} \Xi_L \Gamma \Xi_L \Sigma_{y, L},
\end{equation}
where $\Sigma_{y, \ell}$ and $\Xi_{\ell}$ are defined in
Eqs.~\eqref{eq:Sigma-mu-ell} and~\eqref{eq:Xi-ell}, respectively. The product
$\Sigma_{y, \ell} \Xi_{\ell}$ can be seen to be unitary by noting that both
$\Sigma_{y, \ell}$ and $\Xi_{\ell}$ are Hermitian and involutory. Therefore, the
restriction of Eq.~\eqref{eq:Gamma-IS} to the left half of the chain,
\begin{equation}
  \Gamma_L = \Sigma_{y, L/2} \Xi_{L/2} \Gamma_R \Xi_{L/2} \Sigma_{y, L/2},
\end{equation}
confirms the claimed unitary equivalence between $\Gamma_L$ and $\Gamma_R$.

\subsection{Subsystem entropy}
\label{sec:subsystem-entropy}

Figure~4 in the main text shows the quasiparticle-pair contribution to the
subsystem entropy, which is defined as
\begin{equation}
  \label{eq:S-vN-qp}
  S_{\mathrm{vN}, \ell}^{\mathrm{QP}} = S_{\mathrm{vN}, \ell} - \frac{\ell}{L}
  S_{\mathrm{vN}}^{\mathrm{stat}}.
\end{equation}
In the following, we describe how to calculate the subsystem entropy
$S_{\mathrm{vN}, \ell}$ and the statistical entropy
$S_{\mathrm{vN}}^{\mathrm{stat}} = S_{\mathrm{vN}, L}$, as well as the PTGGE
prediction that captures the late-time behavior of
$S_{\mathrm{vN}, \ell}^{\mathrm{QP}}$. Further, we present a short derivation
that motivates our conjecture for the space-time scaling limit of
$S_{\mathrm{vN}, \ell}^{\mathrm{QP}}$ given in Eq.~(11) in the main text. In
this section, we restrict ourselves to the case $\delta = 0$.

\subsubsection{Subsystem entropy from the covariance matrix}

The von Neumann entropy of a subsystem that consists of $\ell$ contiguous
lattice sites is given by~\cite{Calabrese2005}
\begin{equation}
  \label{eq:S-vN-ell}
  S_{\mathrm{vN}, \ell} = - \tr(\rho_{\ell} \ln(\rho_{\ell})) = \sum_{l = 1}^{\ell} S(\xi_l),
\end{equation}
where
\begin{equation}
  \label{eq:S}
  S(\xi) = - \frac{1 + \xi}{2} \ln \! \left( \frac{1 + \xi}{2} \right) -
  \frac{1 - \xi}{2} \ln \! \left( \frac{1 - \xi}{2} \right),
\end{equation}
and where $\pm \imag \xi_l$ with $0 \leq \xi_l \leq 1$ and
$l \in \{ 1, \dotsc, \ell \}$ are the eigenvalues of the reduced covariance
matrix $\Gamma_{\ell}$ defined in Eq.~\eqref{eq:Gamma-ell}. When $\delta = 0$,
the time-evolved covariance matrix in Eq.~\eqref{eq:Gamma-time-evolved} reduces
to $\Gamma(t) = \Gamma_1(t)$, which, according to Eq.~\eqref{eq:gamma-1-k}, is
exponentially small for $\gamma t \gg 1$. Then, to calculate the subsystem
entropy Eq.~\eqref{eq:S-vN-ell}, we can restrict ourselves to the lowest
nontrivial order in the expansion of $S(\xi)$ in powers of $\xi$,
\begin{equation}
  \label{eq:S-expansion}
  S(\xi) = \ln(2) - \xi^2/2 + O(\xi^4).
\end{equation}
In particular, from the rescaled covariance matrix defined in
Eq.~\eqref{eq:Gamma-prime}, we obtain the spectrum of the reduced rescaled
covariance matrix:
\begin{equation}
  \sigma(\Gamma_{\ell}'(t)) = \{ \pm \imag \xi_l'(t) \} = \{ \pm \imag \e^{4
    \gamma t} \xi_l(t) \}.
\end{equation}
For $\gamma t \gg 1$, the subsystem entropy can be written in terms of the
eigenvalues $\xi_l'(t)$ as
\begin{equation}
  \label{eq:S-vN-ell-late-time}
  S_{\mathrm{vN}, \ell}(t) \sim \ell \ln(2) - \frac{1}{2} \sum_{l = 1}^{\ell}
  \xi_l(t)^2 = \ell \ln(2) - \frac{\e^{-8 \gamma t}}{2} \sum_{l = 1}^{\ell}
  \xi_l'(t)^2.
\end{equation}

\subsubsection{Statistical entropy}

Rather than calculating the statistical entropy
$S_{\mathrm{vN}}^{\mathrm{stat}} = S_{\mathrm{vN}, L}$ numerically for a finite
system, we perform an analytical calculation in the thermodynamic limit as
detailed in the following: The covariance matrix is block-diagonal in momentum
space, with $4 \times 4$ blocks $\mathsf{y}_k$ given in
Eq.~\eqref{eq:y-k}. According to Eqs.~\eqref{eq:y-k-gamma-k}
and~\eqref{eq:gamma-k-gamma--k-spectrum}, the spectrum of $\mathsf{y}_k$ is
given by
$\sigma(\mathsf{y}_k) = \sigma(\gamma_k) \cup \sigma(\gamma_{-k}) = \{ \pm
\xi_{+, k}, \pm \xi_{-, k} \}$. Therefore, in the thermodynamic limit,
\begin{equation}
  \label{eq:S-vN-stat}
  S_{\mathrm{vN}}^{\mathrm{stat}}(t) = \sum_{k \geq 0} \sum_{\sigma = \pm}
  S(\xi_{\sigma, k}(t)) \to L \int_0^{\pi} \frac{d k}{2 \pi}
  \sum_{\sigma = \pm} S(\xi_{\sigma, k}(t)).
\end{equation}
The eigenvalues $\xi_{\pm, k}(t)$ are given in Eq.~\eqref{eq:xi-+-k-xi---k},
where, for $\delta = 0$, $\gamma_{\id, k}(t) = \gamma_{1, \id, k}(t)$ is given
in Eq.~\eqref{eq:gamma-1-id-k-PT-symmetric}, and
$\xi_k(t) = \lvert \boldsymbol{\gamma}_{1, k}(t) \rvert$ can be obtained by
using Eq.~\eqref{eq:xi--}, which yields
\begin{multline}
  \label{eq:xi-k}
  \xi_k(t) = \e^{- 4 \gamma t} \left\{ 1 - \frac{\varepsilon_k^2}{\omega_k^2}
    \sin(\Delta \theta_k)^2 \left[ 2 \left( 1 - \cos(2 \omega_k t) \right)
      \vphantom{\frac{\varepsilon_k^2}{\omega_k^2}} \right. \right. \\
  \left. \left. - \frac{\varepsilon_k^2}{\omega_k^2} \left( 1 - \cos(2 \omega_k
        t) \right)^2 - \sin(2 \omega_k t)^2 \right] \right\}^{1/2},
\end{multline}
where $\Delta \theta_k$ is defined in Eq.~\eqref{eq:Delta-theta-k}. At late
times, when $\gamma t \gg 1$, the eigenvalues
$\xi_{\pm, k}(t) \sim \e^{- 4 \gamma t}$ are exponentially small, and we can use
the expansion in Eq.~\eqref{eq:S-expansion} to simplify
Eq.~\eqref{eq:S-vN-stat}. To make the exponential time dependence explicit, we
define the rescaled quantities
\begin{equation}
  \gamma_{\id, k}'(t) = \e^{4 \gamma t} \gamma_{\id, k}(t), \qquad \xi_k'(t) = \e^{4
    \gamma t} \xi_k(t),
\end{equation}
in terms of which we obtain the statistical entropy in the form
\begin{equation}
  \label{eq:S-vN-stat-expansion}
  S_{\mathrm{vN}}^{\mathrm{stat}}(t) \sim L \left[ \ln(2) - 
    \e^{- 8 \gamma t} \int_0^{\pi} \frac{d k}{2 \pi} \left( \gamma_{\id,
        k}'(t)^2 + \xi_k'(t)^2 \right) \right].
\end{equation}

\subsubsection{Numerical time evolution}

At short times $\gamma t \lesssim 1$, we calculate the covariance matrix for a
finite system numerically according to Eq.~\eqref{eq:Gamma-time-evolved}, and we
use Eqs.~\eqref{eq:S-vN-ell} and~\eqref{eq:S-vN-stat} to obtain the data shown
in Fig.~4 in the main text. For longer times $\gamma t \gg 1$, a numerically
stable procedure that avoids calculations with extremely small quantities is to
take the difference between Eqs.~\eqref{eq:S-vN-ell-late-time}
and~\eqref{eq:S-vN-stat-expansion}, which yields
\begin{equation}
  \label{eq:S-vN-qp-expansion}
  S_{\mathrm{vN}, \ell}^{\mathrm{QP}}(t) \sim - \e^{- 8 \gamma t} \left[
    \frac{1}{2} \sum_{l = 1}^{\ell} \xi_l'(t)^2 - \ell \int_0^{\pi}
    \frac{d k}{2 \pi} \left( \gamma_{\id, k}'(t)^2 + \xi_k'(t)^2 \right) \right],
\end{equation}
where the exponential time dependence is contained in the prefactor that is
determined analytically.

\subsubsection{Late-time behavior}
\label{sec:entangl-entr-late}

The late-time limit that is described by the PTGGE can be obtained from
Eq.~\eqref{eq:S-vN-qp-expansion} by properly taking into account dephasing of
oscillatory contributions. Namely, $\xi_l'(t)$ should be replaced by
$\xi_{d, l}'$, where $\pm \imag \xi_{d, l}'$ are the
eigenvalues of $\Gamma_{\mathrm{PTGGE}}'$ defined in
Eq.~\eqref{eq:Gamma-PTGGE}. In contrast, in the terms that describe the
statistical entropy, oscillatory terms have to be dropped only after taking the
squares of $\gamma_{\id, k}'(t)$ and $\xi_k'(t)$,
\begin{equation}
  \label{eq:S-vN-qp-late-time}
  S_{\mathrm{vN}, \ell}^{\mathrm{QP}}(t) \sim - \e^{- 8 \gamma t}
  \left[ \frac{1}{2} \sum_{l = 1}^{\ell} \xi_{d, l}^{\prime 2} -
    \ell \int_0^{\pi} \frac{d k}{2 \pi} \left( \gamma_{\id, k}^{\prime
        2} + \xi_k^{\prime 2}
    \right)_{d} \right],
\end{equation}
where, according to Eq.~\eqref{eq:gamma-1-id-k-PT-symmetric},
\begin{equation}
  \left( \gamma_{\id, k}^{\prime 2} \right)_{d} = \frac{3}{2 \omega_k^4} \left[
    \unitvec{a}_{\id, k} \cdot \left( \mathbf{a}_k \times \mathbf{b} \right)
  \right]^2,
\end{equation}
and, with Eq.~\eqref{eq:xi-k},
\begin{equation}
  \left( \xi_k^{\prime 2} \right)_{d} = 1 + \frac{6 \varepsilon_k^2
    \gamma^2}{\omega_k^4} \sin(\Delta \theta_k)^2.
\end{equation}

\subsubsection{Space-time scaling limit}

The quasiparticle picture~\cite{Calabrese2005, Alba2017, Alba2018} suggests the
following functional form for the time dependence of the subsystem entropy after
a quench in an isolated integrable system:
\begin{equation}
  \label{eq:S-vN-space-time-scaling}
  S_{\mathrm{vN}, \ell}(t) \sim \int_{-\pi}^{\pi} \frac{d k}{2 \pi} \min(2
  \abs{v_k} t, \ell) s_k,
\end{equation}
where $v_k$ is the quasiparticle velocity, and the weight $s_k$ determines the
contribution of quasiparticles with momentum $k$ to the entropy. The above form
holds in the space-time scaling limit, $\ell, t \to \infty$ with $\ell/t$ fixed,
and does not capture $O(1)$ contributions such as area-law entanglement of the
initial state. To provide a quantitative description of the evolution of the
subsystem entropy for a specific model, concrete expressions for $v_k$ and $s_k$
are required. According to Ref.~\cite{Alba2017}, the weight $s_k$ can be fixed
to reproduce the stationary value of the entropy that is reached for
$t \to \infty$. For free fermionic models, this leads to
\begin{equation}
  \label{eq:s-k}
  s_k = - \langle n_k \rangle_0 \ln(\langle n_k \rangle_0) - \left( 1 - \langle n_k
    \rangle_0 \right) \ln(1 - \langle n_k \rangle_0),
\end{equation}
where the occupation numbers
$\langle n_k \rangle_0 = \braket{\psi_0 | d_k^{\dagger} d_k| \psi_0}$ of
eigenmodes of the postquench Hamiltonian are determined by the initial
state. Further, $v_k$ is given by the group velocity
$v_k = d \varepsilon_k/d k$, where $\varepsilon_k$ is the
dispersion relation of quasiparticles.

To derive Eq.~(11) of the main text, which we conjecture to generalize
Eq.~\eqref{eq:S-vN-space-time-scaling} to PT-symmetric and potentially strongly
dissipative open systems, we first rewrite
Eq.~\eqref{eq:S-vN-space-time-scaling} in terms of the covariance matrix in the
complex basis. Under the assumption that $\abs{v_k} = \abs{v_{-k}}$, which is
satisfied for the Kitaev chain, the range of integration
Eq.~\eqref{eq:S-vN-space-time-scaling} can be restricted as
\begin{equation}
  S_{\mathrm{vN}, \ell}(t) \sim \int_0^{\pi} \frac{d k}{2 \pi} \min(2
  \abs{v_k} t, \ell) \left( s_k + s_{-k} \right).
\end{equation}
Next, by noting that, for $S(\xi)$ given in Eq.~\eqref{eq:S},
\begin{equation}
  s_k = S(2 \langle n_k \rangle_0 - 1),
\end{equation}
and with $2 n_k - 1 = - [d_k, d_k^{\dagger}]$ and $S(\xi) = S(-\xi)$, we obtain
\begin{equation}
  s_k + s_{-k} = \tr(S(g_{\mathrm{GGE}, k})),
\end{equation}
where $g_{\mathrm{GGE}, k}$ is defined in Eq.~\eqref{eq:g-GGE-k}, and we use the
cyclic invariance of the trace. We are thus lead to the desired expression for
the subsystem entropy in the space-time scaling limit in terms of the complex
covariance matrix:
\begin{equation}
  \label{eq:S-vN-space-time-scaling-g-k}
  S_{\mathrm{vN}, \ell} \sim \int_0^{\pi} \frac{d k}{2 \pi} \min(2
  \abs{v_k} t, \ell) \tr(S(g_{\mathrm{GGE}, k})).
\end{equation}
Further, in analogy to our approach for the subsystem parity in
Sec.~\ref{sec:subsystem-parity-space-time-scaling}, to account for modifications
of quasiparticle dynamics and statistics due to strong dissipation, we set
$v_k = d \omega_k/d k$ and we replace $g_{\mathrm{GGE}, k}$ by
$g_{\mathrm{PTGGE}, k}(t)$ given in Eq.~\eqref{eq:g-PTGGE-k}. In particular, to
show explicitly the effect of modified quasiparticle statistics, which are
described by the anticommutation relations in Eq.~\eqref{eq:f-k}, we use the
cyclic invariance of the trace to write
\begin{equation}
  \tr(S(g_{\mathrm{PTGGE}, k}(t))) = \tr \! \left( S \! \left( \zeta_k(t)
      f_k^{-1} \right) \right).
\end{equation}
Finally, we subtract the statistical entropy to single out the contribution due
the propagation of pairs of entangled quasiparticles~\cite{Alba2021,
  Carollo2022, Alba2022}. To that end, we rewrite the statistical entropy in
Eq.~\eqref{eq:S-vN-stat} by using $S(\xi) = S(-\xi)$ as
\begin{equation}
  \sum_{\sigma = \pm} S(\xi_{\sigma, k}(t)) = \tr(S(\gamma_k(t)) =
  \tr(S(g_k(t)).
\end{equation}
In the space-time scaling limit, oscillatory contributions dephase in the
integration over momenta, and, therefore, we have to keep only the
nonoscillatory part, which we denote by $\tr(S(g_k(t))_{d}$. Then,
following Ref.~\cite{Carollo2022}, we obtain the quasiparticle-pair contribution
to the subsystem entropy in the form given in Eq.~(11) in the main text. For
$\gamma t \gg 1$, we can simplify Eq.~(11) by using the expansion of $S(\xi)$
given in Eq.~\eqref{eq:S-expansion}. First, for the subsystem entropy in
Eq.~\eqref{eq:S-vN-space-time-scaling-g-k}, and with Eq.~\eqref{eq:zeta-k}, we
obtain
\begin{equation}
  \tr \! \left( S \! \left( \zeta_k(t) f_k^{-1} \right) \right) \sim \ln(2) -
  \frac{\e^{-8 \gamma t}}{2} \tr \! \left( \left( \zeta_k' f_k^{-1} \right)^2 \right).
\end{equation}
Further, for the statistical entropy, as in Eq.~\eqref{eq:S-vN-qp-late-time}, we
find
\begin{equation}
  \tr(S(g_k(t)))_{d} \sim 
  \ln(2) - \e^{- 8 \gamma t} \left( \gamma_{\id, k}^{\prime 2} + \xi_k^{\prime 2} \right)_{d}.
\end{equation}
By combining these results, we are lead to
\begin{multline}
  S_{\mathrm{vN}, \ell}^{\mathrm{QP}}(t) \sim - \e^{- 8 \gamma t} \int_0^{\pi}
  \frac{d k}{2 \pi} \min(2 \abs{v_k} t, \ell) \\ \times \left(
    \frac{1}{2} \tr \! \left( \left( \zeta_k' f_k^{-1} \right)^2 \right) -
    \left( \gamma_{\id, k}^{\prime 2} + \xi_k^{\prime 2} \right)_{d}
  \right).
\end{multline}
This is the form used to obtain the solid lines in Fig.~4 of the main text for
$t/t_F \gtrsim 0.5$.

\section{Universality of PT-symmetric relaxation}

Generalized thermalization, i.e., relaxation to the GGE, is expected to occur
universally in the quench dynamics of isolated integrable systems. Is relaxation
to the PTGGE, which we have discussed so far for the specific example of a
driven-dissipative Kitaev chain, an equally universal property for PT-symmetric
and integrable driven-dissipative systems? In the following, we address this
question by studying the quench dynamics of a selection of PT-symmetric
models. We demonstrate relaxation to the PTGGE for two broad classes of
noninteracting fermionic many-body systems, and---for a subset of
observables---also for an interacting spin chain that can be mapped to fermions
with quadratic jump operators. For noninteracting bosons, we show that the
relaxation dynamics is captured by a dephased ensemble that generalizes the
PTGGE for fermions while maintaining the key property of conserving an extensive
amount of information about the initial state. Crucially, the same
mechanism---dephasing within a subspace that is spanned by eigenmodes of the
Liouvillian with equal decay rates---underlies relaxation to both the PTGGE for
fermions and the dephased ensemble for bosons. Therefore, while the PTGGE in the
form presented in the main text is not as general for driven-dissipative
integrable systems as the GGE for isolated integrable systems, our results
indicate that the underlying mechanism is indeed very general.

Before we enter the detailed discussion of other models, let us point out that
relaxation to the PTGGE as described in the main text is not tied to the
particular form of the Kitaev chain and the jump operators discussed
there. Instead, our specific choice of Hamiltonian and jump operators is to be
understood as one particular representative of a broad class of models. To see
why this is the case, we recall the necessary conditions for relaxation to the
PTGGE of the driven-dissipative Kitaev chain: (i) PT symmetry of the
Liouvillian; (ii) the existence of a fully PT-symmetric phase; (iii) for
relaxation to the PTGGE to persist to arbitrary times (for simplicity, we
restrict our discussion of other models to this case), the steady state must be
an infinite-temperature state. But these requirements leave considerable freedom
to modify the model. On the one hand, this concerns the Hamiltonian: For
example, relaxation to the PTGGE persists for long-range hopping and
pairing~\cite{Starchl2022}. On the other hand, also the jump operators can be
modified: An infinite-temperature steady state is reached whenever the jump
operators are Hermitian, i.e., when they take the form given in
Eq.~\eqref{eq:L-Majorana-basis} with real coefficients $B_{l, l'}$. However, PT
symmetry poses restrictions on the coefficients $B_{l, l'}$. For example, for
purely local dissipation, the choices $L_l = \sqrt{\gamma} w_{2 l - 1}$ and
$L_l = \sqrt{\gamma} w_{2 l}$ are compatible with PT symmetry. Moreover, for
both choices, a fully PT-symmetric phase exists up to a finite critical value of
$\gamma$. These considerations extend straightforwardly to systems in higher
dimensions.

\subsection{Fermionic SSH model with loss and gain}
\label{sec:fermionic-ssh-model-loss-gain}

Noninteracting fermionic systems can broadly be classified into models with and
without pairing terms in the Hamiltonian, and the Kitaev chain, on which our
discussion has focused so far, is a particular representative of the former
class. In the following, we consider the SSH model~\cite{Su1979, Heeger1988} as
a paradigmatic example of a one-dimensional tight-binding chain without
pairing. We present a short summary of results that demonstrate relaxation to
the PTGGE. A more detailed discussion will be provided
elsewhere~\cite{Starchl2022}.

The SSH model describes hopping along a one-dimensional bipartite lattice with
sublattices $A$ and $B$. For a system that consists of $L$ unit cells, the
Hamiltonian reads
\begin{equation}
  \label{eq:H-SSH}
  H_{\mathrm{SSH}} = \sum_{l = 1}^L \left( J_1 c_{A, l}^{\dagger} c_{B, l} + J_2
    c_{A, l + 1}^{\dagger} c_{B, l} + \hc \right),
\end{equation}
where $J_1, J_2 \in \R$. We find it convenient to parameterize $J_1$ and $J_2$
in terms of the total and relative hopping amplitudes,
\begin{equation}
  J = \frac{1}{2} \left( J_1 + J_2 \right), \qquad \Delta J = \frac{1}{2} \left(
    J_1 - J_2 \right).
\end{equation}
The topologically nontrivial phase of the SSH model is realized for
$\Delta J <0$~\cite{Hasan2010, Qi2011, Chiu2016}.

Isolated quadratic fermionic systems without pairing terms in the Hamiltonian do
not exhibit anomalous correlations, and, therefore, their dynamics is captured
by a closed equation of motion for the covariance matrix which is defined as
\begin{equation}  
  G_{l, l'}^{s, s'} = \langle [c_{s, l}, c_{s', l'}^{\dagger}] \rangle.
\end{equation}
For such systems, a natural choice of dissipation which results again in a
closed equation of motion for the covariance matrix
and can lead to an infinite-temperature steady state, is loss and gain with jump
operators
\begin{equation}
  \label{eq:L-l-L-g}
  L_{l, s, l} =  \sqrt{\gamma_{l, s}} c_{s, l}, \qquad
  L_{g, s, l} = \sqrt{\gamma_{g, s}} c_{s, l}^{\dagger},
\end{equation}
where $\gamma_{l, s}$ and $\gamma_{g, s}$ are the loss and
gain rates on sublattice $s \in \{ A, B \}$, respectively. In particular, the
steady state is at infinite temperature if the loss and gain rates are equal,
$\gamma_{l, s} = \gamma_{g, s}$~\cite{Starchl2022}. Focusing
on this case, we set
$\gamma_{l, A} = \gamma_{g, A} = \gamma + \Delta \gamma$, and
$\gamma_{l, B} = \gamma_{g, B} = \gamma - \Delta \gamma$. The
time evolution of the density matrix is then given by a master equation in
Lindblad form,
\begin{equation}
  \imag \frac{d \rho}{d t} = [H_{\mathrm{SSH}}, \rho] + \imag \left(
    \mathcal{D}_{l} \rho + \mathcal{D}_{g} \rho \right),
\end{equation}
with dissipators $\mathcal{D}_{l}$ and $\mathcal{D}_{g}$ that
describe loss and gain, respectively. To state the resulting equation for the
covariance matrix, it is convenient to introduce $2 L$ fermionic operators $c_l$
such that $c_{A, l} = c_{2 l - 1}$ and $c_{B, l} = c_{2 l}$, and to write the
Hamiltonian in the form
\begin{equation}
  \label{eq:H-SSH-H}
  H_{\mathrm{SSH}} = \sum_{l, l' = 1}^{2 L} c_l^{\dagger} H_{l, l'} c_{l'},
\end{equation}
and the jump operators as
\begin{equation}
  L_{l, l} = \sum_{l' = 1}^{2 L} B_{l, l, l'} c_{l'}, \qquad
  L_{g, l} = \sum_{l' = 1}^{2 L} B_{g, l, l'} c_{l'}^{\dagger}.
\end{equation}
In terms of the coefficient matrices $B_{l}$ and $B_{g}$, we
define the bath matrices
\begin{equation}
  \label{eq:M-l-M-g}
  M_{l} = B_{l}^{\dagger} B_{l}, \qquad
  M_{g} = B_{g}^{\transpose} B_{g}^{*}.
\end{equation}
Finally, we write the covariance matrix as
\begin{equation}
  \label{eq:G}
  G_{l, l'} = \langle [c_l, c_{l'}^{\dagger}] \rangle.
\end{equation}
Then, the equation of motion for the covariance matrix reads
\begin{equation}
  \label{eq:G-SSH-eom}
  \frac{d G}{d t} = - \imag
  \left( Z G - G Z^{\dagger} \right) + 2 \left( M_{l} -
    M_{g} \right),
\end{equation}
where $Z = H - \imag \left( M_{l} + M_{g} \right)$. The matrix
$Z$ has PT symmetry, and a fully PT-symmetric phase is realized for sufficiently
small values of $\gamma$ and $\Delta \gamma$~\cite{Starchl2022}. For equal loss
and gain rates, also the bath matrices are identical,
$M_{l} = M_{g}$, and the inhomogeneity in
Eq.~\eqref{eq:G-SSH-eom} vanishes. Consequently, the covariance matrix decays to
zero for $t \to \infty$, which reflects the fact that the steady state has
infinite temperature.

\begin{figure}
  \includegraphics[width=\linewidth]{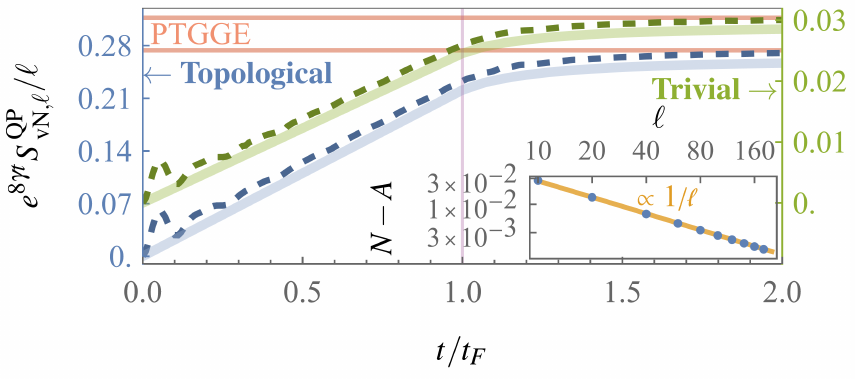}
  \caption{Quasiparticle-pair contribution to the entanglement entropy after
    quenches to the trivial (green, $\Delta J = 0.5 J$) and topological (blue,
    $\Delta J = -0.5J$) PT-symmetric phases of the SSH model for
    $\gamma = \Delta \gamma = 0.1 J$, and $\ell= 20$. The numerical data (dashed
    lines) is close to the analytical prediction (solid lines), which is based
    on a conjecture in analogy to Eq.~(11) of the main
    text~\cite{Starchl2022}. Inset: For the trivial quench and at time
    $t = 2 t_F$, the difference between the numerical data and the
    analytical prediction (blue dots) vanishes as $1/\ell$ (orange line).}
  \label{fig:entropy-SSH}
\end{figure}

In Fig.~\ref{fig:entropy-SSH}, we illustrate relaxation to the PTGGE with the
time evolution of the quasiparticle-pair contribution to the subsystem entropy
following a quench from the ground state with $\Delta J_0 = J$. We consider a
subsystem that consist of $\ell$ unit cells. As shown in the inset of the
figure, the numerical data for the quasiparticle-pair entropy converges with
subsystem size as $1/\ell$ to our analytical prediction. Details of the
underlying analytical and numerical computations will be presented in a
forthcoming publication~\cite{Starchl2022}.

Finally, let us emphasize that, as we have already discussed for the Kitaev
chain, relaxation to the PTGGE does not require the specific forms of the
Hamiltonian and the jump operators given in Eqs.~\eqref{eq:H-SSH}
and~\eqref{eq:L-l-L-g}, respectively. Instead, both can be modified as long as
the requirements (i)--(iii) stated above are satisfied. Therefore, the behavior
shown in Fig.~\ref{fig:entropy-SSH} is representative for a whole class of
models.

\subsection{Bosonic SSH model with loss and gain}
\label{sec:bosonic-ssh-model-loss-gain}

Next, we consider a bosonic version of the SSH model with loss and gain. As
anticipated above, the PT-symmetric relaxation dynamics of noninteracting bosons
is described by a dephased ensemble that differs from the PTGGE for
fermions. The nature and origin of the modifications for bosons can be
understood as follows: The Gaussian states which describe the noninteracting
systems we consider are fully determined by two-point correlations. However, the
decomposition of a two-point function such as
$\langle c_l c_{l'}^{\dagger} \rangle$ into correlation- and statistical
contributions differs for fermions and bosons. In both cases we can write
$c_l c_{l'}^{\dagger} = \left( [c_l, c_{l'}^{\dagger}] + \{ c_l,
  c_{l'}^{\dagger} \} \right) \! /2$.
Fermionic statistics are encoded in the anticommutator,
$\{ c_l, c_{l'}^{\dagger} \} = \delta_{l, l'}$, and correlations in the
expectation value of the commutator which, therefore, features in the covariance
matrix Eq.~\eqref{eq:G}. For the fermionic SSH model, we have chosen equal loss
and gain rates such that the steady state is the infinite-temperature state,
$\rho_{\mathrm{SS}} = \rho_{\infty} = 1/2^{2 L}$. Therefore, taking the
expectation value of any operator in the steady state reduces to taking the
trace of that operator. But the trace of a commutator vanishes, and,
consequently, the covariance matrix vanishes in the steady state,
$\langle [c_l, c_{l'}^{\dagger}] \rangle_{\mathrm{SS}} = \tr \! \left( [c_l,
  c_{l'}^{\dagger}] \right) = 0$.
In the PT-symmetric phase, the decay of the covariance matrix is described by a
single exponential factor, $G(t) = \e^{- 4 \gamma t} G'(t)$, and dephasing among
PT-symmetric modes induces relaxation of $G'(t)$ according to
$G'(t) \to G'_{\mathrm{PTGGE}}$. In contrast, for bosons, statistics are
contained in the commutator, $[a_l, a_{l'}^{\dagger}] = \delta_{l, l'}$, and
correlations are described by the covariance matrix defined as the expectation
value of the anticommutator,
\begin{equation}
  \label{eq:G-bosons}
  G_{l,l'} = \braket{\{a_l,a_{l'}^{\dagger}\}}.
\end{equation}
However, the anticommutator $\{ a_l, a_l^{\dagger} \} = 1 + 2 a_l^{\dagger} a_l$
is positive-definite, and its expectation value is generically nonzero. This
leads to a covariance matrix of the general form
\begin{equation}
  \label{eq:G-bosons-general-solution}
  G(t) = \e^{-2 \delta t} G'(t) + G_{\mathrm{SS}},
\end{equation}
with a nonvanishing steady-state contribution $G_{\mathrm{SS}}$. While the fact
that this constant term is nonzero for any choice of dissipation represents a
fundamental difference to fermionic systems, the structure of the time-dependent
term is the same for both fermions and bosons: PT symmetry leads to decay with a
single exponential factor, and, also for bosons, relaxation of
$G'(t) \to G_{d}'$ is induced by dephasing. Therefore, the late-time
relaxation dynamics of a PT-symmetric bosonic system is described by a dephased
ensemble,
\begin{equation}
  \label{eq:G-d-bosons}
  G(t) \sim G_{d}(t) = \e^{-2 \delta t} G_{d}' +
  G_{\mathrm{SS}}.
\end{equation}
In analogy to the fermionic PTGGE, information about the initial state is
contained in the bosonic dephased ensemble in the term
$G_{d}'$. Additionally, information about the final state enters
through both $G_{d}'$ and $G_{\mathrm{SS}}$.

To illustrate relaxation to this dephased ensemble, we study quench dynamics of
a bosonic SSH model that is described by the Hamiltonian
\begin{equation}
  \label{eq:H-SSH-bosons}
  H_{\mathrm{BSSH}} = \sum_{l = 1}^L \left( J_1 a_{A, l}^{\dagger} a_{B, l} + J_2
    a_{A, l + 1}^{\dagger} a_{B, l} + \hc \right),
\end{equation}
where $a_{s, l}$ and $a_{s, l}^{\dagger}$ with $s \in \{ A, B \}$ are bosonic
annihilation and creation operators, respectively, and obey the canonical
commutation relations
$[ a_{s, l}, a_{s', l'}^{\dagger}] = \delta_{s, s'} \delta_{l,l'}$ and
$[a_{s, l}, a_{s', l'}] = 0$. As in the fermionic case, we consider local loss
and gain, described by the jump operators
\begin{equation}
  L_{l, s, l} = \sqrt{\gamma_{l, s}} a_{s, l}, \qquad
  L_{g, s, l} = \sqrt{\gamma_{g, s}} a_{s, l}^{\dagger}.
\end{equation}
For fermions, we have focused exclusively on the case of equal loss and gain
rates, $\gamma_{l, s} = \gamma_{g, s}$. Here we consider the
more general case of unequal loss and gain rates on both sublattices,
$\gamma_{l/g,A} = \gamma_{l/g} + \Delta
\gamma_{l/g}$
and
$\gamma_{l/g,B} = \gamma_{l/g} - \Delta
\gamma_{l/g}$.
We find it convenient to express the loss and gain rates in terms of the four
parameters $\gamma$, $\delta$, $\Delta \gamma$, and $\Delta$, which are defined
as follows:
\begin{equation}
  \begin{aligned}
    \gamma &= \frac{\gamma_{l} + \gamma_{g}}{2},
    & \delta & = \gamma_{l} - \gamma_{g} \\
    \Delta \gamma & = \frac{\Delta\gamma_{l} +
      \Delta\gamma_{g}}{2}, & \Delta & = \Delta\gamma_{l} -
    \Delta\gamma_{g}.
  \end{aligned}
\end{equation}
The equation of motion for the bosonic covariance matrix defined in
Eq.~\eqref{eq:G-bosons} reads~\cite{Starchl2022}
\begin{equation}
  \label{eq:G-SSH-eom-bosons}
  \frac{dG}{dt} = -\imag\left(Z G - GZ^{\dagger} \right) + 2
  \left( M_{l} + M_{g} \right),
\end{equation}
with $Z = H - \imag \left( M_{l} - M_{g} \right)$ and where the Hamiltonian
matrix $H$ and the bath matrices $M_{l}$ and $M_{g}$ are defined in
Eqs.~\eqref{eq:H-SSH-H} and~\eqref{eq:M-l-M-g}, respectively. The only but
important difference from the fermionic version in Eq.~\eqref{eq:G-SSH-eom} is
the sign of the gain matrix $M_{g}$. Due to this sign change, the inhomogeneity
in Eq.~\eqref{eq:G-SSH-eom-bosons}, which is the sum of two
positive-semidefinite matrices, is always nonzero for nonvanishing
dissipation. In particular, it is nonzero for equal loss and gain rates,
$\delta = \Delta = 0$, when the inhomogeneity in the fermionic equation of
motion~\eqref{eq:G-SSH-eom} vanishes. This reflects the fact that, as explained
above, the covariance matix is generically nonzero in the steady
state. Nevertheless, in the fully PT-symmetric phase, which is realized for
sufficiently small values of loss and gain rates, all eigenvalues $\lambda_l$ of
$Z$ have constant imaginary part, which leads to the general form of the
covariance matrix given in Eq.~\eqref{eq:G-bosons-general-solution}. Dephasing
among modes with different real parts $\Re(\lambda_l)$ induces relaxation to an
ensemble that is determined by $G_{d}(t)$ defined in
Eq.~\eqref{eq:G-d-bosons}. As we proceed to show, relaxation of
$G'(t) \to G_{d}'$ is clearly visible in quasiparticle-pair contribution to the
subsystem entropy.

\begin{figure}
  \includegraphics[width=\linewidth]{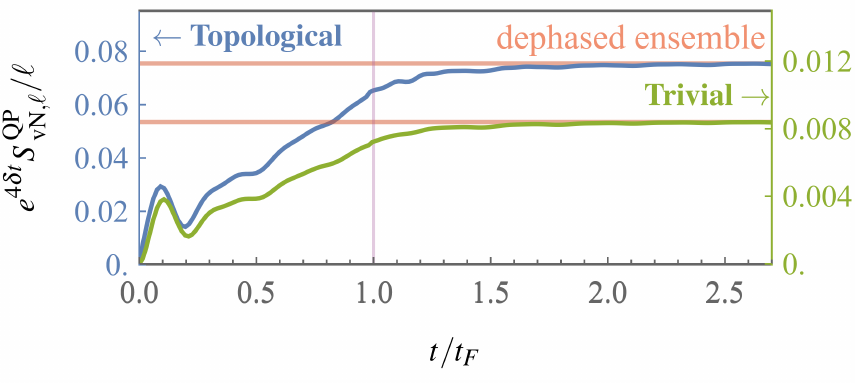}
  \caption{Quasiparticle-pair contribution to the entanglement entropy after
    quenches of the bosonic SSH model with loss and gain to trivial (green,
    $\Delta J = 0.5 J$) and topological (blue $\Delta J = -0.5J$) PT-symmetric
    phases for $\gamma = \Delta \gamma = 0.15 J$, $\delta = \Delta = 0.12 J$ and
    $\ell= 10$. In the initial state in Eq.~\eqref{eq:rho-0}, we have set
    $\mu_A = 1$, $\mu_B = 10$. The red horizontal lines indicate the asymptotic
    prediction given in Eq.~\eqref{eq:S-vN-qp-bosons}.}
  \label{fig:entropy-SSH-bosons}
\end{figure}

We thus consider quench dynamics and choose a mixed initial state $\rho_0$ as
follows: For $J_2 = 0$, the Hamiltonian in Eq.~\eqref{eq:H-SSH-bosons} is
diagonal when it is expressed in terms of bosonic modes $b_{s, l}$ that are
defined as
\begin{equation}
  \begin{pmatrix}
    b_{A, l} \\ b_{B, l}
  \end{pmatrix}
  = \frac{1}{\sqrt{2}}
  \begin{pmatrix}
    1 & 1 \\ 1 & - 1
  \end{pmatrix}
  \begin{pmatrix}
    a_{A, l} \\ a_{B, l}
  \end{pmatrix}.
\end{equation}
The mixed initial state $\rho_0$ is determined by two parameters $\mu_A$ and
$\mu_B$,
\begin{equation}
  \label{eq:rho-0}
  \rho_0 = \frac{1}{Z_0} \e^{- \sum_{s = A, B} \mu_s \sum_{l = 1}^L b_{s,
      l}^{\dagger} b_{s, l}},
\end{equation}
where the normalization $Z_0$ ensures that $\tr(\rho_0) = 1$.

Further, the loss and gain rates are chosen so as to lead to a particularly
simple steady state. The steady-state occupation of a single bosonic lattice
site is determined by the ratio between loss and gain rates. Therefore, to
achieve a spatially uniform distribution of particles, we choose equal ratios on
both sublattices,
$\gamma_{l, A}/\gamma_{g, A} = \gamma_{l,
  B}/\gamma_{g, B}$
or, equivalently, $\gamma/\Delta \gamma = \delta/\Delta$. The resulting uniform
distribution is also stationary under unitary evolution generated by the SSH
Hamiltonian in Eq.~\eqref{eq:H-SSH-bosons}. The covariance in the steady state
is thus given by $G_{\mathrm{SS}} = \left( \delta/\gamma \right) \id$.

As for the fermionic case, the entanglement entropy can be calculated from the
covariance matrix~\cite{Simon1994, Adesso2007}. In particular, for a system
without anomalous correlations, the von Neumann entanglement entropy of a
subsystem that consists of $\ell$ unit cells is given by
\begin{equation}
  S_{\mathrm{vN}, \ell} = \sum_{l = 1}^{2 \ell} S(\xi_l),
\end{equation}
where $\xi_l \geq 1$ are the $2 \ell$ eigenvalues of the covariance matrix
$G_{\ell}$ restricted to a subsystem that consists of $\ell$ unit cells, and
\begin{equation}
  \label{eq:S-bosons}
  S(\xi) = \frac{1 + \xi}{2} \ln \! \left( \frac{1 + \xi}{2} \right) -
  \frac{1 - \xi}{2} \ln \! \left( \frac{1 - \xi}{2} \right).
\end{equation}
The function $S(\xi)$ for bosons differs from the fermionic version given in
Eq.~\eqref{eq:S} only by the sign of the first term. A procedure similar to that
described in Sec.~\ref{sec:subsystem-entropy} for fermions leads to the
following result for the late-time asymptotics of the quasiparticle-pair
contribution to the subsystem entropy:
\begin{multline}
  \label{eq:S-vN-qp-bosons}
  S_{\mathrm{vN}, \ell}^{\mathrm{QP}}(t) \sim \frac{S_2}{2} \e^{- 4 \delta t}
  \left[ \sum_{l = 1}^{2 \ell} \xi_{d, l}^{\prime 2} - 2 \ell
    \int_{-\pi}^{\pi} \frac{d k}{2 \pi} \left( g_{\id, k}^{\prime 2} +
      \xi_k^{\prime 2} \right)_{d} \right],
\end{multline}
where $S_2 = d^2 S/d \xi^2$ is the second derivative of Eq.~\eqref{eq:S-bosons},
evaluated at $\xi = \delta/\gamma$, and $\xi_{d, l}'$ are the eigenvalues of
$G_{d}'$ in Eq.~\eqref{eq:G-d-bosons}. Further, $G'(t)$ in
Eq.~\eqref{eq:G-bosons-general-solution} is a block Toeplitz matrix with
$2 \times 2$ blocks $g_l'(t)$, which we decompose in momentum space as
$g_k'(t) = g_{\id, k}'(t) \id + \mathbf{g}_k'(t) \cdot
\boldsymbol{\sigma}$.
This decomposition defines both $g_{\id, k}'(t)$ and
$\xi_k'(t) = \abs{\mathbf{g}_k'(t)}$. As in Eq.~\eqref{eq:S-vN-qp-late-time},
the subscript ``\textit{d}'' in the last term in Eq.~\eqref{eq:S-vN-qp-bosons}
indicates that oscillating terms should be dropped after taking the square.

In Fig.~\ref{fig:entropy-SSH-bosons}, the quasiparticle-pair contribution to the
entanglement entropy is shown for quenches to the trivial and topological
PT-symmetric phases. The characteristic signatures of the quasiparticle picture
in the fermionic model transfer well to the bosonic case, consequently
reinforcing particular universal aspects of a PT-symmetric ensemble
description. Accordingly, after rescaling with $\e^{4 \delta t}$, the bosonic
entropy shows initial ballistic growth, followed by saturation roughly at the
characteristic Fermi time $t_F$, which depends on the modified
quasiparticle velocity. The late-time asymptotic saturation of the entropy is
well predicted by the dephased ensemble through Eq.~\eqref{eq:S-vN-qp-bosons}.

\subsection{Dimerized XY model with dephasing}
\label{sec:dimerized-xy-model-dephasing}

In the study of quench dynamics of isolated integrable systems, an important
role is played by one-dimensional spin chains, or, equivalently, lattice models
of hard-core bosons. Paradigmatic examples are the XX model~\cite{Rigol2007,
  Calabrese2011, Calabrese2012I, Calabrese2012II, Vidmar2016} and the transverse
field Ising model~\cite{Rigol2007, Calabrese2011, Calabrese2012I,
  Calabrese2012II, Vidmar2016}. For these models of interacting spins or bosons,
analytical progress is facilitated by mapping them to noninteracting fermions
through the Jordan-Wigner transformation. For example, pertinent GGEs can be
expressed in terms of occupation numbers of free fermionic modes. In contrast,
for interacting systems that cannot be mapped to noninteracting systems, the
construction of the GGE is much more complicated~\cite{Essler2016}. Therefore,
to address the question whether our results for PT-symmetric relaxation of
driven-dissipative integrable systems also apply to interacting systems,
attempting to generalize these results to interacting systems that can be mapped
to noninteracting fermions~\cite{Vidmar2016} presents itself as a natural first
step. However, applying the Jordan-Wigner transformation to driven-dissipative
spin chains with jump operators that are linear in spin operators does not lead
to noninteracting fermionic models with jump operators that are linear in
fermionic operators. Instead, the Jordan-Wigner transformation in
Eq.~\eqref{eq:Jordan-Wigner} maps spin flips corresponding to the jump operators
$\sigma_l^{\pm}$ to nonlocal products of fermionic operators. Dephasing, which
is described by local Hermitian jump operators,
\begin{equation}
  \label{eq:L-dephasing-spins}
  L_l = \frac{\sqrt{\gamma}}{2} \sigma^z_l,
\end{equation}
takes a much simpler but still quadratic form in fermionic operators. Thus, in
the following, we will study quench dynamics of a dimerized XY spin
chain~\cite{Giamarchi2004} with dephasing as a canonical and experimentally
relevant form of dissipation, which occurs naturally and can also be induced
deliberately, e.g., in quantum simulations of the XY model with trapped
ions~\cite{Monroe2021}. We emphasize that there is a fundamental difference
between master equations with linear and quadratic Hermitian jump operators such
as the driven-dissipative Kitaev chain and the XY model with dephasing,
respectively: In the former case, the Gaussianty of an initially Gaussian state
is preserved; in the latter case, an initially Gaussian state evolves into a
mixture of Gaussian states~\cite{Horstmann2013}, and, therefore, nontrivial
higher-order connected correlations are generated. Nevertheless, a master
equation with quadratic Hamiltonians and quadratic Hermitian jump operators
still leads to a closed equation of motion for the covariance matrix. As we will
show, the relaxation of the covariance matrix---and, therefore, of all spin
observables that can be expressed in terms of the covariance matrix---is
described by a PTGGE even though the state of the system is not Gaussian.

The dimerized XY spin chain~\cite{Giamarchi2004} is described by the following
Hamiltonian:
\begin{equation}
  \label{eq:H-XY}
  H_{\mathrm{XY}} = \frac{1}{2} \sum_{l = 1}^{2 L} \left( J - \left( -1 \right)^l \Delta J
  \right) \left( \sigma^x_l \sigma^x_{l + 1} + \sigma^y_l \sigma^y_{l + 1}
  \right),
\end{equation}
where $\sigma_l^{x, y, z}$ are Pauli matrices that act on the spin at site $l$
with $l \in \{ 1, \dotsc, 2 L \}$. We choose the spatially uniform component of
the XY couplings as antiferromagnetic, $J > 0$. The spin-Peierls term with
dimerization strength $\Delta J$ describes a modulation of the XY couplings
across successive bonds between the values
$J \pm \Delta J$~\cite{Giamarchi2004}. This modulation results in dimer order in
the ground state, which is detected by a nonvanishing expectation value of the
operator~\cite{Haldane1982, *Haldane1982a}
\begin{equation}
  \label{eq:d-l}
  d_l = \sigma^+_l \sigma^-_{l + 1} - \sigma^+_{l - 1} \sigma^-_l + \hc
\end{equation}
In particular, for $\Delta J = J$, the spin chain is fully dimerized and the
ground state is given by a product of singlets across the odd bonds, which leads
to the dimerization order parameter taking the value
$\langle d_{2 l - 1} \rangle_0 = - 1 = - \langle d_{2 l} \rangle_0$.

With dephasing as given in Eq.~\eqref{eq:L-dephasing-spins}, the master equation
for the XY chain reads
\begin{equation}
  \label{eq:master-equation-XY}
  \imag \frac{d \rho}{d t} = \mathcal{L}_{\mathrm{XY}} \rho =
  [H_{\mathrm{XY}}, \rho] - \imag \sum_{l = 1}^{2 L} [L_l, [L_l, \rho]].
\end{equation}
Since the jump operators are Hermitian, the steady state is again at infinite
temperature. Further, the Liouvillian $\mathcal{L}_{\mathrm{XY}}$ of the XY
model with dephasing has PT symmetry~\cite{Prosen2012a}.

To numerically study quench dynamics, we switch to a representation of the XY
model in terms of fermions. This representation can be obtained through the
Jordan-Wigner transformation in Eq.~\eqref{eq:Jordan-Wigner}, which maps the XY
Hamiltonian in Eq.~\eqref{eq:H-XY} to
\begin{equation}
  H_{\mathrm{XY}} = - \sum_{l = 1}^{2 L} \left( J - \left( -1 \right)^l
    \Delta J \right) \left( c_l^{\dagger} c_{l + 1} + \hc \right).
\end{equation}
That is, the Hamiltonian takes the form of the SSH model in
Eq.~\eqref{eq:H-SSH}, with hopping amplitudes
$J_{1, 2} = - \left( J \pm \Delta J \right).$
Further, the jump operators for dephasing in Eq.~\eqref{eq:L-dephasing-spins}
are quadratic in fermionic operators,
\begin{equation}
  \label{eq:L-dephasing}
  L_l = \sqrt{\gamma} c_l^{\dagger} c_l,
\end{equation}
where we omit a constant contribution that cancels due to the commutator
structure in Eq.~\eqref{eq:master-equation-XY}.

\begin{figure}
  \includegraphics[width=\linewidth]{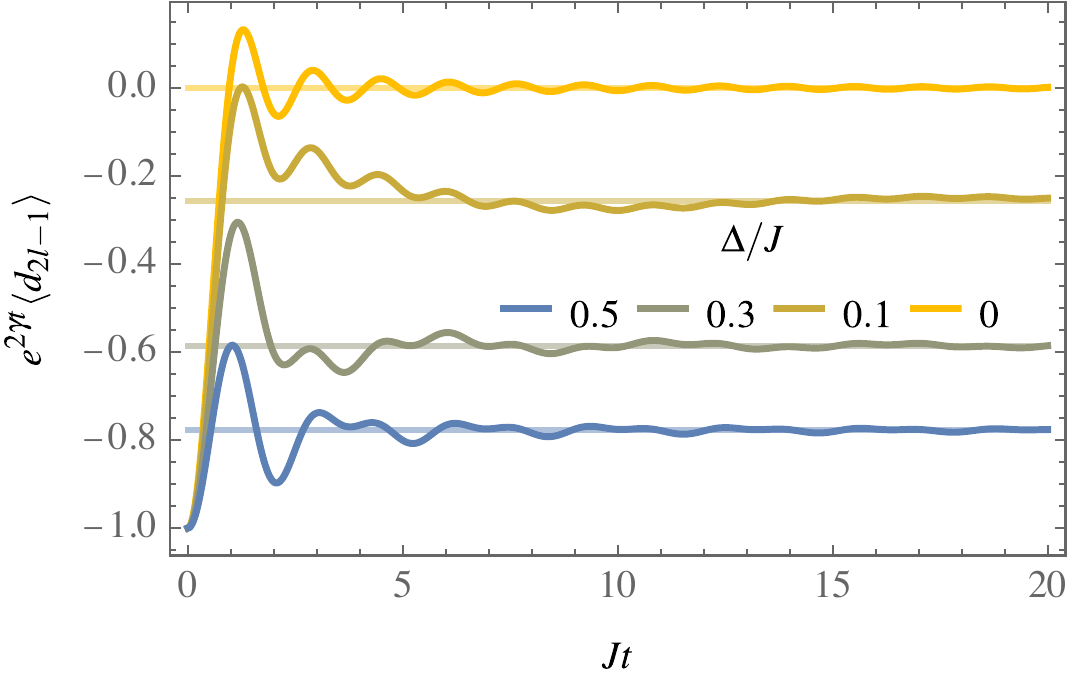}
  \caption{Dimerization order parameter $\langle d_{2 l - 1} \rangle$ across odd
    bonds. The initial value for a fully dimerized chain is
    $\langle d_{2 l - 1} \rangle_0 = - 1$. After rescaling with
    $\e^{2 \gamma t}$, the dimerization order parameter approaches a stationary
    value that is determined by the PTGGE and indicated by horizontal lines in
    the figure.}
  \label{fig:dimerization-XY}
\end{figure}

In analogy to Eq.~\eqref{eq:G} for the SSH model with linear dissipation, we
define the covariance matrix as
\begin{equation}
  G_{l, l'} = \langle [c_l, c_{l'}^{\dagger}] \rangle,
\end{equation}
where $l, l' \in \{ 1, \dotsc, 2 L \}$. To formulate the equation of motion for
the covariance matrix, we write the Hamiltonian and the jump operators in the
form
\begin{equation}
  H_{\mathrm{XY}} = \sum_{l, l' = 1}^{2 L} c_l^{\dagger} H_{l, l'} c_{l'}, \quad
  L_l = \sum_{m, m' = 1}^{2 L} c_m^{\dagger} B_{l, m, m'} c_{m'},
\end{equation}
where, for dephasing as described in Eq.~\eqref{eq:L-dephasing}, the elements of
the matrix $B_l$ are given by
$B_{l, m, m'} = \sqrt{\gamma} \delta_{l, m} \delta_{l, m'}$. The covariance
matrix evolves in time according to
\begin{equation}
  \label{eq:equation-of-motion-covariance-matrix-quadratic-jump-ops}
  \frac{d G}{d t} = - \imag [H, G] - \sum_{l = 1}^L [B_l,
  [B_l, G]].
\end{equation}
To illustrate the effect of dephasing most transparently, we decompose the
covariance matrix into diagonal and offdiagonal elements as
$G = G_{\mathrm{OD}} + G_{\mathrm{OD}}$. Then,
Eq.~\eqref{eq:equation-of-motion-covariance-matrix-quadratic-jump-ops} reduces
to
\begin{equation}
  \frac{d G}{d t} = - \imag [H, G] - 2 \gamma
  G_{\mathrm{OD}},
\end{equation}
which shows that dephasing affects only offdiagonal elements. This equation can
be simplified further by taking advantage of the inversion symmetry of the SSH
model: One can show that for translationally invariant and inversion-symmetric
initial states, the diagonal component $G_{\mathrm{D}}$ vanishes at all
times~\cite{Starchl2022}. This applies, in particular, to the ground state of
the SSH Hamiltonian. Then, with $G = G_{\mathrm{OD}}$, we obtain
\begin{equation}
  \frac{d G}{d t} = - \imag \left( Z G - G Z^{\dagger}
  \right),
\end{equation}
where $Z = H - \imag \gamma \id$; that is, time evolution with dephasing differs
from the dynamics of isolated system only by an overall decay of $G$ at the rate
$2 \gamma$. Consequently, the PTGGE which describes relaxation of the covariance
matrix and, therefore, of all expectation values of spin observables that can be
expressed in terms of the covariance matrix, is obtained simply by multiplying
the Lagrange multipliers in the GGE by $\e^{- 2 \gamma t}$. In contrast to the
combination of loss and gain discussed in
Sec.~\ref{sec:fermionic-ssh-model-loss-gain}, the dynamics and statistics of
elementary excitations are not affected by dephasing.

\begin{figure}
  \includegraphics[width=\linewidth]{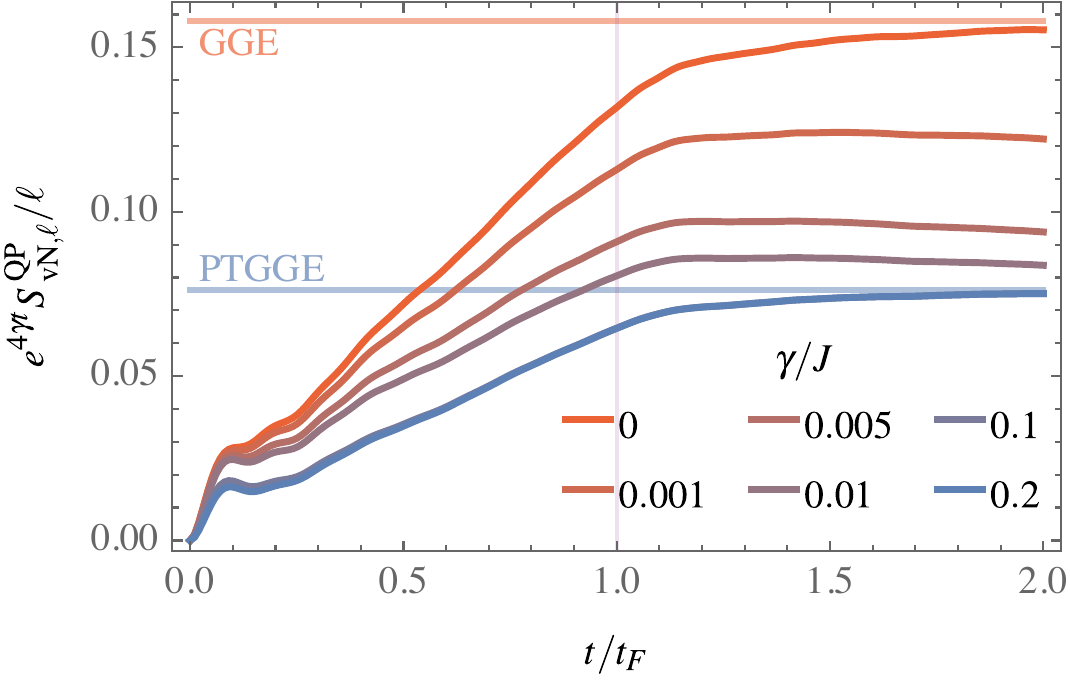}
  \caption{Quasiparticle-pair contribution to the subsystem entropy, calculated
    from the covariance matrix and thus neglecting deviations from Gaussianity,
    after quenches of the dimerized XY model with dephasing for
    $\Delta J = 0.3 J$ and and $\ell = 20$. For $\gamma = 0$, the
    quasiparticle-pair entropy reduces to the entanglement entropy, and its
    stationary value is determined by the conventional GGE; for any
    $\gamma > 0$, the rescaled quasiparticle-pair entropy approaches a
    stationary value that is independent of $\gamma$ and determined by the
    PTGGE. The PTGGE prediction is reached at $\gamma t \gg 1$ for very small
    values of $\gamma$ with $\gamma < 1/t_F \approx 0.07 J$. In
    contrast, for $\gamma > 1/t_F$, the curves collapse, and converge
    towards the PTGGE value at $t \gtrsim t_F$. The value of
    $t_F$ does not depend on $\gamma$, which reflects the fact that
    dephasing does not modify the dynamics of quasiparticle excitations.}
  \label{fig:entropy-XY}
\end{figure}

To illustrate the resulting dynamics, we consider quenches from the ground state
for $\Delta J_0 = J$ with covariance matrix
\begin{equation}
  G_0 = \bigoplus_{l = 1}^L \sigma_x,
\end{equation}
to postquench values $\Delta J \neq \Delta J_0$. Relaxation of the covariance
matrix to the PTGGE is shown in Fig.~\ref{fig:dimerization-XY} with the time
evolution of the dimerization order parameter Eq.~\eqref{eq:d-l}, which can be
expressed as $\langle d_l \rangle = \Re( G_{l - 1, l} - G_{l, l + 1})$. As
stated below Eq.~\eqref{eq:d-l}, the dimerization order parameter takes the
value $\langle d_{2 l - 1} \rangle_0 = - 1$ in the initial ground state. Due to
dephasing, it decays exponentially, but after rescaling with $\e^{2 \gamma t}$,
it approaches a stationary value as predicted by the PTGGE.

Since the state of the system after the quench is not Gaussian, the entanglement
entropy is not fully determined by the covariance matrix. However, it is still
informative to neglect deviations from Gaussianity and to study the time
evolution of the quasiparticle-pair entropy computed from the covariance matrix
alone~\cite{Alba2021}. In Fig.~\ref{fig:entropy-XY}, we show the time evolution
of the rescaled quasiparticle-pair entropy, which is defined here as
$\e^{4 \gamma t} S_{\mathrm{vN}, \ell}^{\mathrm{QP}}(t)$ and can be calculated
from the covariance matrix as outlined in
Sec.~\ref{sec:fermionic-ssh-model-loss-gain}. The fact that dephasing does not
affect the dynamics and statistics of quasiparticle excitations has two notable
consequences that can be seen in the figure: First, the PTGGE prediction for the
rescaled quasiparticle-pair entropy is independent of the value of $\gamma$; and
second, also the Fermi time, which marks the transition from the initial linear
increase of the rescaled entropy to the much slower approach to the stationary
value, does not depend on $\gamma$.

\appendix*

\section{Time evolution induced by a non-Hermitian $2 \times 2$ Hamiltonian}
\label{sec:time-evol-nH}

In this short technical appendix, we summarize some relations that allow us to
determine the time evolution of the components of the covariance matrix in
Eq.~\eqref{eq:gamma-1-k-gamma-2-k}. We consider a non-Hermitian $2 \times 2$
Hamiltonian $x$ that is given by
\begin{equation}
  x = x_{\id} \id + \mathbf{x} \cdot \boldsymbol{\sigma}, \qquad \mathbf{x} =
  \mathbf{a} + \imag \mathbf{b},
\end{equation}
where $\mathbf{a}, \mathbf{b} \in \R^3$. Further, we define
\begin{equation}
  \omega = \sqrt{\mathbf{x} \cdot \mathbf{x}} = \sqrt{\varepsilon^2 + \imag 2 \mathbf{a}
    \cdot \mathbf{b} - b^2},
\end{equation}
where $\varepsilon = \abs{\mathbf{a}}$, and $b = \abs{\mathbf{b}}$. In general,
$\omega \in \C$. Finally, we define a vector $\unitvec{x} =
\mathbf{x}/\omega$.
In terms of these quantities, the exponential of $x$ an be written as
\begin{equation}
  \e^{-\imag x t} = \e^{-\imag x_{\id} t} \left( \cos(\omega t) \id - \imag
    \sin(\omega t) \unitvec{x} \cdot \boldsymbol{\sigma} \right).
\end{equation}
Using this relation, and with
\begin{equation}
  \boldsymbol{\sigma} \left( \mathbf{x} \cdot \boldsymbol{\sigma} \right) =
  \mathbf{x} \id + \imag \mathbf{x} \times \boldsymbol{\sigma},
\end{equation}
it is straightforward to show that
\begin{multline}
  \boldsymbol{\tau} = \e^{- \imag x t} \boldsymbol{\sigma} \e^{\imag x^{\dagger}
    t} = \e^{2 \Im(x_{\id}) t} \left\{ \abs{\cos(\omega t)}^2 \boldsymbol{\sigma}
  \right. \\ - \imag \sin(\omega t) \cos(\omega^{*} t) \left(
    \unitvec{x} - \imag \unitvec{x} \times \boldsymbol{\sigma} \right) \\
  + \imag \sin(\omega^{*} t) \cos(\omega t) \left( \unitvec{x}^{*} + \imag
    \unitvec{x}^{*} \times \boldsymbol{\sigma} \right) \\ \left.  +
    \abs{\sin(\omega t)}^2 \left[ \left( \unitvec{x} \cdot \boldsymbol{\sigma}
      \right) \unitvec{x}^{*} + \unitvec{x} \left( \unitvec{x}^{*} \cdot
        \boldsymbol{\sigma} \right) - \imag \unitvec{x} \times \unitvec{x}^{*} -
      \left( \unitvec{x} \cdot \unitvec{x}^{*} \right) \boldsymbol{\sigma}
    \right] \right\}.
\end{multline}
Next, we consider the scalar product of $\boldsymbol{\tau}$ and
$\mathbf{a}_0 \in \R^3$, which we denote by $\gamma$:
\begin{equation}
  \gamma = \mathbf{a}_0 \cdot \boldsymbol{\tau} = \gamma_{\id} \id +
  \boldsymbol{\gamma} \cdot \boldsymbol{\sigma}.
\end{equation}
We obtain
\begin{multline}
  \gamma_{\id} = 2 \e^{2 \Im(x_{\id}) t} \left[ \vphantom{\abs{\frac{\sin(\omega
          t)}{\omega}}^2} \Im \! \left( \sin(\omega t) \cos(\omega^{*} t)
      \mathbf{a}_0 \cdot \unitvec{x} \right) \right. \\ \left. -
    \abs{\frac{\sin(\omega t)}{\omega}}^2 \mathbf{a}_0 \cdot \left( \mathbf{a}
      \times \mathbf{b} \right) \right],
\end{multline}
and
\begin{multline}
  \boldsymbol{\gamma} = \e^{2 \Im(x_{\id}) t} \left[ \left( \abs{\cos(\omega t)}^2 -
      \abs{\sin(\omega t)}^2 \unitvec{x} \cdot \unitvec{x}^{*} \right)
    \mathbf{a}_0 \right. \\ - 2 \Re \! \left( \sin(\omega t) \cos(\omega^{*} t)
    \mathbf{a}_0 \times \unitvec{x} \right) \\
  \left. + 2 \abs{\sin(\omega t)}^2 \Re \! \left( \left(
        \mathbf{a}_0 \cdot \unitvec{x} \right) \unitvec{x}^{*} \right) \right].
\end{multline}
In the main text, we are primarily concerned with the case
$\mathbf{a} \cdot \mathbf{b} = 0$ and $\varepsilon > b$, such that
\begin{equation}
  \omega = \sqrt{\varepsilon^2 - b^2} \in \R_{>0},
\end{equation}
and $\mathbf{a}_0 \cdot \mathbf{b} = 0$. These conditions lead to the following
simplifications of the above expressions for $\gamma_{\id}$ and
$\boldsymbol{\gamma}$:
\begin{equation}
  \label{eq:gamma-id-PT-symmetric}
  \gamma_{\id} = - \frac{\e^{2 \Im(x_{\id}) t}}{\omega^2} \left( 1 - \cos(2 \omega t) \right)
  \mathbf{a}_0 \cdot \left( \mathbf{a} \times \mathbf{b} \right),
\end{equation}
and
\begin{multline}
  \label{eq:gamma-vector-PT-symmetric}
  \boldsymbol{\gamma} = \e^{2 \Im(x_{\id}) t} \\ \times \left[ \mathbf{a}_0 -
    \frac{\varepsilon^2}{\omega^2} \left( 1 - \cos(2 \omega t) \right)
    \mathbf{a}_{\perp} + \frac{\varepsilon}{\omega} \sin(2 \omega t)
    \mathbf{a}_o \right],
\end{multline}
where, with $\unitvec{a} = \mathbf{a}/\varepsilon$, the parallel, perpendicular, and
out-of-plane components are given by
\begin{equation}
  \mathbf{a}_{\parallel} = \left( \mathbf{a}_0 \cdot
    \unitvec{a} \right) \unitvec{a}, \qquad \mathbf{a}_{\perp} =
  \mathbf{a}_0 - \mathbf{a}_{\parallel}, \qquad \mathbf{a}_o =
  - \mathbf{a}_0 \times \unitvec{a}.
\end{equation}
Finally, we calculate the norm of the vector $\boldsymbol{\gamma}$, which we
denote by $\xi = \abs{\boldsymbol{\gamma}}$. To that end, we use that
$\mathbf{a}_{\parallel}$, $\mathbf{a}_{\perp}$, and $\mathbf{a}_o$
are mutually orthogonal, and that
\begin{equation}
  \mathbf{a}_0 \cdot \mathbf{a}_{\perp} = \abs{\mathbf{a}_{\perp}}^2 =
  \abs{\mathbf{a}_o}^2 = a_0^2 - \left( \mathbf{a}_0 \cdot
    \unitvec{a} \right)^2.
\end{equation}
We further assume that $\mathbf{a}_0 = \unitvec{a}_0$ is a unit vector, such
that $a_0 = \abs{\mathbf{a}_0} = 1$. Then,
\begin{equation}
  \unitvec{a}_0 \cdot \mathbf{a}_{\perp} = \abs{\mathbf{a}_{\perp}}^2 =
  \abs{\mathbf{a}_o}^2 = 1 - \left( \unitvec{a}_0 \cdot
    \unitvec{a} \right)^2 = \sin(\Delta \theta)^2,
\end{equation}
where $\Delta \theta$ is the angle between $\unitvec{a}_0$ and $\unitvec{a}$,
\begin{equation}
  \cos(\Delta \theta) = \unitvec{a}_0 \cdot \unitvec{a},
\end{equation}
and we obtain
\begin{multline}
  \label{eq:xi--}
  \xi^2 = \e^{4 \Im(x_{\id}) t} \left\{ 1 - \frac{\varepsilon^2}{\omega^2}
    \sin(\Delta \theta)^2 \left[ 2 \left( 1 - \cos(2 \omega t) \right)
      \vphantom{\frac{\varepsilon^2}{\omega^2}} \right. \right. \\
  \left. \left. - \frac{\varepsilon^2}{\omega^2} \left( 1 - \cos(2 \omega t)
      \right)^2 - \sin(2 \omega t)^2 \right] \right\}.
\end{multline}


%